\documentclass[a4paper,11pt]{article}
\usepackage{jheppub}
\usepackage{mathrsfs}
\usepackage{amsfonts}
\usepackage{setspace}
\usepackage{cellspace}
\usepackage{amsmath,amssymb,bm}
\usepackage[colorlinks=true,linkcolor=blue]{hyperref}
\usepackage{xcolor}
\usepackage{epsfig}
\usepackage{slashed}
\usepackage{caption}
\usepackage{hhline,multirow,tabularx}  
\usepackage{dcolumn}    
\usepackage{url}        
\usepackage{braket} 
\definecolor{cyan}{rgb}{0.0, 1.0, 1.0}
\definecolor{applegreen}{rgb}{0.55, 0.71, 0.0}
\definecolor{arylideyellow}{rgb}{0.91, 0.84, 0.42}
\definecolor{bananayellow}{rgb}{1.0, 0.88, 0.21}
\definecolor{burlywood}{rgb}{0.87, 0.72, 0.53}
\definecolor{buff}{rgb}{0.94, 0.86, 0.51}
\definecolor{blond}{rgb}{0.98, 0.94, 0.75}
\definecolor{bisque}{rgb}{1.0, 0.89, 0.77}
\definecolor{bananamania}{rgb}{0.98, 0.91, 0.71}
\definecolor{apricot}{rgb}{0.98, 0.81, 0.69}
\definecolor{almond}{rgb}{0.94, 0.87, 0.8}

\title{Renormalon-like factorial enhancements to power expansion/OPE in a super-renormalizable 2D $O(N)$ quartic model}
\author[a]{Yizhuang Liu}

\affiliation[a]{Institute of Theoretical Physics,
Jagiellonian University, 30-348 Kraków, Poland}

\emailAdd{yizhuang.liu@uj.edu.pl}

\abstract {In this work, we investigate the power-expansion/OPE in super-renormalizable QFTs. In comparison to ``marginal'' short distance asymptotics, the power-expansion in super-renormalizable theory has a higher level of correlation with the coupling constant expansion, and the fixed-point perturbation theory for the coefficient functions suffers increasing level of IR divergences when the order of power-expansion/coupling constant expansion increases. This increasing IR sensitivity anticipates the presence of vacuum condensates of non-trivial local operators in the OPE, which not only cancel the IR divergences for coefficient functions through their UV divergences, but also shape the structure of logarithms at each power in the power-expansion. This UV-IR correspondence has been used to argue for the renormalon cancellation between coefficient functions and operators in marginal asymptotics, where the residue IR sensitivity of coefficient functions to be canceled by the operators manifests through high-order factorial growths after renormalization subtraction. 

However, in supper-renomalizable theory, high-order behavior in coupling constants in the coefficient functions is simultaneously large-order behavior in the power expansion, and one may wonder if there are also factorial growths for high-order/power terms in the coefficient functions, as they also require large-numbers of IR subtractions. In this work, we address the issue regarding the asymptotic behavior of power-expansion/OPE in supper-renormalizable theory. Using an $O(N)$ quartic model at the next-to-leading order in the large-$N$ expansion, we show that there are indeed factorial enhancements to high power terms in the coefficient functions. In the fixed-point perturbation theory, the factorial enhancements are introduced by the IR subtractions. Moreover, there are also factorial enhancements to the operator condensates, and the factorial enhancements cancel between coefficient functions and operators only {\it off-diagonally} across different powers.  The observed factorial enhancements imply the divergence of the momentum-space power expansion.  }
\date{\today}

\begin{document}
\maketitle
\flushbottom
\section{Introduction}
Operator product expansion (OPE)~\cite{Wilson:1970ag} is among one of the major properties concerning short distance behavior of correlation functions in local QFTs. In the ``simplest'' local QFTs, the conformal field theories (CFTs), the existence of OPE in the sense of asymptotic expansion is sufficient to imply the absolute convergence of OPE in the coordinate-space analyticity region~\cite{Luscher:1974ez,Mack:1976pa,Kravchuk:2021kwe}. In massive QFTs with CFTs as their short-distance fixed-points, on the other hand, although still exist at worst in the sense of asymptotic expansion in the Euclidean region, the behavior of OPE is much less controlled due to the existence of explicit mass scales. Not only derivative operators are much harder to be resumed, the coefficient functions for individual operators are no longer homogeneous functions and will contain non-trivial $z^2m^2$ dependencies~\cite{Novikov:1984rf,Zamolodchikov:1990bk,Lukyanov:1996jj} that couple the short distance and large distance scales. 

In super-renormalizable QFTs, namely, QFTs that are obtained by perturbing the short-distance fixed points through {\it relevant} operators, the $z^2m^2$ dependencies in the coefficient functions of {\it short-distance-scheme operators} can still be separated into well-defined powers and calculated power by power in fixed-point perturbation theory (also called CFT perturbation theory) with appropriate IR subtractions~\cite{Chetyrkin:1982zq,Chetyrkin:1983qlc,Zamolodchikov:1990bk}. If the degrees of IR divergences are non-integer valued, then the IR subtractions could be performed by analytic continuations without introducing actual counter terms~\cite{Zamolodchikov:1990bk}. However, when the degrees are integer-valued (called ``resonance situation'' in~\cite{Zamolodchikov:1990bk}), there could be logarithmic singularities that can not be avoided by analytic continuation. In these cases, IR subtractions in the coefficient functions are mandatory and introduce scale-scheme dependencies to the coefficient functions.  Such scale-scheme dependencies are then canceled power-wisely in the OPE, by the scale-scheme dependencies introduced to the operators during there logarithmic UV renormalization. The net effect is the appearance of logarithmic terms in the power expansion that couple the large and short distance scales, in addition to purely powers. And the numbers in the logarithms usually increase when the power increases. 

The role of logarithms in renormalizable QFTs had attracted attentions since early days of QFTs, since in merely renormalizable perturbative expansions, the logarithms are not suppressed by powers and could even affect the well-posedness of the theory~\cite{PhysRev.95.1300}. It was later realized~\cite{PhysRevD.2.1541,PhysRevLett.30.1343,PhysRevLett.30.1346} how such logarithms could be absorbed into running couplings and how the renormalizable perturbative expansions could also be assigned non-perturbative meanings~\cite{PhysRevLett.34.1227,Gawedzki:1984pj,Gawedzki:1985ic,Gawedzki:1985ez,Feldman:1986ax,Slade_2015}. In particular, in asymptotically-free QFTs, it is commonly accepted that Schwinger functions at short distances allow OPE, with perturbative asymptotic expansions for the coefficient functions at (inverse) logarithmic levels. In such cases, the asymptotic behavior of perturbative expansions are known to be factorially affected by the IR-renormalons~\cite{tHooft:1977xjm,Parisi:1978az,Parisi:1978bj} caused by running-coupling-type logarithms in the integrands. Moreover, it has long been pointed out that the IR renormalons anticipated the existences of non-trivial operator condensates in the OPE~\cite{Parisi:1978az,Novikov:1980uj, Shifman:1978bx,David:1983gz,David:1985xj, Novikov:1984rf,Mueller:1984vh,Ji:1994md,Beneke:1998ui,Beneke:1998eq,Braun:2004bu,Shifman:2013uka,Dunne:2015eoa}: IR renormalons of coefficient functions at lower powers are canceled by UV renormalons of operators at higher powers. And the evidences for such cancellation pattern are enormous, ranging from numerical lattice perturbation theory~\cite{Bauer:2011ws,Bali:2014fea} to explicit bubble-chain calculations and 2D large-$N$ computations~\cite{David:1982qv,David:1983gz,David:1985xj,Beneke:1998eq,Braun:2004bu,Marino:2024uco,Liu:2024omb}. 

It is in fact not very hard to see, that the phenomenon of renormalon cancellation in QCD-like theories is similar to the cancellation of logarithmic singularities between coefficient functions and operators in super-renormalizable theories: in these theories, interactions only generate operator mixing from high-dimensional to lower-dimensional operators multiplied by positive mass powers. Thus, the logarithmic cancellations are not between an operator and its own coefficient function, but usually between an operator and higher-power terms in the coefficient functions of lower dimensional operators. This fact has been noticed before~\cite{Novikov:1984rf} and becomes a source of intuition for the renormalon cancellation pattern in QCD-like theories. On the other hand, another essential feature of QCD-like renormalons: factorial enhancements generated by momentum integrations due to logarithms in the integrands, were less well known to be existing in super-renormalizable QFTs as well.  A renormalon has previously been found~\cite{Marino:2019fvu} in the coupling constant expansion of free-energy in a 2D $O(N)$ quartic model, due to the ``massless particles'' after expanding around the false vacuum in the negative mass formulation. But in the context of OPE/power expansion,  neither the important role played by the increasing numbers of logarithms nor the possible existence of renormalon-like factorial enhancements were seriously investigated in super-renormalizable QFTs.

It is the task of this work to investigate the issue of asymptotic behavior of power expansion/OPE in a super-renormalizable 2D QFT that is populated by logarithms that can be further amplified by bubble-chains. 
We performed a careful investigation of the large $p^2$ expansion of a scalar-scalar two-point function, in the same large-$N$ $O(N)$ quartic model as in~\cite{Marino:2019fvu}
at the next-to-leading order in the large-$N$ expansion. 
We show that because the large-$p^2$ logarithms of the individual bubbles can be amplified by bubble-chains, there are 
factorial enhancements to the power expansion. More precisely, at the $n$-th power in the large $p^2$ expansion, a typical contribution to the coefficient function of the identity operator takes the following form (see Eqs.~(\ref{eq:hardMellin}, \ref{eq:hard00mellin}) for more details)
\begin{align}
{\cal H}_{0,0,s_1=n}(p^2)\sim \frac{1}{(p^2)^n}\left(\frac{g}{4\pi}\right)^{n+1} \frac{d^n}{dt^n} \bigg[{\color{red}\left(\frac{m^2}{p^2}\right)^{t}} \psi(1-n-t)-\frac{1}{t}\bigg]_{t=0} \ .
\end{align}
As such, due to the high-order derivatives, when $n$ becomes large, the $t=\pm1,\pm2$ poles of the digamma function lead to factorial enhancements that are {\it power changing}
\begin{align}
{\cal H}_{0,0,s_1=n}(p^2)\bigg|_{n\rightarrow \infty}\rightarrow -n!\frac{g^{n+1}}{(4\pi)^{n+1}}\frac{m^2}{(p^2)^{n+1}}+n!\frac{(-1)^ng^{n+1}}{(4\pi)^{n+1}}\frac{1}{m^2(p^2)^{n-1}}+ ...\ .
\end{align}
We show that such factorial enhancements are canceled between the coefficient functions and operators only off-diagonally across different powers. Restricted to any given power, the factorial enhancements are no-longer canceled.  {\it The large-$p^2$ power expansion is divergent}.  To make the OPE convergent, one must subtract all the factorial growths from operators and add back to coefficient functions. This requires infinitely many redefinitions of operators and can not be achieved through a single ``Borel prescription''. On the other hand, in the coordinate space expansion, due to the overall $\frac{1}{(n!)^2}$ suppression, the $n!$ introduced by the IR subtraction is insufficient to modify the convergence properties of the power-expansion/OPE.

The paper is organized as follows. \begin{enumerate}
    \item In the Sec.~\ref{sec:OPEintro}, we provide a worked example of the OPE calculation up to all the non-trivial coefficient functions up to ${\cal O}(z^4)$, for the scalar two-point function in the 2D one-component $\phi^4$ theory. The section is largely for pedagogical purposes to convince unfamiliar readers the fact that the coefficient functions in super-renormalizable QFTs can still be calculated through IR-renormalized massless Feynman integrals regulated in dimensional regularization (DR). This example demonstrates well the logarithmic UV-IR correspondence between coefficient functions and operators in supper-renormalizable QFTs.
    \item In the Sec.~\ref{sec:OPEdiverge}, we start to investigate the asymptotic behavior of the large-$p^2$ expansion in the large-$N$ model. We first perform the calculation through IR renormalized massless bubble-chain diagrams, for the diagrams with zero and one mass insertions. Technically, the factorial enhancements are generated through a novel $\frac{1}{\epsilon} \times \epsilon$ effect that is absent in 4D bubble-chain calculations. We demonstrate in detail how the factorial enhancements cancel between the coefficient functions and the operator condensates. We then double-check with the short-distance calculation by introducing Mellin-Barnes representations for the full massive bubble chain diagram and systematically generate the power-expansion by shifting the contours. The Mellin-Barnes representation not only confirms the short-distance calculation, but also allows to effectively obtain the factorial asymptotics for both the coefficient functions and the operators. Our work  provides a concrete example showing the ``OPE divergence'' phenomenon~\cite{Shifman:1994yf} as asked in~\cite{Marino:2023epd}.
    \item Finally, we comment and conclude in Sec.~\ref{sec:conclu}. Certain technical details and additional analyses for condensates are collected in the four appendices.

\end{enumerate}

\section{OPE in supper-renormalizable theory: $\phi^4_2$ as an example }\label{sec:OPEintro}
In this section, we provide a worked example for OPE calculation both in coordinate and momentum spaces, in the massive $\phi^4_2$ theory for the scalar two-point function. We always stay in Euclidean signature in this work. Although this section may look familiar to experts, the author still suggests reading this section, which could also serve as a motivation for later sections. 

We would like to stress that this example demonstrates well how the severe IR divergences in super-renormalizable massless perturbation theory could be handled and how such perturbation theory could be made meaningful: they are not Borel summable to the Schwinger functions of the full theory\footnote{In case of $\phi_2^4$, the full theory is the Borel re-summation of the massive $\phi_2^4$ perturbative series~\cite{eckmann1975decay}, and in more general cases should be constructed non-perturbatively through scaling-limits of lattice models or through exact form-factor expansions in the presence of integrability.}, but represent consistent algorithms to compute the correct small $z^2$ or large $p^2$ expansions of the OPE coefficient functions for the short-distance-scheme operators in the full theory~\cite{Zamolodchikov:1990bk,Lukyanov:1996jj}. In appropriate IR regularization schemes such as the dimensional regularization, power IR-divergences in the coefficient function computations are removed by analytic continuation, while the logarithmic IR singularities match exactly to the logarithmic UV singularities of the local operators~\cite{Novikov:1984rf,Zamolodchikov:1990bk} and can be consistently renormalized together with the operators. Attaching the renormalized coefficient functions to the operator condensates, one generates the correct small $z^2$ or large $p^2$ expansions for the full theory's Schwinger functions up to an arbitrary finite power parametrized by a finite number of condensates.

As a reminder, the Euclidean action of the the one-component scalar $\phi_2^4$ is given by
\begin{align}
S=\mu_0^{D-2}\int d^Dx \bigg( \frac{1}{2}(\nabla \phi)^2+\frac{m^2}{2}\phi^2+\frac{g}{4!}:\phi^4: \bigg) \ .
\end{align}
Throughout the paper we use DR with $D=d=2-2\epsilon$ for all the UV/IR divergences and $\mu_0^2=\mu^2\frac{e^{\gamma_E}}{4\pi}$ are the standard renormalization scales. Notice that the Wick-ordering is with respect to the free Gaussian field with mass $m$. This defines the mass parameter $m$, as well as the bare mass
\begin{align}
m_0^2=m^2-\frac{g\mu_0^{2-D}}{2}\int\frac{d^Dk}{(2\pi)^D}\frac{1}{k^2+m^2}\equiv m^2-\frac{g}{2}I_0 \ .
\end{align}
In the $\epsilon \rightarrow 0$ limit, all correlation functions of the $\phi$ field at separated points are finite. Their Borel-summable small-$g$ expansions can be perturbatively calculated directly at $D=2$ using massive propagators and the standard Feynman rules, with the neglection of the one-loop tadpoles. Notice that the parameters $m$ and $g$ are $\epsilon$ independent. 

The theory is well-defined and serves as the most general parameterization of 2D one-component stable quartic scalar field theory with $Z_2$ symmetry. In particular, one can set $m^2>0$ without losing of generality. This is because for $m^2<0$, the Wick-ordering must be performed with another mass $M^2>0$ without destroying reality, and one can always find $\tilde m^2>0$ such that
\begin{align}
m_0^2=m^2-\frac{g}{8\pi}\ln \frac{\mu^2}{M^2}-\frac{g}{8\pi \epsilon}+{\cal O}(\epsilon)=\tilde m^2-\frac{g}{8\pi}\ln \frac{\mu^2}{\tilde m^2}-\frac{g}{8\pi \epsilon}+{\cal O}(\epsilon) \ ,
\end{align}
as far as $M^2>0$ even when $m^2<0$. The above implies that in the $\epsilon \rightarrow 0$ limit, the theory is equivalent to the standard one with $\tilde m^2>0$ at least order by order in the $\frac{g}{\tilde m^2}$ expansion. It is normally conjectured~\cite{Chang:1976ek} that,  when $g$ increase to $\frac{g}{m^2}=\frac{g_c}{m^2}>0$, there is a phase transition in the 2D Ising universality class from $Z_2$ unbroken to $Z_2$ broken phase, since $m_0^2$ becomes ``too negative'' at large $g$. The large distance behavior of the theory will be very different in the two phases, but if one is interested in the small distance behavior at $z^2g\ll 1, z^2m^2\ll 1$, then the two phases should share the same structure of the OPE with identical small distance asymptotics power by power for the coefficient functions. 

Below we show how the OPE calculation using DR regulated massless integrals works for the two point function 
\begin{align} \label{eq:defS}
S\left(z^2m^2,\frac{g}{m^2}\right)=\langle \phi(z)\phi(0) \rangle=\int \frac{d^2p}{(2\pi)^2}e^{ip\cdot z}S\left(p^2\right) \ ,
\end{align}
in this theory, up to all the coefficient functions contributing to ${\cal O}(z^4)$. Since in the coefficient functions there are no inverse powers in $m^2$ and $g$, only $g^2$ (two-loop) is required for this calculation. To show that the OPE is correct, it is helpful to show that the coefficient functions combined with the operator condensates reproduce the correct small-$z^2$ expansion of tree-level and two-loop massive diagrams of $S$. In the appendix~\ref{sec:sunrise}, we show using explicit calculations that for the tree-level and two-loop two-point function
\begin{align}
S_{g^2}\left(z^2m^2,\frac{g}{m^2}\right)=\int \frac{d^2p}{(2\pi)^2}\frac{e^{ip\cdot z}}{(p^2+m^2)}+\frac{g^2}{6}\int \frac{d^2p}{(2\pi)^2}\frac{e^{ip\cdot z}}{(p^2+m^2)^2}\Sigma(p^2) \ , \\
\Sigma(p^2)=\int \frac{d^2k d^2l}{(2\pi)^4}\frac{1}{(k^2+m^2)(l^2+m^2)((p-k-l)^2+m^2)} \ , \label{eq:defsigma}
\end{align}
the $z^2m^2$ expansion up to $(z^2m^2)^2$ reads
\begin{align}\label{eq:fullexpansion1}
&S_{g^2}\left(z^2m^2,\frac{g}{m^2}\right)=\nonumber \\&-\frac{1}{4\pi}L+\frac{7\zeta_3}{24(4\pi)^3}\frac{g^2}{m^4}\nonumber \\
& +\bigg(\frac{2-L}{4\pi}-\frac{7\zeta_3}{8(4\pi)^3}\frac{g^2}{m^4}\bigg)\frac{z^2m^2}{4} \nonumber \\ 
&+ \bigg(\frac{3-L}{16\pi}+\frac{g^2}{6(4\pi)^3m^4}\left(-\frac{L^3}{4}+\frac{9L^2}{4}-\frac{69L}{8}+\frac{27}{2}-\frac{49\zeta_3}{16}\right)\bigg)\left(\frac{z^2m^2}{4}\right)^2+{\cal O}(z^6m^6) \ .
\end{align}
Here we have defined the coordinate-space logarithm
\begin{align}
L=L_{m}=\ln \frac{z^2m^2e^{2\gamma_E}}{4}, \ L_{\mu}=\ln \frac{z^2\mu^2e^{2\gamma_E}}{4} \ .
\end{align}
If we consider large Euclidean momentum expansion, then the result reads
\begin{align}
& S_{g^2}(p^2)=\frac{1}{p^2}-\frac{m^2}{(p^2)^2}+\frac{m^4}{(p^2)^3}+\frac{g^2}{32\pi^2(p^2)^3}l^2 +{\cal O}\left(\frac{1}{(p^2)^4}\right)\ , \\
& l=l_{m}=\ln \frac{p^2}{m^2} \ , l_{\mu}=\ln \frac{p^2}{\mu^2} \  ,
\end{align}
where we also defined the momentum-space logarithms. In the rest of this section we show how the OPE calculation correctly reproduces the above and allows its generalization with high-order corrections to operator condensates.

\begin{figure}[htbp]
    \centering
    \includegraphics[height=6.0cm]{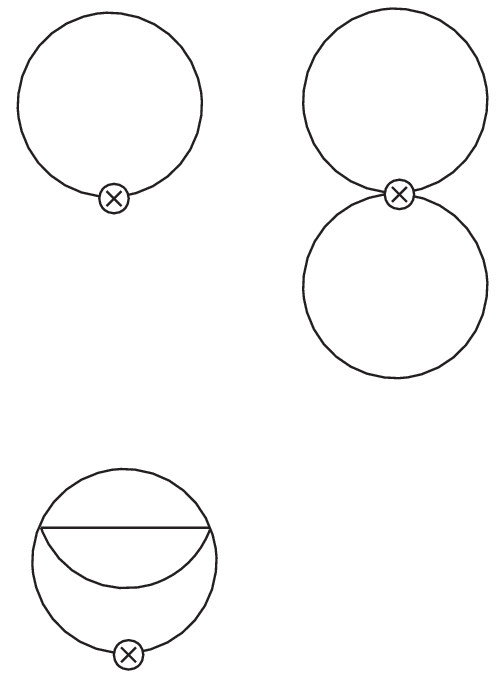}
    \caption{The tree-level condensates for $\phi^2$, $\phi(i\partial)^2\phi$, $\phi\partial^4\phi$ and $\phi^4$ (upper). The order ${\cal O}(g^2)$ condensates for $\phi^2, \phi (i\partial)^2\phi$ and $\phi \partial^4\phi$ (lower). Operator insertions are denoted as crossed circles. }
    \label{fig:condensates} 
\end{figure}

Before moving to the OPE calculation, it is time to determine the $g^2$ corrections to the operator condensates. Here we only need the condensates for $\phi^2$, $\phi(i\partial)^2\phi$ and $\phi(i\partial)^4\phi$. Notice that in this paper we use the notation $(i\partial)^{2k}\equiv (-\partial^2)^k$. The diagrams are shown in Fig.~\ref{fig:condensates}. To calculate these diagrams, one needs only the following three-loop vacuum diagram
\begin{align}
&I=\frac{1}{(2\pi)^6}\int \frac{d^2kd^2ld^2q}{(k^2+m^2)(l^2+m^2)(q^2+m^2)((k+l+q)^2+m^2)}\nonumber \\ 
&=\frac{1}{(4\pi)^3m^2}\int_{0}^{\infty} \frac{d\alpha_1d\alpha_2d\alpha_3d\alpha_4}{\alpha_1\alpha_2(\alpha_3+\alpha_4)+\alpha_3\alpha_4(\alpha_1+\alpha_2)}e^{-\alpha_1-\alpha_2-\alpha_3-\alpha_4} \nonumber \\ 
&=\frac{7\zeta_3}{(4\pi)^3m^2} \ . \label{eq:defI}
\end{align}
To obtain the last result, one splits the denominator using Mellin-Barnes. 
Thus, one has
\begin{align}
& \langle \phi^2\rangle_{g^2}=-\frac{g^2}{6}\frac{1}{4}\partial_{m^2}I=\frac{g^2}{6}\frac{7\zeta_3}{(4\pi)^34(m^2)^2} \ , \label{eq: O12}\\
& \langle \phi (i\partial)^2\phi\rangle_{g^2}=\frac{g^2}{6}I-m^2\langle \phi^2\rangle_{g^2}=\frac{g^2}{6}\frac{3}{4}I \label{eq:O22} \ .
\end{align}
The above are just the two terms for the $\langle \phi^2\rangle_{g^2}$ and $\langle\phi \partial^2\phi\rangle_{g^2}$ contributions. One also needs the divergent condensate $\langle\phi \partial^4\phi\rangle_{g^2}$ regularized in DR. One has
\begin{align}
&\langle \phi\partial^4\phi \rangle_{g^2}= \frac{g^2}{6}I_0^3-2m^2\langle \phi (i\partial)^2\phi\rangle_{g^2}-m^4\langle \phi^2\rangle_{g^2}\nonumber \\ 
&=\frac{g^2}{6}I_0^3-\frac{g^2}{6}\frac{7\zeta_3}{(4\pi)^3}\bigg(\frac{1}{4}+\frac{3}{2}\bigg)+{\cal O}(\epsilon) \ , \label{eq:O4}
\end{align}
where
$I_0=\frac{\mu_0^{2-d}}{(2\pi)^D}\int \frac{d^D k}{k^2+m^2}$
is just the one-loop tadpole. Clearly, the first two operators are already finite at $g^2$ order, while the $\phi(\partial)^4\phi$ contain an $\frac{1}{\epsilon^3}$ UV singularity that we will see cancel exactly with the IR singularities for the coordinate-space coefficient functions at the power $z^4$.

\begin{figure}[htbp]
    \centering
    \includegraphics[height=2.0cm]{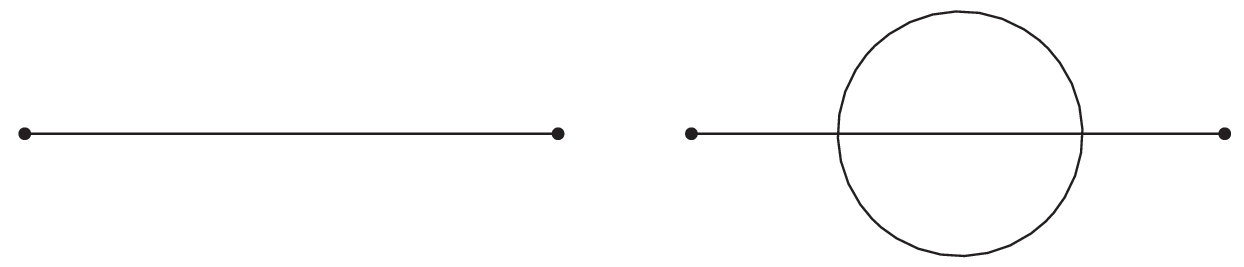}
    \caption{The diagrams in the full theory for two-point function up to ${\cal O}(g^2)$. The solid lines are massive propagators with mass $m^2$.   }
    \label{fig:full}
\end{figure}

\begin{figure}[htbp]
    \centering
    \includegraphics[height=6.0cm]{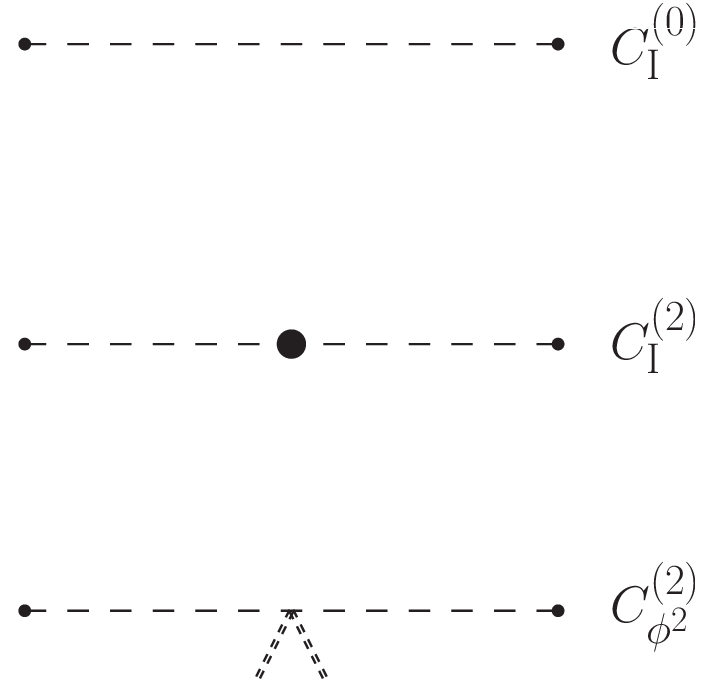}
    \caption{The coefficient functions for various operators at the order ${\cal O}(z^0)$ and ${\cal O}(z^2)$. The dashed lines are massless propagators. The big black circles are insertion of the bare mass $-m_0^2$. The double lines denote the ``external soft fields'' from condensates attached to the coefficient functions. They will be amputated in the coefficient function. The tree-level coefficient $1$ for $\phi^2$ and $-\frac{z^2}{2d}$ for $\phi(i\partial)^2\phi$ are omitted in the figure, but are crucial for coordinates space expansion.  }
    \label{fig:opeLPandNLP}
\end{figure}

\subsection{Leading power ($z^0$) and next-to-leading power ($z^2$) }
At the leading power, in the momentum space there is only a single term $\frac{1}{p^2}$\footnote{Notice that the momentum-space two point function in $D$-dimension in our convention has an overall factor $\mu_0^{2-D}$. This overall factor will not participate in the momentum-space renormalization and will be omitted in this work. In the coordinate space it participates in the momentum integral and can not be neglected.}. In the coordinate space, on the other hand, there is already a log
\begin{align}
S_{g^2}^{(0)}\left(z^2m^2,\frac{g}{m^2}\right)=-\frac{1}{4\pi}\ln \frac{z^2m^2e^{2\gamma_E}}{4}+\frac{7\zeta_3}{24(4\pi)^3}\frac{g^2}{m^4} \ .
\end{align}
Can one explain the logarithm here? Clearly, the condensate $\langle \phi^2\rangle_0$ is required. One has the bare coefficient function for the identity operator
\begin{align}
C_{\rm I}^{(0)}(z^2\mu^2,d)=\mu^{2-d}_0\int \frac{d^dp}{(2\pi)^d}\frac{e^{ip\cdot z}}{p^2}=\frac{1}{(4\pi)^{\frac{d}{2}}}\Gamma\left(\frac{d}{2}-1\right)\left(\frac{z^2\mu_0^2}{4}\right)^{1-\frac{d}{2}} \ ,
\end{align}
and the tree-level bare condensate
\begin{align}
\langle \phi^2\rangle_0=I_0=\mu_0^{2-d}\int \frac{d^dp}{(2\pi)^d}\frac{1}{p^2+m^2}=\left(\frac{\mu_0^2}{m^2}\right)^{1-\frac{d}{2}} \frac{1}{(4\pi)^{\frac{d}{2}}}  \Gamma \left(1-\frac{d}{2}\right) \ .
\end{align}
Clearly, the bare OPE in the $d\rightarrow 2$ limit reproduces the full logarithm
\begin{align}
C_{\rm I}^{(0)}(z^2\mu^2,d)+\langle \phi^2\rangle_0 \bigg|_{d\rightarrow 2} \rightarrow -\frac{1}{4\pi}\ln \frac{z^2m^2e^{2\gamma_E}}{4} \ . 
\end{align}
The above suggests to define the minimally subtracted condensate with $d=2-2\epsilon$, as well as renormalized coefficient functions that are separately finite as $d\rightarrow 2$:
\begin{align}
\langle \phi^2\rangle_0=\langle \phi^2\rangle_r+\frac{1}{4\pi \epsilon} \ , \label{eq:constants1} \\
C_{{\rm I}, r}^{(0)}(z^2\mu^2)=C_{\rm I}^{(0)}(z^2\mu^2,d)+\frac{1}{4\pi \epsilon}\bigg|_{d\rightarrow2} \rightarrow -\frac{1}{4\pi}\ln \frac{z^2\mu^2e^{2\gamma_E}}{4}=-\frac{1}{4\pi}L_\mu \ . 
\end{align}
Clearly, the $\frac{1}{4\pi \epsilon}$ pole of the bare operator exactly cancels the IR divergence of the bare coefficient function, and it can also be regarded as the  {\it IR subtraction} or {\it IR counter-term } for the coefficient function.  An important fact of the $\overline{\rm MS}$-scheme operators is that their renormalization constants depend only on short distance fluctuations ($\frac{1}{\epsilon}$ poles), therefore these operators are extremely point-like and suitable for short distance expansions. Moreover, the above allows the generalization to include high-order corrections to the operator condensates:
\begin{align}
\langle \phi^2\rangle_r=\frac{1}{4\pi}\ln \frac{\mu^2}{m^2}+\langle :\phi^2:\rangle \ , \\ 
S^{(0)}\left(z^2m^2,\frac{g}{m^2}\right)=C^{(0)}_{{\rm I},r}(z^2\mu^2)+\langle \phi^2\rangle_r=-\frac{1}{4\pi}\ln \frac{z^2m^2e^{2\gamma_E}}{4}+\langle :\phi^2:\rangle \ .
\end{align}
Thus, the $\mu^2$ determines the separation point between short distance and large distance fluctuations in the $\overline{\rm MS}$ scheme. The $\langle :\phi^2:\rangle$ start to contribute at $g^2$ and is given in Eq.~(\ref{eq: O12}). It exactly reproduces the $\zeta_3$ term at $z^0$ in the first line of Eq.~(\ref{eq:fullexpansion1}).  Higher order corrections will not contribute any new logarithms at the leading power, but only modifies this condensate. Notice that the bare coefficient function completely determines the $\frac{1}{4\pi \epsilon}$ for the operator renormalization.

We then consider the nexet-to-leading power. At this order, the expansion is still simple
\begin{align}
S_{g^2}^{(2)}\left(z^2m^2,\frac{g}{m^2}\right)=\frac{z^2m^2}{4} \times \frac{2-L}{4\pi} -\frac{z^2}{4}\frac{7\zeta_3}{8(4\pi)^3}\frac{g^2}{m^2} \ .
\end{align}
There are two condensates $\phi^2$ and $\phi (i\partial)^2\phi$ in the OPE. The bare coefficient functions follow from figure~\ref{fig:opeLPandNLP}
\begin{align}
& C_{\rm I}^{(2)}=-m_0^2\mu_0^{2-d}\int\frac{d^dp}{(2\pi)^d}\frac{e^{ip\cdot z}}{(p^2)^2}=- \frac{m_0^2z^2}{4(4\pi)^{\frac{d}{2}}}\left(\frac{z^2\mu_0^2}{4}\right)^{1-\frac{d}{2}}\Gamma \left(\frac{d}{2}-2\right)\ , \\ 
& C_{\phi^2}^{(2)}=-\frac{g}{2}\mu_0^{2-d}\int\frac{d^dp}{(2\pi)^d}\frac{e^{ip\cdot z}}{(p^2)^2}= - \frac{gz^2}{8(4\pi)^{\frac{d}{2}}}\left(\frac{z^2\mu_0^2}{4}\right)^{1-\frac{d}{2}}\Gamma \left(\frac{d}{2}-2\right)\ , \\
& C_{\phi(i\partial)^2\phi}^{(2)}=-\frac{z^2}{2d} \ .
\end{align}
In the momentum space, one has 
\begin{align}
S^{(2)}(p^2)(p^2)^2=-m_0^2-\frac{g}{2}\langle\phi^2 \rangle_0=-m_0^2-\frac{g}{8\pi \epsilon}-\frac{g}{2}\langle\phi^2\rangle_r \ ,
\end{align}
where we have used the renormalization constant Eq.~(\ref{eq:constants1}). By requiring its finiteness, one determines the renormalization of the bare mass 
\begin{align}
m_0^2=m_r^2-\frac{g}{8\pi \epsilon} \ .
\end{align}
Here we have introduced {\it short distance renormalized mass} $m_r^2$ that is $\epsilon$ dependent, but is finite in the $\epsilon \rightarrow 0$ limit. It can be regarded as the bare mass subtracted only the short-distance contribution to the mass rernomalization, but without including any effects at the IR scale $m$.  Due to this, it is the most natural mass that appears in the short distance calculation and in the renormalized coefficient functions.

Here, we have seen that by requiring the IR/UV finiteness of renormalized coefficient functions automatically determines the mass renormlaization. This also resolves the ``puzzle'' concerning the consistency between the DR rule that throws away scaleless integrals and the fact that the $\overline{\rm MS}$ renormalization is short distance in nature: the mass renormalization in $\phi_2^4$ comes purely from the one-loop tadpole which becomes scaleless in the massless theory. How this renormalization can be determined without computing any massive integrals? Clearly, the approach here is the way: by requiring IR finiteness of the leading-power coefficient function $C_{\rm I,r}^{(0)}$ uniquely determines the operator renormalization for $\langle\phi^2\rangle_0$, the latter then determines the mass renormalization by requiring UV finiteness of the next-to-leading power coefficient function. In fact, this pattern will persist to all powers: requiring $\epsilon\rightarrow0$ finiteness of the coefficient functions automatically determines the UV poles of operators~\footnote{In IR regularization schemes with IR scales, non-trivial matching procedures are required in order for the IR-subtracted coefficient functions to match with the specific UV renormalization schemes of the operators. For example, for the $m\bar\psi\psi$ perturbation to the Ising CFT, one can put the theory on a finite cylinder with a radius $R$ and absorb all the power IR divergences into the small-$mR$ expansions of the condensates. This allows to generate the small-$mz$ expansion~\cite{Wu:1975mw} of the $\langle\sigma(z)\sigma(0)\rangle$ correlator perturbatively.  }. 

Here we move back to the coordinate space. To determine the renormalization of $\phi \partial^2\phi$, one can write 
\begin{align} \label{eq:tree1bare}
\langle \phi (i\partial)^2\phi\rangle_0=a_1m_r^2+a_2g\langle\phi^2\rangle_r +\langle \phi (i\partial)^2 \phi\rangle_r \ ,
\end{align}
with $a_1, a_2$ are $\frac{1}{\epsilon}$ poles. Notice that in this paper, we call purely $\frac{1}{\epsilon}$ coefficients such as $a_1$, $a_2$ that express bare operators in terms of renormalized ones as {\it UV poles} of the operators. They are simultaneously {\it IR counter-terms} for the coefficient functions.  Now, requiring finiteness of renormalized coefficient functions
\begin{align}
& C_{{\rm I},r}^{(2)}=C_{\rm I}^{(2)}+\frac{1}{4\pi \epsilon}C_{\phi^2}^{(2)}-\frac{z^2}{2d}m_r^2a_1 \rightarrow {\cal O}(1) \ , \\
& C_{\phi^2,r}^{(2)}=C_{\phi^2}^{(2)}-\frac{z^2}{2d}ga_2 \rightarrow{\cal O}(1) \ .
\end{align}
This uniquely determines the operator poles
\begin{align} \label{eq:constants2}
a_1=-\frac{1}{4\pi\epsilon} \ , a_2=-\frac{1}{8\pi \epsilon} \ .
\end{align}
Thus, we obtains 
\begin{align}
& C_{{\rm I},r}^{(2)}=\frac{2-L_\mu}{4\pi}\frac{z^2m_r^2}{4} \ , \\
& C_{\phi^2,r}^{(2)}=\frac{2-L_\mu}{4\pi}\frac{z^2g}{8} \ , \\
& C_{\phi(i\partial)^2\phi,\ r}^{(2)}=-\frac{z^2}{4} \ .
\end{align}
Alternatively, one has for the ```tree-level'' condensate
\begin{align}
\langle \phi (i\partial)^2\phi\rangle_0=-m^2I_0=-\frac{m_r^2}{4\pi \epsilon}-\frac{g}{8\pi\epsilon}I_r-I_r\left(m_r^2+\frac{g}{2}I_r\right) \ .
\end{align}
The above is consistent with $m_0^2=m^2-\frac{g}{2}I_0$ and $I_0=I_r+\frac{1}{4\pi \epsilon}$. Thus, the renormalization constants can also be read from the above equation and agrees with the one determined from the IR-UV cancellation requirement. It is easy to show that the above coefficient functions combined with the tree-level condensates exactly reproduces the full expansion at $z^2$ and ${\cal O}(g^0)$. 
Furthermore, from the above the RGE for the renormalized operator condensate $\langle\phi\partial^2\phi\rangle_r$ and the general solution is 
\begin{align}
&\frac{d}{d \ln \mu^2}\langle \phi(i\partial)^2\phi\rangle_r=-\frac{m_r^2}{4\pi}-\frac{g}{8\pi}\langle\phi^2 \rangle_r \ , \\  
&\langle \phi(i\partial)^2\phi\rangle_r=-I_r\left(m_r^2+\frac{g}{2}I_r\right)-\frac{g}{8\pi}\ln\frac{\mu^2}{m^2}\langle :\phi^2:\rangle+{\cal R}\langle \phi (i\partial)^2\phi \rangle \ .
\end{align}
The general expansion at the power $z^2$ is therefore after setting $\mu=m$
\begin{align}
S^{(2)}\left(z^2m^2,\frac{g}{m^2}\right)=\frac{2-L}{4\pi}\frac{z^2}{4}\left(m^2+\frac{g}{2}\langle :\phi^2:\rangle\right)-\frac{z^2}{4}{\cal R}\langle \phi (i\partial)^2\phi \rangle \ . 
\end{align}
Again, the high-order contribution only modifies the two condensates, but no new form of $z$ dependency will be generated.  The order $g^2$ value of ${\cal R}\langle \phi \partial^2\phi\rangle$ is given in Eq.~(\ref{eq:O22}) and exactly reproduces the $\zeta_3$ term at the power $z^2$ after combining with the coefficient function $-\frac{z^2}{4}$.

\subsection{The OPE at ${\cal O}(z^4)$}
In this subsection we check the expansion at the power $z^4$ or $\frac{1}{(p^2)^3}$
with OPE calculations. There are three non-trivial operators require at this order: $O_1=\phi^2$, $O_2=\phi^4$ and $O_3=\phi(\partial)^4\phi$. The last operator $O_3$ with $\partial^4$ only contributes in the coordinate space expansion.  The relevant diagrams for the coefficient functions and operators are shown in Fig.~\ref{fig:ope} and Fig.~\ref{fig:condensates}.

\subsubsection{Momentum space OPE}

\begin{figure}[htbp]
    \centering
    \includegraphics[height=7.0cm]{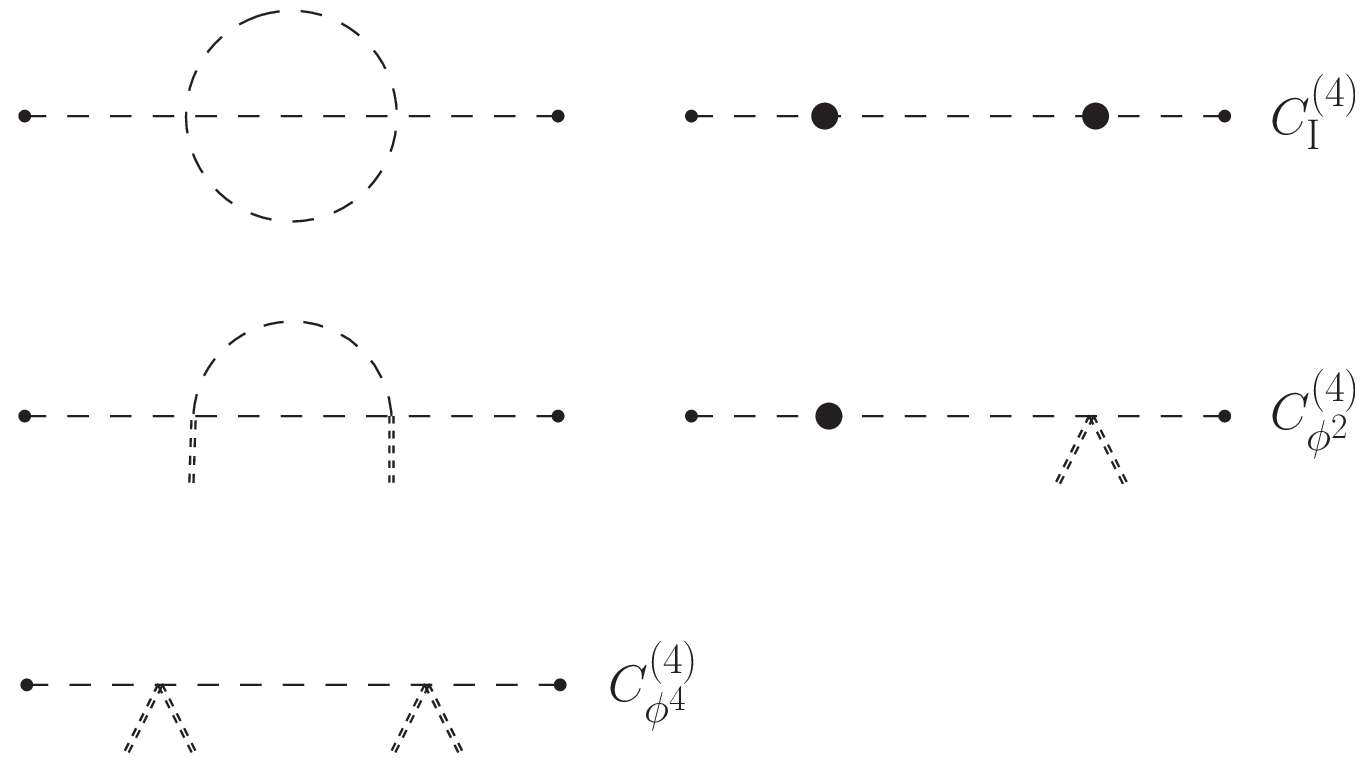}
    \caption{The coefficient functions for various operators at the order ${\cal O}(z^4)$. The dashed lines are massless propagators. The big black circles are insertion of the bare mass $-m_0^2$. The double lines denote the ``amputated external soft fields'' to keep track the operator content of the condensates. The tree-level coefficient $\frac{3z^4}{24d(d+2)}$ for $\phi\partial^4\phi$ are omitted in the figure,  but crucial for coordinate space expansion.  }
    \label{fig:ope}
\end{figure}
To check the expansion against the OPE, one needs the coefficient functions. In the momentum space, one needs the following functions
\begin{align}
&C_0\left(\frac{p^2}{\mu^2},d\right)=\mu_0^{4-2d}\int \frac{d^dkd^dl}{(2\pi)^{2d}}\frac{1}{k^2l^2(p-k-l)^2}\nonumber \\ 
&=\frac{1}{p^2}\left(\frac{\mu_0^2}{p^2}\right)^{2-d}\frac{8^{1-d} \pi ^{\frac{3}{2}-d} \Gamma \left(\frac{d}{2}-1\right)^2}{\Gamma \left(\frac{d-1}{2}\right) \Gamma \left(\frac{3 d}{2}-3\right)\sin \pi d} \ , \\
&C_1\left(\frac{p^2}{\mu^2},d\right)
=\mu_0^{2-d}\int\frac{d^dk}{(2\pi)^d}\frac{1}{k^2(p-k)^2}\nonumber \\ &=\frac{1}{p^2}\left(\frac{\mu_0^2}{p^2}\right)^{1-\frac{d}{2}}\frac{2^{3-2 d} \pi ^{\frac{1}{2}-\frac{d}{2}} \Gamma \left(2-\frac{d}{2}\right) \Gamma \left(\frac{d}{2}-1\right)}{\Gamma \left(\frac{d-1}{2}\right)} \ , \label{eq: C1}\\ 
& I_0\left(\frac{\mu^2}{m^2},d\right)=\left(\frac{\mu_0^2}{m^2}\right)^{1-\frac{d}{2}} \frac{1}{(4\pi)^{\frac{d}{2}}} \Gamma \left(1-\frac{d}{2}\right) \equiv \langle \phi^2\rangle_0 \ , \\ 
& \langle \phi^4 \rangle_0= 3I_0^2 \ .
\end{align}
Notice that $C_0$ and $C_1$ are just the massless self-energy diagrams with zero and two external field insertions without symmetry factors, see Fig.~\ref{fig:ope}. One can show that 
\begin{align}\label{eq:bareexp1}
C_0\left(\frac{p^2}{\mu^2},d\right)+3C_1\left(\frac{p^2}{\mu^2},d\right)I_0\left(\frac{\mu^2}{m^2},d\right)+\frac{3}{p^2}I_0^2\left(\frac{\mu^2}{m^2},d\right)\bigg|_{d\rightarrow2 }\rightarrow\frac{3}{p^2(4\pi)^2}\ln^2\frac{p^2}{m^2} \ ,
\end{align}
which is exactly the expansion of the massive sunrise self-energy $\Sigma(p^2)$ defined in Eq.~(\ref{eq:defsigma}). Clearly, this can be regarded as follows:  $\Sigma(p^2)$  at the power $\frac{1}{p^2}$ contains the following soft sub-regions: there are three ways for one of the three propagators in the sunrise to become soft and form the quadratic condensate $\langle \phi^2\rangle_0=I_0$. There are also three ways for two of them to becomes soft and form the quartic condensate $\langle \phi^4\rangle_0=3I_0$. Combining with the region where all three propagators are hard, one indeed reproduces the expansion for $\Sigma(p^2)$ at $\frac{1}{p^2}$.  

We would like to proceed further to obtain the OPE for the two-point function. For this one must notice that the $\langle\phi^4\rangle_0$ insertion for the self-energy is not all of the $\langle \phi^4\rangle_0=3I_0^2$ contribution to the two-point function. The reason is that there is also a ``reducible'' term that should cancel with the $\frac{1}{\epsilon^2}$ for the double $m_0^2$ insertion. 
Indeed, if one compute the last diagram in Fig.~(\ref{fig:ope}) as a correlation function between $\phi(-p)\phi(p)$ and four amputated zero-momentum scalar fields, then one can see the coefficient function for $\phi_0^4$ insertion should be $\frac{g^2}{4p^2}$. The contribution to the self-energy is $\frac{g^2}{6p^2}\times 3I_0^2$, while there is another $\frac{1}{12p^2}\times 3I_0^2$ that should cancel with mass insertions. Similarly, for the $\langle \phi^2\rangle_0$ insertion one should also include the $m_0^2\times g$ order diagram shown in Fig.~\ref{fig:ope}.
Therefore, the bare OPE for the two-point function reads
\begin{align}
S^{(4)}(p^2)(p^2)^2=\left(\frac{g^2}{6}C_0+\frac{m_0^4}{p^2}\right)+\left(\frac{g^2}{2}C_1+\frac{g m_0^2}{p^2}\right)\langle \phi^2\rangle_0+\left(\frac{g^2}{6p^2}+\frac{g^2}{12p^2}\right)\langle\phi^4\rangle_0 \ .
\end{align}
Then, one introduces the renormalized operators in the $\overline{MS}$ scheme as
\begin{align}
\langle\phi^2\rangle_0=\langle \phi^2\rangle_r+\frac{1}{4\pi \epsilon} \ , \\
\langle\phi^4\rangle_0=\langle \phi^4\rangle_r+\frac{3}{2\pi \epsilon}\langle\phi^2\rangle_r+\frac{3}{(4\pi)^2\epsilon^2} \ .
\end{align}
Notice that $\langle \phi^4 \rangle_r = 3\langle \phi^2\rangle_r^2$ at tree-level, despite possible $\frac{1}{\epsilon} \times\epsilon$ effects. The point is that $\langle \phi^2\rangle_r$ and $\langle \phi^4\rangle_r$ are finite in the $\epsilon \rightarrow 0$ limit, but in the $\epsilon \rightarrow 0$ limit they can still contain ${\cal O}(\epsilon)$ terms. 

Given the above, by collecting coefficients for renormalized operators, one has 
\begin{align}
&S^{(4)}(p^2)(p^2)^2=C_{{\rm I},r}^{(4)}(p)+C_{\phi^2,r}^{(4)}(p)\langle \phi^2\rangle_r+C_{\phi^4,r}^{(4)}(p)\langle \phi^4\rangle_r \ , \\ 
&C_{{\rm I},r}^{(4)}(p)=\frac{g^2}{6}C_0+\frac{g^2}{8\pi \epsilon}C_1+\frac{g^2}{2p^2}\frac{1}{(4\pi)^2\epsilon^2}+\frac{m_r^4}{p^2} \ , \\
&C_{\phi^2,r}^{(4)}(p)=\frac{g^2}{2}C_1+\frac{g^2}{2p^2}\frac{1}{2\pi \epsilon}+\frac{gm_r^2}{p^2} \ , \\
&C_{\phi^4,r}^{(4)}(p)=\frac{g^2}{4p^2} \ . 
\end{align}
In the $\epsilon \rightarrow 0$ limit, all of the above are finite. In fact, requiring that the $\epsilon \rightarrow 0$ limit exist uniquely determines all the $\frac{1}{\epsilon}$ poles in the mixing, once the tree-level mixing constant $Z_{10}^{(0)}=\frac{1}{4\pi \epsilon}$ for $\phi^2$ is determined from the coordinate space expansion at $z^0$ order, and the one-loop level mixing $m_0^2=m_r^2-\frac{g}{8\pi\epsilon}$ is fixed from the $z^2$ order consideration. In particular
\begin{align}
& C_{{\rm I},r}^{(4)}(p)=\frac{g^2}{32\pi^2 p^2}\ln^2 \frac{p^2}{\mu^2}+\frac{m_r^4}{p^2} \ , \\
& C_{\phi^2,r}^{(4)}(p)=\frac{g^2}{4\pi p^2}\ln \frac{p^2}{\mu^2}+\frac{gm_r^2}{p^2} \ . 
\end{align}
Notice that all the $m_0^2$ insertions in $C_{\rm I}$ again combine with the operator UV poles for $\phi^{2n}$ inserted along the same propagator to form the renormalized mass $m_r^2$. Anticipating that, in the coefficient function calculation, one can instead use $m_r^2$ and neglect many operator UV poles corresponding to ``reducible'' insertions of $(\phi^2)^n$ along the same propagator. This is particularly convenient in the large $N$ calculation, since in such cases different contractions between the $(\phi^{2})^n$ usually have different large $N$ power-counting.

Given the OPE,  to reproduce the full result, notice in the current case one has $m_r^2=m^2-\frac{g}{2}\langle \phi^2\rangle_r=m^2-\frac{g}{8\pi}\ln \frac{\mu^2}{m^2}+{\cal O}(\epsilon)$, and the tree-level operator condensates in the $\epsilon \rightarrow 0$ limit read
\begin{align}
&\langle \phi^2\rangle_r=\frac{1}{4\pi}\ln \frac{\mu^2}{m^2} \ , \label{eq:treesquare} \\ 
&\langle \phi^4\rangle_r=\frac{3}{16\pi^2}\ln^2 \frac{\mu^2}{m^2}  \ . \label{eq:treequartic}
\end{align}
Combining all above, one has the $\mu$-independent result $S_{g^2}^{(4)}(p^2)(p^2)^2=\frac{g^2}{32\pi^2p^2}\ln^2 \frac{p^2}{m^2}+\frac{m^4}{p^2}=C_{{\rm I},r}^{(4)}(\mu^2=m^2)$. This again reveals an important fact: if only tree-level condensates are required, then one can set $\mu=m$ to remove them and keep only the coefficient function for the identity operator. This is because at tree-level, $\langle\phi^{2n}\rangle$ always factorize into $\frac{(2n)!}{(n!)2^n}\langle\phi^2\rangle^n$, therefore after renormalization they always vanish at $\mu=m$. This is also true if some of the $\phi^2$s are replaced by $\phi(\partial^{2k}\phi)$, but only self-contractions are present, and only the ``tadpole-type'' high-order corrections to these operators are allowed in terms of the $\frac{1}{k^2+m_r^2}$ propagators. Later in Sec.~(\ref{sec:OPEdiverge}) we will use this observation.

Here we should stress that if one only keeps the naive perturbative result $\frac{g^2}{6}C_0$ in $C_{{\rm I}, r}$ without introducing additive counter-terms, then not only the $\frac{1}{\epsilon}$ poles no longer cancel, but the double log term in the finite part would also have the wrong coefficient $\frac{g^2}{16\pi^2p^2}\ln^2\frac{p^2}{\mu^2}$.
\begin{align}
\frac{g^2}{6}C_0=\frac{g^2}{32 \pi ^2 p^2 \epsilon^2}-\frac{g^2}{16 \pi ^2 p^2 \epsilon }\ln\frac{p^2}{\mu^2}+\frac{ g^2}{16\pi ^2 p^2}\ln^2\frac{p^2}{\mu^2}-\frac{g^2}{32 \pi^2p^2}\zeta_2+{\cal O}(\epsilon) \ .
\end{align}
In the ``additive counter-term'' $\frac{g^2}{8\pi \epsilon}C_1$, the $\frac{1}{\epsilon}\times \epsilon$ terms will generate another $-\frac{g^2}{32\pi^2p^2}\ln \frac{p^2}{\mu^2}$ :
\begin{align}
\frac{g^2}{8\pi \epsilon}C_1=-\frac{g^2}{16\pi^2 p^2\epsilon^2}+\frac{g^2}{16 \pi ^2 p^2 \epsilon }\ln\frac{p^2}{\mu^2}-\frac{g^2}{32\pi^2p^2}\ln^2\frac{p^2}{\mu^2}+\frac{g^2}{32\pi^2p^2}\zeta_2+{\cal O}(\epsilon) \ .
\end{align}
It is clear now that all the $\frac{1}{\epsilon}$ poles cancel, the $\frac{1}{\epsilon^2}$ poles add to cancel the remaining $\frac{g^2}{2p^2}\frac{1}{(4\pi)^2\epsilon^2}$. The $\zeta_2$ term all cancel and the double-log terms combine to reproduce the full result for $C_{{\rm I},r}$. Therefore, in the DR calculation, it is important to keep the $d$ dependency of the bare coefficient functions to the end, and the $\epsilon \times \frac{1}{\epsilon}$ effects can be crucial.

Beyond the order $g^2$, the condensates will be modified, but the full expansion will be scale invariant using the RGE
\begin{align}
\frac{d}{d\ln \mu^2}\phi_r^2=\frac{1}{4\pi} \ , \\ 
\frac{d}{d\ln \mu^2}\phi_r^4=\frac{3}{2\pi}\phi_r^2 \ .
\end{align}
Thus, one has
\begin{align}
&\langle \phi^2\rangle_r=\frac{1}{4\pi}\ln \frac{\mu^2}{m^2}+\langle:\phi^2:\rangle \ , \\ 
&\langle \phi^4\rangle_r=\frac{3}{16\pi^2}\ln^2 \frac{\mu^2}{m^2}+\frac{3}{2\pi}\ln \frac{\mu^2}{m^2}\langle:\phi^2:\rangle+\langle:\phi^4:\rangle \ .
\end{align}
It is not difficult to show that here $\langle:\phi^2:\rangle$ and $\langle:\phi^4:\rangle$ are just the vacuum condensates for $\mu$-independent Wick-polynomials.
Given the $\mu$-independent condensates, the expansion will be modified as
\begin{align}
S^{(4)}(p^2)(p^2)^2=\frac{g^2}{32\pi^2p^2}\ln^2 \frac{p^2}{m^2}+\frac{m^4}{p^2}+\bigg(\frac{gm^2}{p^2}+\frac{g^2}{4\pi p^2}\ln\frac{p^2}{m^2}\bigg)\langle:\phi^2:\rangle+\frac{g^2}{4p^2}\langle:\phi^4:\rangle\ .
\end{align}
Here we note that the above is the generic form of the large $p^2$ expansion at the next-to-next-leading order in $\frac{1}{p^2}$ for $\phi_2^4$, where $\langle:\phi^2:\rangle$ and $\langle :\phi^4:\rangle$ can be non-perturbative. At small $g$, condensates for these two Wick-polynomials are of orders ${\cal O}(\frac{g^2}{m^4})$ and ${\cal O}\left(\frac{g}{m^2}\right)$.  The double logarithmic dependency is always purely determined perturbatively.

\subsubsection{Coordinate space OPE}
We now move back to the coordinate space. The coefficient functions can be Fourier transformed into the $z$ space. Here one needs the following functions
\begin{align}
C_0(z^2\mu^2,d)&= \mu_0^{6-3d}\int \frac{d^dpd^dkd^dl}{(2\pi)^{3d}}\frac{e^{ip\cdot z}}{(p^2)^2k^2l^2(p-k-l)^2}\nonumber \\ 
&=z^4(z^2\mu_0^2)^{\frac{6-3d}{2}}\frac{2^{-d-5} \pi ^{-\frac{3}{2}(d-1)}  \Gamma \left(\frac{d}{2}-1\right)^2}{(9d^2-54d+80) \Gamma (5-d) \Gamma \left(\frac{d-1}{2}\right)\sin \pi d} \ , \\ 
C_1(z^2\mu^2,d)&=\mu_0^{4-2d}\int \frac{d^dpd^dk}{(2\pi)^{2d}}\frac{e^{ip\cdot z}}{(p^2)^2k^2(p-k)^2}\nonumber \\ 
&=-z^4(z^2\mu_0^2)^{2-d}\frac{2^{-d-5} \pi ^{\frac{3}{2}-d}   \Gamma (d-4)}{\Gamma \left(4-\frac{d}{2}\right) \Gamma \left(\frac{d-1}{2}\right)\sin\left(\frac{\pi  d}{2}\right)} \ , \\
C_2(z^2\mu^2,d)&=\mu_0^{2-d}\int \frac{d^dp}{(2\pi)^d}\frac{e^{ip\cdot z}}{(p^2)^3}=z^4(z^2\mu_0^2)^{1-\frac{d}{2}}\frac{\pi ^{-\frac{d}{2}} }{128} \ \Gamma \left(\frac{d}{2}-3\right) \ , \\ 
C_3(z^2,d)&=\frac{3z^4}{24d(d+2)} \ .
\end{align}
Notice that they are already for the two-point function rather than the self-energy. 
Given the above, similar to the expansion Eq.~(\ref{eq:bareexp1}) one has
\begin{align}
&C_0(z^2\mu^2,d)+3C_1(z^2\mu^2,d)I_0\left(\frac{\mu^2}{m^2},d\right)+3C_2(z^2\mu^2,d)I_0^2\left(\frac{\mu^2}{m^2},d\right)+C_3(z^2,d)I_0^3\left(\frac{\mu^2}{m^2},d\right) \nonumber \\ 
&\bigg|_{d\rightarrow2}\rightarrow \frac{z^4}{16(4\pi)^3}\bigg(-\frac{L^3}{4}+\frac{9 L^2}{4}-\frac{69L}{8}+\frac{27}{2}\bigg)  \ , 
\end{align}
which is exactly the ``hard'' contribution in Eq.~(\ref{eq:fullexpansion1}), including the constant $\frac{27}{2}$. Notice that the last term is due to the UV divergent $I_0^3$ contribution in the $\langle \phi\partial^4\phi\rangle_{g^2}$ as shown in Eq.~(\ref{eq:O4}), and the coefficient $\frac{3z^4}{24d(d+2)}$ is obtained by Taylor expanding
\begin{align}
\langle \phi \partial^{i}\partial^j\partial^k\partial^l\phi\rangle=\frac{\delta^{ij}\delta^{kl}+\delta^{ik}\delta^{jl}+\delta^{il}\delta^{jk}}{d(d+2)}\langle \phi \partial^4\phi \rangle \ .
\end{align}
In terms of the functions above, the bare coefficient functions read
\begin{align}
&C_{\rm I}^{(4)}=\frac{g^2}{6}C_0(z^2\mu^2,d)+m_0^4C_2(z^2\mu^2,d) \ , \\
&C_{\phi^2}^{(4)}=\frac{g^2}{2}C_1(z^2\mu^2,d)+gm_0^2C_2(z^2\mu^2,d) \ , \\ 
&C_{\phi^4}^{(4)}=\left(\frac{g^2}{6}+\frac{g^2}{12}\right)C_2(z^2\mu^2,d) \ , \\
&C_{\phi(\partial)^4\phi}^{(4)}=C_3(z^2,d) \ .
\end{align}
We now rewrite the OPE in the coordinate space in terms of renormalized operators. For this purpose one needs the renormalization of $\phi\partial^4\phi$, which can be read from below with $I_r=I_0-\frac{1}{4\pi \epsilon}=\langle \phi^2\rangle_r$
\begin{align}
m^4I_0+\frac{g^2I_0^3}{6}=&\frac{m_r^4}{4\pi \epsilon}+\frac{g^2}{384 \pi ^3 \epsilon^3}+\bigg(\frac{g^2}{32 \pi ^2 \epsilon^2}+\frac{g m_r^2}{4 \pi  \epsilon}\bigg)\langle\phi^2\rangle_r+\frac{g^2}{16\pi \epsilon}\langle \phi^4 \rangle_r\nonumber \\ &+\frac{g^2 I_r^3}{6}+I_r\left(m_r^2+\frac{g}{2}I_r\right)^2+{\cal O}(\epsilon) \ ,
\end{align}
which defines the UV poles and the renormalized operator. Defining the operators 
\begin{align}
O_0=1 \ , \ O_1=(\phi^2)_r \ , \ O_2=(\phi^4)_r \ , \ O_3=(\phi\partial^4\phi)_r \ ,
\end{align}
then for zero momentum transfer, the RGE are two-loop exact and read 
\begin{align}
&\frac{\partial O_i}{\partial \ln \mu^2}=\gamma_{ij}O_j \ ,  \\ 
&\gamma_{0i}=0 \ , \ \gamma_{10}=\frac{1}{4\pi} \ , \gamma_{21}=\frac{3}{2\pi} \ , \ \gamma_{30}=\frac{m_r^4}{4\pi} \ ,  \ \gamma_{31}=\frac{gm_r^2}{4\pi} \ ,  \ \gamma_{32}=\frac{g^2}{16\pi} \ .
\end{align}
Given the above, one has the coefficient functions for renormalized operators 
\begin{align}
& C_{{\rm I},r}^{(4)}=\frac{g^2}{6}C_0+\frac{g^2}{8\pi \epsilon}C_1+\frac{g^2}{(4\pi \epsilon)^2}\frac{C_2}{2}+m_r^4C_2+\left(\frac{g^2}{384 \pi ^3 \epsilon^3}+\frac{ m_r^4}{4 \pi  \epsilon}\right)C_3 \ , \\
& C_{\phi^2,r}^{(4)}=\frac{g^2}{2}C_1+\frac{g^2}{2\pi \epsilon}\frac{C_2}{2}+gm_r^2C_2+\bigg(\frac{g^2}{32 \pi ^2 \epsilon^2}+\frac{g m_r^2}{4 \pi  \epsilon}\bigg)C_3 \ , \\
& C_{\phi^4,r}^{(4)}=\frac{g^2}{4}C_2+\frac{g^2}{16 \pi  \epsilon}C_3 \ , \\
& C_{\phi(i\partial)^4\phi,r}^{(4)}=C_3 \ .
\end{align}
In the $\epsilon \rightarrow 0$ limit, all the above are finite. The explicit results read
\begin{align}
&C_{{\rm I},r}^{(4)}=\frac{1}{(4\pi)^3}\bigg(-\frac{L_\mu^3}{24}+\frac{3 L_\mu^2}{8}-\frac{23L_\mu}{16}+\frac{9}{4}\bigg)\left(\frac{z^2g}{4}\right)^2+\frac{(3-L_\mu)}{16 \pi }\left(\frac{z^2m_r^2}{4}\right)^2 \ , \\ 
&C_{\phi^2,r}^{(4)}=\frac{1}{(4\pi)^2}\bigg(\frac{ L_\mu^2}{8}-\frac{3L_\mu}{4}+\frac{23}{16}\bigg)\left(\frac{z^2g}{4}\right)^2+\frac{ (3-L_\mu) g}{16 \pi }\left(\frac{z^2m_r}{4}\right)^2 \ , \\
&C_{\phi^4,r}^{(4)}=\frac{1}{4}\frac{(3-L_\mu)}{16 \pi }\left(\frac{z^2g}{4}\right)^2 \ , \\
&C_{\phi(i\partial)^4\phi,r}^{(4)}=\frac{z^4}{64} \ .
\end{align}
Here in the $L_{\mu}$, $m$ is replaced by the $\overline{\rm MS}$ scheme $\mu$. In fact, by requiring the cancellation of $\frac{1}{\epsilon}$ poles, the three series of UV poles for $O_3\rightarrow O_i$ mixing 
\begin{align}
& Z_{30}=\frac{g^2}{384 \pi ^3 \epsilon^3}+\frac{ m_r^4}{4 \pi  \epsilon} \ , \\
& Z_{31}=\frac{g^2}{32 \pi ^2 \epsilon^2}+\frac{g m_r^2}{4 \pi  \epsilon} \ ,  \\ 
& Z_{32}=\frac{g^2}{16\pi \epsilon} \ ,
\end{align}
can all be uniquely determined,  once the renormalization of $m_0$ and $\langle \phi^2 \rangle$ are fixed by $z^0$ and $z^2$ power considerations. 

It is easy to see that the OPE coefficients above satisfies the RGE $\frac{d}{d\ln \mu}C_i=-C_j\gamma_{ji}$ and the full expansion is RGE invariant. In particular, to order $g^2$, the expansion of $S_{g^2}(z^2m^2,g^2)$ at the order $z^4m^4$ in Eq.~(\ref{eq:fullexpansion1}) 
can be reproduced by using the tree-level condensates Eqs.~(\ref{eq:treesquare}, \ref{eq:treequartic}) for $\langle \phi^2\rangle_r$, $\langle \phi^4\rangle_r$, and the $g^2$ order condensate $m^4I_r+\frac{g^2}{6}I_r^3-\frac{g^2}{6}\frac{49\zeta_3}{4(4\pi)^3}$ for $\langle \phi(\partial^4)\phi \rangle_r$. Beyond the $g^2$ order, all contributions are included in the three non-trivial condensates. In particular, one has 
\begin{align}
\langle \phi \partial^4\phi\rangle_r=&m^4I_r+\frac{g^2 I_r^3}{6}+gI_r\left(m^2+\frac{gI_r}{2}\right)\langle :\phi^2:\rangle+\frac{g^2I_r}{4}\langle:\phi^4:\rangle+{\cal R}\langle \phi \partial^4 \phi\rangle \ ,
\end{align}
where $I_r=\frac{1}{4\pi}\ln \frac{\mu^2}{m^2}+{\cal O}(\epsilon)$ and ${\cal R}(\phi \partial^4 \phi)=-\frac{g^2}{6}\frac{49\zeta_3}{4(4\pi)^3}+{\cal O}(g^3)$.  In terms of the above, the full expansion of the two point function at ${\cal O}(z^4)$ reads
\begin{align}
S^{(4)}\left(z^2m^2,\frac{g}{m^2}\right) =&\frac{1}{(4\pi)^3}\bigg(-\frac{L^3}{24}+\frac{3 L^2}{8}-\frac{23L}{16}+\frac{9}{4}\bigg)\left(\frac{z^2g}{4}\right)^2+\frac{(3-L)}{16 \pi }\left(\frac{z^2m^2}{4}\right)^2 \nonumber \\ 
&+\bigg[\frac{1}{(4\pi)^2}\bigg(\frac{ L^2}{8}-\frac{3L}{4}+\frac{23}{16}\bigg)\left(\frac{z^2g}{4}\right)^2+\frac{ (3-L) g}{16 \pi }\left(\frac{z^2m}{4}\right)^2\bigg]\langle:\phi^2:\rangle \nonumber \\ 
&+\frac{1}{4}\frac{(3-L)}{16 \pi }\left(\frac{z^2g}{4}\right)^2 \langle :\phi^4: \rangle +\frac{z^4}{64} {\cal R}\langle \phi(i\partial)^4\phi \rangle \ .
\end{align}
It is interesting to notice that by combining the short distance expansions at the orders $z^0$, $z^2$ and $z^4$, the parameters $m^2$, $g$ and all the condensates can be determined. On the other hand, if one only have access to the large distance correlation functions, then it is very difficult to determine these quantities. The fact that $g$, $m^2$ and the condensates are obtained from short distance measurements indicates that they are intrinsically short distance quantities, despite the fact that condensates are also sensitive to large distance fluctuations.  It is also interesting to notice that by Fourier transforming the coordinate space expansion term by term, one obtains the momentum-space expansion. On the other hand, from the momentum-space expansion, one can not Fourier-transforming back, unless additional subtraction terms at $\vec{p}=0$ are included to make the expansion tempered.

\section{Asymptotic behavior of power-expansion/OPE in a large-$N$ quartic $O(N)$ model: factorial enhancements and their cancellation}\label{sec:OPEdiverge}
Given the example in the previous section, we can see that the renormalized coefficient functions for various operators can be calculated in massless fixed-point perturbation theory with IR subtractions.  One of the natural questions is, what is the high-order behavior of the coefficient functions, when regarded as an asymptotic expansion in $z^2g$? 

For example, let's consider the coefficient function $C_{\rm I}$ for the identity operator that arises from $\phi(z)\times \phi(0)$. The $C_{\rm I}$ has an expansion
\begin{align}
C_{\rm I}=\sum_{n=0}^{\infty}z^{2n}g^n\sum_{k=0}^n \left(\frac{m_r^2}{g}\right)^kC_{n,k}(L_\mu) \ ,
\end{align}
where $C_{n,k}$ depends only on $L_\mu$.  Here we focus on the terms $C_{n,k}$. These terms are nothing but the IR renormalized two-point functions in the massless perturbation theory at the order $n-k$ with $k$ mass insertions. Heuristically,  one may argue that the total numbers of Feynman diagrams grow as $(n)!$, while being the $n$-th power in the small $z^2$ expansion introduces another factor of $\frac{1}{(n!)^2}$. Thus $C_{n,k}$ may still be suppressed by $\frac{1}{n!}$, suggesting an infinite convergence radius for the power expansion. However, the fact that $C_{n,k}$ requires IR-renormalization, may introduce additional factorial growth, in a way similar to the famous IR-renormalon problem~\cite{tHooft:1977xjm,Parisi:1978az,Parisi:1978bj} for the logarithmic asymptotics in marginal theory. 

To investigate this issue, one may again use the large-$N$ expansion trick. Consider the following $O(N)$ generalization of the scalar $\phi^4$ theory in 2D~\footnote{In~\cite{Marino:2019fvu}, this model has been studied in the $O(N)$-broken false vacuum and a renormalon has been observed in the coupling-constant expansion of the free-energy. In our work, we are interested in the OPE in the physical unbroken phase, and it remains to see if there are connections between the renormalon found in~\cite{Marino:2019fvu} and the factorial enhancements observed in our work.}
\begin{align}
S=\mu_0^{D-2}\int d^Dx \bigg(\frac{1}{2}\partial_{i}\Phi^a\partial_{i}\Phi^a+\frac{m^2+\delta m^2}{2}\Phi^a\Phi^a+\frac{g}{8N}:(\Phi^a\Phi^a): :(\Phi^a\Phi^a): \bigg)\ .
\end{align}
Here $a=1,2,3,...N$ with summation convention implied, and we have use again the bare mass $m_0^2=m^2+\delta m^2-\frac{g}{2}I_0$. Here, the Wick-ordering is performed with respect to the free Gaussian field with mass $m$. Notice that $\delta m^2={\cal O}\left(\frac{g}{N}\right)$ which we chose to cancel the one-loop self-energy correction at $\frac{1}{N}$. Equivalently, one may write in terms of the overall Wick-ordering $:(\Phi^a\Phi^a)^2:$. Also notice that the normalization for the quartic term is different from that of the $N=1$. Unlike the $N=1$ case, for $N\ge 3$ it is normally conjectured in a way similar to the lattice $O(N)$ model~\cite{Polyakov:1975rr} that there is no phase transition as one increases the coupling constant. 
To simplify notation, from now on we use 
\begin{align}
\Phi^2\equiv \sum_{a=1}^N\Phi^a\Phi^a \ , \ \Phi(i\partial)^{2n}\Phi \equiv\sum_{a=1}^N \Phi^a(-\partial^2)^n\Phi^a \ , 
\end{align}
 to denote $O(N)$-contracted quadratic operators.

Here we consider the coefficient function for the identity operator in the OPE exapsion in the short distance limit  $z^2g\ll1, z^2m^2\ll1$, for again the two point function $\langle\Phi^a(z)\Phi^b(0)\rangle \rightarrow \delta^{ab} C_{\rm I}(z) + ....$.  The IR structure is irrelevant for this limit. 
Clearly, one has the large $N$ expansion $C_{\rm I}=C_{\rm I,L}+\frac{1}{N}C_{\rm I,NL}+...$ . To calculate $C_{\rm I,{\rm NL}}$, notice that we only need to consider a massless bubble chain and subtract out the IR sub-divergences to obtain the IR renormalized coefficient functions. To further amplify the factorial enhancements, and to make contact with previous investigations, in the following we perform the calculation in the momentum space expansion, but will comment on the features of the coordinate space expansion.

\subsection{Coefficient functions from IR renormalized bubble chain}
\begin{figure}[htbp]
    \centering
    \includegraphics[height=3.5cm]{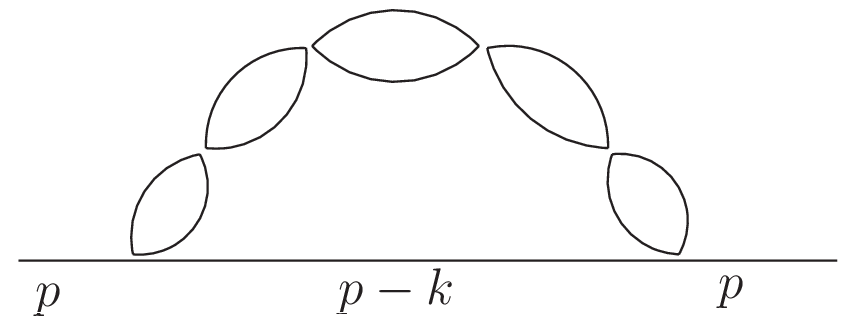}
    \caption{The massive bubble-chain diagram with $n=5$ bubbles. Solid lines are massive scalar propagators. Notice in the figure we have separated the quartic interaction vertices slightly apart just to show the contractions among the $O(N)$ indices, no point-spiting-type regularization has been implied. At the order $\frac{1}{N}$ for the two-point function $S(p^2)$, the $O(N)$ indices only contract within each bubbles.}
    \label{fig:bubblefull}
\end{figure}

To be more precise, we consider the $\frac{1}{N}$ order two-point function in the momentum-space
\begin{align}
& S_{\rm NL}(p^2)(p^2+m^2)^2=\frac{g}{N}\int \frac{d^2k}{(2\pi)^2}\frac{1}{(p-k)^2+m^2}\frac{g F(k^2)}{1+gF(k^2)} +\frac{g}{2N}\langle :\Phi^2: \rangle\ ,  \label{eq:fulloneverN}\\
& F(k^2)=\frac{1}{2}\int\frac{d^2l}{(2\pi)^2}\frac{1}{(l^2+m^2)((k-l)^2+m^2)}=\frac{1}{4\pi k^2}\frac{1}{A(k^2)}\ln \frac{A(k^2)+1}{A(k^2)-1} \ , \label{eq:defF}\\
& A(k^2)=\sqrt{1+\frac{4m^2}{k^2}} \ .
\end{align}
Clearly, all contributions are absolutely convergent. The factor $\frac{gF}{1+gF}$ clearly has the interpretation of a massive chain of bubbles starting with bubble number $n=1$. The $n=0$ contribution is logarithmically divergent and subtracted by $\delta m^2$. We would like to examine the OPE of the above. First notice that the last term in Eq.~(\ref{eq:fulloneverN}) is due to the $\frac{1}{N}$ order tadpole contribution and contains no factorial behavior at all, so we focus on the the first term. See Fig.~\ref{fig:bubblefull} for a depiction of the massive bubble chain diagram with $n=5$.

\begin{figure}[htbp]
    \centering
    \includegraphics[height=5.0cm]{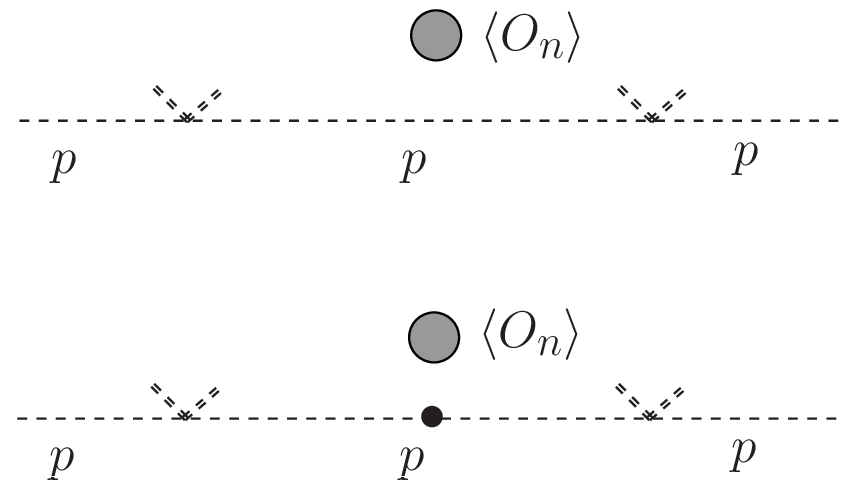}
    \caption{The contribution of the operator $O_n$ due to the soft $k^2$ region, in the presence of zero (upper) and one (lower) mass-insertions after expanding the $\frac{1}{(p-k)^2+m^2}$ at small $k$.
    The dashed line represents massless scalar propagator, black dot denotes the mass insertion and the gray-blob denotes the operator condensate $\langle O_n \rangle$. As the case of $\phi_2^4$, double-lines represent ``amputated soft external fields'' to keep track the operators associated to the coefficient functions.  }
    \label{fig:operatorOn}
\end{figure}

By investigating the regions, the operators beside the identity operator that we will need are the following:
\begin{enumerate}
    \item Tree-level operators $\prod_{l}\frac{1}{N}\Phi(i\partial)^{2k_l}\Phi$. At the order $\frac{1}{N}$, these operators are responsible for the IR sub-contributions within the bubble chain when $k^2$ itself is in the hard region, as well as the IR contribution on the $(p-k)$ line when $p-k$ becomes soft (this means $k$ is hard). Clearly, the ``tree-level'' insertions within the bubble-chain are mutually independent and can be performed separately within the individual bubbles. Indeed, when $k$ is hard, only one line within each bubble can become soft, and the Taylor expansion for that soft momentum do not affect other bubbles. To perform the insertion responsible for the small $p-k$ region, one must Taylor expand the momentum flowing into the bubble-chain, but the condensates formed by making  $p-k$ soft are still the tree-level condensates and do not correlate with the tree-level condensates formed within the bubble chain. At the external legs and any hard internal lines, there can also be an arbitrary number of tree-level $\frac{1}{N}\Phi^2$ insertions that always combine together with the $m_r^2$ insertion to form the $m^2$.  All these tree-level condensates can be removed by simply setting $\mu^2=m^2$. 
    \item The non-trivial operators $O_n=\frac{g^2}{4N}\Phi^2 (i\partial)^{2n-2}\Phi^2$ with $n\ge 1$\footnote{Notice that for $O_1$ there are ``disconnected contributions'' $\frac{g^2}{4N}\langle \Phi^2 \rangle^2$ formed by contracting each $\Phi^2$s. Only connected contributions contribute to the bubble-chain diagram, while other contributions correspond to reducible $\Phi^2$ insertions along the propagator. They combine to form the full $O_1$ operator contribution as the case of $\phi_2^4$. However, up to the next to leading order in $\frac{1}{N}$, disconnected contributions always contain a tree-level tadpole and vanish at $\mu^2=m^2$. Since this point do not affect $n\ge 2$, we will not distinguish connected and disconneced contributions in the text.}.  Clearly, these operators are responsible for the contributions when $k$ becomes soft. On the other hand, their coefficient functions are relatively simple and can be obtained by Taylor expanding the $\frac{1}{(p-k)^2+m^2}$ at small $k$. For example, with zero and one mass insertion, one has in symmetric states after angular averaging over $k$: 
    \begin{align} 
&\frac{1}{(p-k)^2}=\sum_{n=0}^{\infty}\frac{k^{2n}}{(p^2)^{n+1}}\frac{\Gamma (1-\epsilon) \Gamma (\epsilon+n+1)}{\Gamma (1+\epsilon) \Gamma (-\epsilon+n+1)} \ ,\label{eq:Oninsert1} \\
-&\frac{m^2}{(p-k)^4}=-\frac{m^2}{p^2}\sum_{n=0}^{\infty}\frac{k^{2n}}{(p^2)^{n+1}}\frac{(n+1)\Gamma (1-\epsilon) \Gamma (\epsilon+n+2)}{\Gamma (2+\epsilon) \Gamma (-\epsilon+n+1)} \ ,\label{eq:Oninsert2} 
\end{align}
   and with more mass insertions one can generalize without difficulty. In Appendix.~\ref{sec:angular}, we explain how such angular averages can be performed. These coefficient functions are clearly absent from any factorial growth. See Fig.~\ref{fig:operatorOn} for a depiction of these operator contributions. 
\end{enumerate}

Given the above and using the fact that all tree-level operators can be removed simply by setting $\mu=m$, the expansion at $\mu=m$ takes a very simple form
\begin{align}\label{eq:expansiongenral}
NS_{\rm NL}(p^2)-\frac{g}{2(p^2+m^2)^2}\langle :\Phi^2:\rangle=C_{{\rm I},{\rm NL}}(p^2)+\sum_{n=1}^{\infty}\frac{\langle O_n\rangle}{(p^2)^{n}}\frac{\, _2F_1\left(n,n;1;-\frac{m^2}{p^2}\right)}{(p^2+m^2)^2} \ ,
\end{align}
where $\langle O_n\rangle$ denotes the renormalized operator condensates at $\mu^2=m^2$. Notice that the disconnected contributions to $O_1$ also vanish at $\mu^2=m^2$ to the desired $\frac{1}{N}$ order. To obtain the last term in Eq.~(\ref{eq:expansiongenral}), one needs to expand the $\frac{1}{(p-k)^2+m^2}$ to the order $k^{2n-2}$ and perform the angular average, see Appendix.~\ref{sec:angular} for more details.  For other values of $\mu$, there will be many tree-level operators as discussed above, but the coefficient function $C_{\rm I; {\rm NL}}$ remains essentially the same form with $m_r^2$ replacing $m^2$ in the pre-coefficients, while $\mu^2$ replacing $m^2$ in the logarithms. For coefficient functions of $O_n$, one simply replacing $m$ with $m_r$. From the above, it is clear that the coefficient functions for $O_n$ contain no factorial growths at all. Thus, to search for factorial growths, it is sufficient to start from the coefficient function for the identity operator. We can write the $\frac{1}{N}$ order coefficient function of the identity operator for $S_{\rm NL}(p^2)$ in the general form as 
\begin{align}
C_{{\rm I},{\rm NL}}(p^2)=\frac{1}{(p^2+m_r^2)^2}\sum_{n=1}^{\infty} C^{(n)}_{\rm I,NL}(p^2) \ , \\
C^{(n)}_{\rm I,NL}(p^2)=\frac{g^{n+1}}{(p^2)^{n}}\sum_{k=0}^{n-1}\left(\frac{m_r^2}{g}\right)^k C_{n,k}(l_{\mu}) \ ,
\end{align}
where $C_{n,k}$ is a polynomial on $l_{\mu}=\ln \frac{p^2}{\mu^2}$. Notice $C_{n,n}=0$. Since merely mass insertion will not cause the factorial growth, as can be checked from simple calculations in lower-order diagrams, we start from $C_{n,0}$.

\subsubsection{Calculation of $C_{n,0}$: factorial growth from $\frac{1}{\epsilon} \times \epsilon$}

\begin{figure}[htbp]
    \centering
    \includegraphics[height=7.0cm]{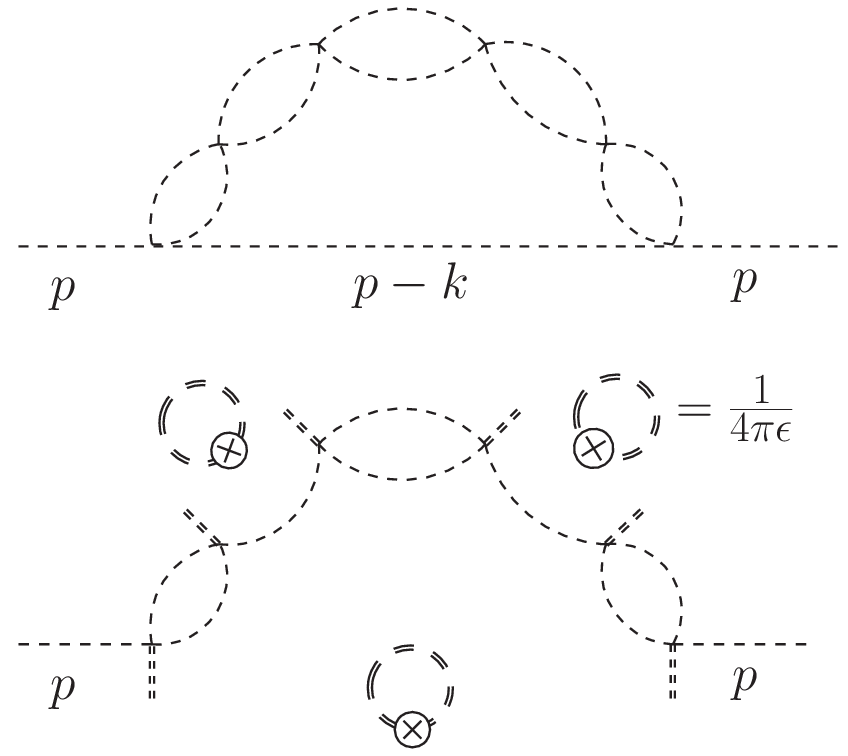}
    \caption{The massless bubble chain diagram contributing to $C_{n,0}$ with $n=5$ without mass insertions (upper) and the IR subtractions (lower) excluding the one for $O_n$. Since the contraction of $O(N)$ indices is clear, we no longer separate the quartic interactions in the figure. The IR subtractions on multiple bubbles factorize and can be performed separately in each bubble. Similarly, the IR counter-term for the $p-k$ line also factorizes.}
    \label{fig: masslessbub}
\end{figure}

\begin{figure}[htbp]
    \centering
    \includegraphics[height=5.0cm]{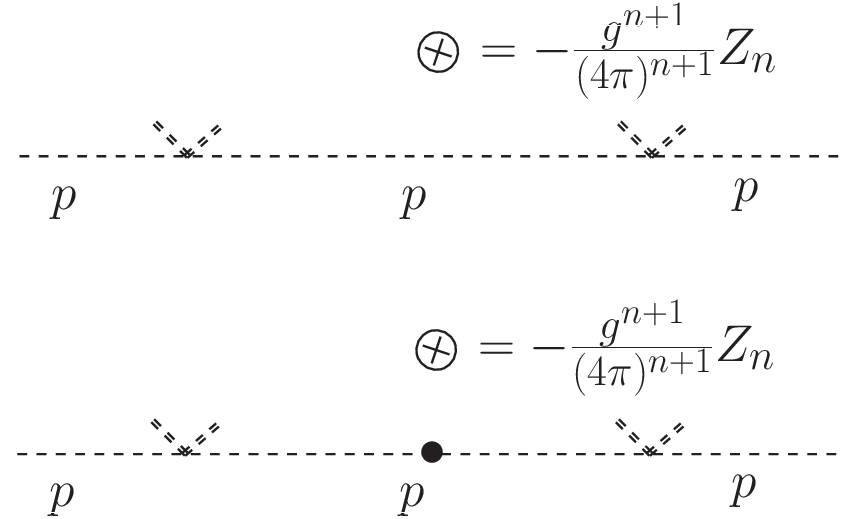}
    \caption{The IR counter-terms due to UV poles of the operator $O_n$, in the presence of zero (upper) and one (lower) mass-insertions after expanding the $\frac{1}{(p-k)^2+m^2}$ at small $k$.
    The crossed circle denotes the $g^{n+1}$ order operator UV poles of the $O_n \rightarrow {\rm I}$ mixing. These contributions should be added back to $C_{n,0}$ and $C_{n+1,1}$ respectively.}
    \label{fig: operator-counter}
\end{figure}

Given the above, let's consider the calculation detail. For individual bubble chain, one needs the building block that already appeared earlier in Eq.~(\ref{eq: C1})
\begin{align}
\frac{1}{2}C_1=\frac{\mu_0^{2-d}}{2}\int \frac{d^dk}{(2\pi)^d}\frac{1}{k^2(p-k)^2}=\frac{1}{p^2}\left(\frac{\mu^2}{p^2}\right)^{\epsilon}f(\epsilon)\times \frac{-1}{4\pi \epsilon} \ , 
\end{align}
where
\begin{align}\label{eq:fe}
f(\epsilon)=\frac{\sqrt{\pi } 2^{2 \epsilon} e^{\gamma_E  \epsilon} \Gamma (1-\epsilon) \Gamma (1+\epsilon)}{\Gamma \left(\frac{1}{2}-\epsilon\right)}=1-\frac{\pi^2}{12}\epsilon^2+{\cal O}(\epsilon^3) \ .
\end{align}
Given the above, the bare coefficient function (in the below we use hat to denote bare or un-subtracted coefficient functions) corresponding to the $n$-bubble chain diagram (depicted in the upper of Fig.~\ref{fig: masslessbub} with $n=5$) reads
\begin{align}
&\frac{g^{n+1}}{(p^2)^{n}}\hat C_{n,0}=(-1)^{n+1}(-1)^n\frac{g^{n+1}}{(4\pi)^n}\frac{f^n(\epsilon)}{\epsilon^n}\times \mu^{2\epsilon}\times\left(\frac{e^{\gamma_E}}{4\pi}\right)^{\epsilon}\int\frac{d^dk}{(2\pi)^d}\left(\frac{\mu^2}{k^2}\right)^{n\epsilon}\frac{1}{(k^2)^n}\frac{1}{(p-k)^2} \nonumber \\ 
&=-\frac{g^{n+1}}{(4\pi)^n}\frac{f^n(\epsilon)}{\epsilon^n}\times\left(\frac{e^{\gamma_E}}{4\pi}\right)^{\epsilon}\times \mu^{2(n+1)\epsilon}\frac{1}{(4\pi)^{1-\epsilon}}\int_0^\infty \frac{\rho}{\rho^{1-\epsilon}} d\rho\int_0^1dx \frac{(\rho x)^{n+n\epsilon-1}}{\Gamma(n+n\epsilon)}e^{-p^2\rho x(1-x)} \nonumber \\ 
&=-\frac{g^{n+1}}{(4\pi)^{n+1}}\frac{f^n(\epsilon)}{\epsilon^n}e^{\epsilon\gamma_E}\times \frac{1}{(p^2)^n}\left(\frac{\mu^2}{p^2}\right)^s\frac{\Gamma (-\epsilon)  \Gamma (1-n-s) \Gamma (n+s)}{\Gamma (1-n-\epsilon-s) \Gamma (n+s-\epsilon)} \ ,
\end{align}
where we have introduced the factor $s=(n+1)\epsilon$. The bared coefficient function thus can be expressed as 
\begin{align}
 \hat C_{n,0}=  -\frac{1}{(4\pi)^{n+1}}\frac{G(n,\epsilon,s)}{\epsilon^{n+2}(n+1)} \ , 
\end{align}
where
\begin{align}
G(n,\epsilon,s)=f^{\frac{s}{\epsilon}-1}(\epsilon)e^{\epsilon\gamma_E}\left(\frac{\mu^2}{p^2}\right)^s\frac{\epsilon\Gamma (-\epsilon)  s\Gamma (1-n-s) \Gamma (n+s)}{\Gamma (1-n-\epsilon-s) \Gamma (n+s-\epsilon)}  \ .
\end{align}
The crucial property of the above is that it is analytic at $s=0$ and $\epsilon=0$. Now, the IR subtraction for each of the bubble can be performed by adding the operator UV poles $\frac{1}{4\pi \epsilon}$ in Eq.~(\ref{eq:constants1}) for $\frac{1}{N}\Phi^2$ insertions replacing one of the two propagators in $C_1$ (clearly there are two ways, thus the subtraction is $2\times \frac{1}{2}\times\frac{1}{4\pi \epsilon}\times\frac{1}{p^2}$, where $\frac{1}{p^2}$ is another propagator Taylor-expanded to the zeroth order):
\begin{align}
\frac{1}{2}C_1 \rightarrow \frac{1}{p^2}\times \frac{-1}{4\pi \epsilon }\left(\left(\frac{\mu^2}{p^2}\right)^{\epsilon}f(\epsilon)-1\right) \ .
\end{align}
Therefore, the bubble-chain-partly-renormalized diagram can be obtained by replacing $s\rightarrow (n+1-k)\epsilon$ with $k$ ranging from $0$ to $n$ and summing over $k$ using the binomial theorem: 
\begin{align}\label{eq:Rbc}
R_{b} \hat C_{n,0}=-\frac{1}{(4\pi)^{n+1}}\sum_{k=0}^n(-1)^k\binom{n}{k}\frac{G(n,\epsilon,(n+1-k)\epsilon)}{\epsilon^{n+2}(n+1-k)} \ .
\end{align}
Notice we used the notation $R_bF$ to denote the {\it bubble-chain-partly-renormalized} quantity $F$, for which the IR sub-divergences within the bubble chain (not including the entire bubble-chain) have been subtracted.  See Fig.~\ref{fig: masslessbub} for a depiction of the bubble-chain diagram and IR subtractions for $C_{n,0}$ with $n=5$.
Now, one needs the standard summation formulas~\cite{Palanques-Mestre:1983ogz,Beneke:1994sw,Beneke:1995pq}, as well as the new one (this is actually related to the second Stirling number $S(n+1,n)$)
\begin{align}
\sum_{k=0}^n(-1)^k\binom{n}{k}(n-k+1)^{n+1}=\frac{(n+2)!}{2} \ .
\end{align}
To show this, one writes
\begin{align}
&\sum_{k=0}^n(-1)^k\binom{n}{k}(n-k+x)^{n+1}\nonumber \\ 
&=x(n+1)\sum_{k=0}^n(-1)^k\binom{n}{k}(n-k)^{n}+\sum_{k=0}^n(-1)^k\binom{n}{k}(n-k)^{n+1} \nonumber \\ 
&=x(n+1)n!+n!\binom{n+1}{2}=\frac{n!(n+1)(n+2x)}{2} \ .
\end{align}
This is due to the fact that $\sum_{k=0}^n(-1)^k\binom{n}{k}(n-k)^{n+1} $ is nothing but $n!S(n+1,n)$ where $S(n+1,n)$ is the Stirling number of the second kind counting the partition of $n+1$ objects into $n$ unlabeled sets. Given this summation formula, one obtains up to all constant terms, 
\begin{align} \label{eq:partlyCn0}
& R_{b}\hat C_{n,0}\nonumber \\ 
&=-\frac{1}{(4\pi)^{n+1}}\bigg(\frac{(-1)^n}{(n+1)\epsilon^{n+2}}G(n,\epsilon,0)+\frac{1}{(n+1)\epsilon}\frac{d^{n+1}}{ds^{n+1}}G(n,\epsilon,s)+\frac{1}{2}\frac{d^{n+2}}{ds^{n+2}}G(n,\epsilon,s)\bigg)_{s=0} \ .
\end{align}
The term that causes the trouble is the second one. We need 
\begin{align}
& G(n,\epsilon,s)\nonumber \\ 
&=-s \left(\frac{\mu^2}{p^2}\right)^s+\frac{\epsilon}{12}\left(\frac{\mu^2}{p^2}\right)^s\bigg(-24s\gamma_E+\pi^2s^2-12s\psi(1-n-s)-12s\psi(n+s)\bigg)+{\cal O}(\epsilon^2) \ .
\end{align}
Now, the first term will contribute to a $\ln p^2$-dependent $\frac{1}{\epsilon}$ pole,
\begin{align}
-\frac{1}{(n+1)\epsilon}\times \frac{d^{n+1}}{ds^{n+1}} s \left(\frac{\mu^2}{p^2}\right)^s\bigg|_{s=0}=-\frac{1}{\epsilon}\ln^{n}\frac{\mu^2}{p^2} \ .
\end{align}
This term must be canceled by subtracting the IR contribution from the $p-k$ propagator. Since here the IR counter term is again the UV pole $\frac{1}{4\pi \epsilon}$ for the $\frac{1}{N}\Phi^2$ operator, one simply needs to multiply this UV pole  with the IR-subtracted bubble chain at the momentum $k=p$. Therefore, one has
\begin{align}\label{eq:contertermpminusk}
T_{p-k}R_b \hat C_{n,0}=-\frac{1}{4\pi \epsilon}\times \frac{1}{(4\pi)^{n}\epsilon^n}\left(f(\epsilon)\left(\frac{\mu^2}{p^2}\right)^{\epsilon}-1\right)^n \ .
\end{align}
Here, $T_{p-k}F$ denotes the quantity obtained from $F$ such that the $p-k$ line have been replaced by the appropriate IR counter-term. Expanding the bracket using the binomial theorem, after using the summation formulas one reaches
\begin{align}\label{eq:TRC0}
T_{p-k}R_b \hat C_{n,0}=-\frac{1}{(4\pi)^{n+1}}\bigg(\frac{1}{\epsilon}\frac{d^n}{ds^n}H(\epsilon,s)\bigg|_{s=0}+\frac{n+2}{2}\frac{d^{n+1}H(\epsilon,s)}{ds^{n+1}}\bigg|_{s=0}\bigg), \nonumber \\ 
H(\epsilon,s)=f^{\frac{s}{\epsilon}-1}(\epsilon)\bigg(\frac{\mu^2}{p^2}\bigg)^{s-\epsilon} \ .
\end{align}
To further simplify, one needs the small $\epsilon$ epansion
\begin{align}
H(\epsilon,s)=\left(\frac{\mu^2}{p^2}\right)^s-\frac{\epsilon}{12} \left(\frac{\mu^2}{p^2}\right)^s \left(12 \ln \left(\frac{\mu^2}{p^2}\right)+\pi ^2 s\right)+{\cal O}(\epsilon^2) \ .
\end{align}
The $p$-dependent divergent terms are now 
\begin{align}
\frac{1}{\epsilon}\frac{d^{n}}{ds^{n}} \left(\frac{\mu^2}{p^2}\right)^s\bigg|_{s=0}=\frac{1}{\epsilon}\ln^n \frac{\mu^2}{p^2} \ ,
\end{align}
which indeed cancel with that for $R_b\hat C_{n,0}$. 
Given the above, the {\it partly renormalized} $C_{n,0}$, including all IR subtractions except the one for the operator $O_{n}$ reads
\begin{align}
R_p\hat C_{n,0}=R_b\hat C_{n,0}+T_{p-k}R_b \hat C_{n,0} \ .
\end{align}
It remains to add the operator poles for $O_n=\frac{g^2}{4N}\Phi^2(i\partial)^{2n-2}\Phi^2$. Clearly, since all the high-order derivatives in $s$ are finite in $R_p\hat C_{n,0}$, the operator poles must cancel the remaining $p$-independent divergences in the first term of $R_b\hat C_{n,0}$ after the equality in Eq.~(\ref{eq:partlyCn0}). 
Therefore, one writes fully renormalized $C_{n,0}$ as 
\begin{align}\label{eq:Cn0def}
C_{n,0}=R_p\hat C_{n,0}-\frac{1}{(4\pi)^{n+1}}Z_n\left(\frac{1}{\epsilon}\right)\frac{\Gamma(1-\epsilon)\Gamma(n+\epsilon)}{\Gamma(1+\epsilon)\Gamma(n-\epsilon)} \ ,
\end{align}
where $-\frac{Z_ng^{n+1}}{(4\pi)^{n+1}}$ denotes operator UV poles for the $O_n \rightarrow{\rm I}$ mixing that are at the order $g^{n+1}$. See Fig.~\ref{fig: operator-counter} for a depiction of the IR subtractions corresponding to UV poles of $O_n$. Clearly, since the dimension of $O_n$ is $4+2n-2=2(n+1)$, this term has the highest power in $g$ in the UV poles of $O_n$. 
We further separate the high-order derivatives in $s$ and constant terms in $C_{n,0}$:
\begin{align}
C_{n,0}=C_{n,0}^A+C_{n,0}^{B} \ , 
\end{align}
where $C_{n,0}^B$ contains all the high-order derivatives in $s$ and is finite (its final form will be given in Eq.~(\ref{eq:CBn})), while 
\begin{align}
C_{n,0}^A=-\frac{1}{(4\pi)^{n+1}}\bigg(\frac{(-1)^n}{(n+1)\epsilon^{n+2}}G(n,\epsilon,0)+Z_n\left(\frac{1}{\epsilon}\right)\frac{\Gamma(1-\epsilon)\Gamma(n+\epsilon)}{\Gamma(1+\epsilon)\Gamma(n-\epsilon)}\bigg) \ .
\end{align}
Now, for $n\in Z_{\ge 1}$ one can simplifies the above as
\begin{align}\label{eq:Cnasimp}
C_{n,0}^A=-\frac{1}{(4\pi)^{n+1}}\frac{\Gamma(1-\epsilon)\Gamma(n+\epsilon)}{\Gamma(1+\epsilon)\Gamma(n-\epsilon)}\bigg(Z_n+\frac{1}{\epsilon^{n+1}}\frac{2 (-1)^n  \sin (\pi  \epsilon) \Gamma (-2 \epsilon)}{(n+1)\pi \Gamma^2 (1-\epsilon)}\bigg) \ .
\end{align}
Clearly, the above uniquely determines the $Z$ as the pole part of
\begin{align}\label{eq:Z}
Z_n=-\text{poles of} \ \bigg(\frac{1}{\epsilon^{n+1}}\frac{2 (-1)^n  \sin (\pi  \epsilon) \Gamma (-2 \epsilon)}{(n+1)\pi \Gamma^2 (1-\epsilon)}\bigg) \ ,
\end{align}
and the renormalized constant parts are therefore
\begin{align}\label{eq:CnA}
C_{n,0}^A=-\frac{1}{(4\pi)^{n+1}}\text{finite of} \ \bigg(\frac{1}{\epsilon^{n+1}}\frac{2 (-1)^n  \sin (\pi  \epsilon) \Gamma (-2 \epsilon)}{(n+1)\pi \Gamma^2 (1-\epsilon)}\bigg) \ . 
\end{align}
Clearly, this part is free from any factorial growth since the function $\frac{\sin (\pi  \epsilon) \Gamma (-2 \epsilon)}{\Gamma^2 (1-\epsilon)}$ is analytic in a neighborhood of $\epsilon=0$, thus the Taylor expansion coefficients can only growth exponentially. In Appendix.~\ref{sec:operatorms}, we show that the $Z_n$ obtained here completely agrees with direct analysis of the $\epsilon$-dependent operator condensates

Now, one collects all the terms that are due to high order derivatives in $s$. One has
\begin{align}
C_{n,0}^B=-\frac{1}{(4\pi)^{n+1}}\bigg(\frac{1}{n+1}\frac{d^{n+1}}{ds^{n+1}}{\cal H}_1(s)+\frac{d^n}{ds^n}{\cal H}_2(s)+\frac{1}{2}\frac{d^{n+2}}{ds^{n+2}}{\cal H}_3(s)+\frac{n+2}{2}\frac{d^{n+1}}{ds^{n+1}}{\cal H}_4(s)\bigg)_{s=0} \ .
\end{align}
Here
\begin{align}
&{\cal H}_1(s)=\frac{1}{12}\left(\frac{\mu^2}{p^2}\right)^s\bigg(-24s\gamma_E+\pi^2s^2-12s\psi(1-n-s)-12s\psi(n+s)\bigg) \ , \\
&{\cal H}_2(s)=-\frac{1}{12} \left(\frac{\mu^2}{p^2}\right)^s \left(12 \ln \left(\frac{\mu^2}{p^2}\right)+\pi ^2 s\right) \ , \\
&{\cal H}_3(s)=-s\left(\frac{\mu^2}{p^2}\right)^s \ , \\
&{\cal H}_4(s)=\left(\frac{\mu^2}{p^2}\right)^s \ .
\end{align}
Notice that the first two terms arise due to $\frac{1}{\epsilon} \times \epsilon$ effects. For $n=1$, one obtains $C_{1,0}^A=0$, while
\begin{align}
C_{1,0}^B=C_{1,0}=\frac{1}{(4\pi)^2 }\frac{3}{2}\ln^2\frac{p^2}{\mu^2} \ ,
\end{align}
which agrees exactly with the previous calculation. For $n=2$, one has 
\begin{align}
C_{2,0}=-\frac{1}{(4\pi)^{2+1}}\bigg(\frac{4 l_{\mu}^3}{3}-2 l_{\mu}^2-4l_{\mu}+\frac{10 \zeta_3}{3}-4\bigg) \ ,
\end{align}
where $l_{\mu}=\ln \frac{p^2}{\mu^2}$. It is clear from now that $C_{n,0}^A$ and ${\cal H}_2$ to ${\cal H}_4$ terms will not contribute to factorial growth at all. However, the ${\cal H}_1$ contribution contains the very dangerous $s\psi(1-n-s)+s\psi(n+s)$ term, which can contribute to factorial growth.

To determine the leading factorial behavior, one notices that the contribution $C_{n,0}^B$ can be simplified after many cancellations:
\begin{align}\label{eq:CBn}
C_{n,0}^B(l_{\mu})=\frac{n!}{(4\pi)^{n+1}}\oint_{|s|=r<1}\frac{ds}{2\pi i s^{n+1}}e^{-sl_{\mu}}\bigg(2\gamma_E-l_{\mu}+\psi(1-n-s)+\psi(n+s)\bigg) \ .
\end{align}
Notice that in this paper, the contour integrals $\oint$ are always counter-clockwise.  Given the above, it is clear that the leading non-Borel summable large $n$ asymptotics can be extracted by shifting the contours outside and pick up the negative residues at $s=1$
\begin{align}\label{eq:Cnasym1}
C_{n,0}\bigg|_{n\rightarrow\infty,+1}=-\frac{1}{(4\pi)^{n+1}}\times n!\times \bigg(\frac{\mu^2}{p^2}\bigg)\times\left(1+\frac{\mu^2}{2^{n+1}p^2}+..\right) \ .
\end{align}
Notice that here we introduced the following notation to distinguish alternating and non-alternating asymptotics in $n$: for alternating asymptotics of a quantity $F(n)$ we write $F(n)|_{n\rightarrow\infty,(-1)^n}$, while for non-alternating one we write $F(n)|_{n\rightarrow\infty,+1}$. 
It is interesting to notice that the leading factorial growth is again one more power higher. Here there are two possibilities. The first possibility is that they cancel within the same coefficient function for the identity operator but with mass-insertions, if $\mu^2$ is identified as the mass. We will rule out this possibility after calculating the diagrams with mass-insertions. The second possibility is that they cancel with the operator contributions for $O_{n+1}=\frac{g^2}{4N}\Phi^2(i\partial)^{2n}\Phi^2$ and $O_{n+2}$...., but with the same bubble-chain number, because this diagram contains power-divergences for these operators. This would be consistent with the interpretation that the non-alternating asymptotics in Eq.~(\ref{eq:Cnasym1}) can be regarded as an IR renormalon.

It is also interesting to notice that there are also alternating factorial growths
\begin{align}\label{eq:Cnasym2}
C_{n,0}\bigg|_{n\rightarrow \infty,(-1)^n}=\frac{(-1)^n}{(4\pi)^{n+1}}\times n!\times \bigg(\frac{p^2}{\mu^2}\bigg)\times\left(1+\frac{p^2}{2^{n+1}\mu^2}+...\right) \ .
\end{align}
Later we will show that they cancel with factorial growths for operators $O_{n-1},O_{n-2}...$ but with the same numbers of bubbles.

\subsubsection{Mass insertion: $C_{n,1}$}
To rule out the possibility that the factorial enhancements above could cancel with the diagrams with mass insertions, let's consider $C_{n+1,1}$, namely coefficient functions with one mass insertions. There are three possibilities. Either the insertion is within the bubbles, or it is on the $p-k$ line, or it is on the external lines. Since the last type of contributions are trivial and has already been factorized out in our definition, here we only focus on the first two types of mass-insertions. In order to label the mass insertions, here we introduced the new notation: $C_{n;l,k}$ denotes the contribution to $C_{n,l+k}$ where $l$ mass insertions are on the $p-k$ line, while $k$ mass insertions are within the bubble chain. One clearly has $C_{n,k}=\sum_{l=0}^kC_{n;l,k-l}$. 
\begin{figure}[htbp]
    \centering
    \includegraphics[height=7.0cm]{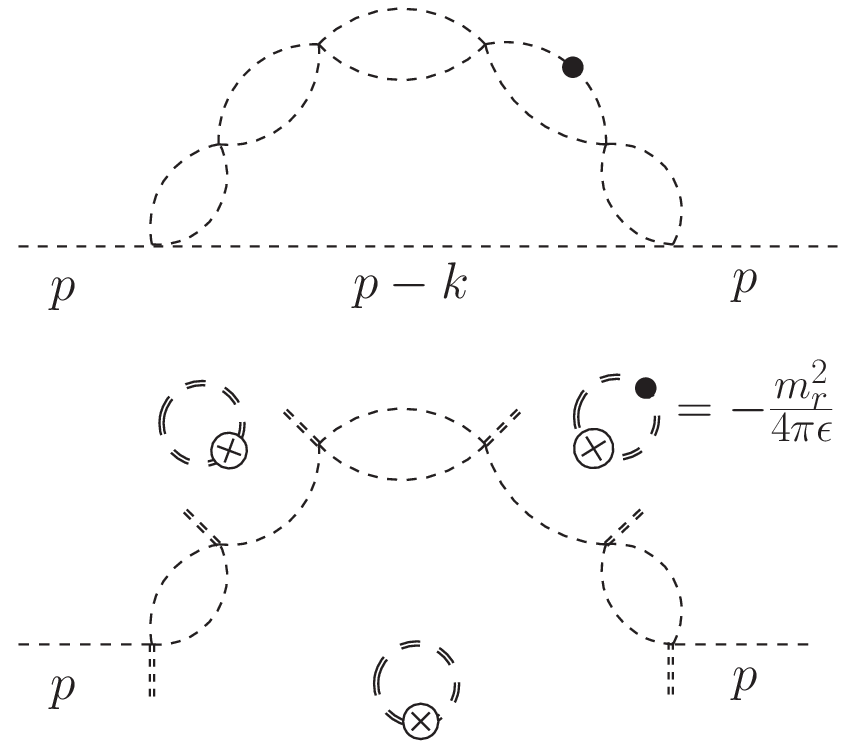}
    \caption{The massless bubble chain diagram contributing to $C_{n+1;0,1}$, with one mass insertion within the bubbles for $n=5$ (upper) and the related IR subtractions (lower). Notice that the IR subtraction for the mass-inserted propagator is due to the $a_1$ in Eq.~(\ref{eq:constants2}).  }
    \label{fig:insert1}
\end{figure}

We first consider the mass insertion within the bubble chain, see Fig.~\ref{fig:insert1} for a depiction. Notice that the mass-insertion diagram can be placed in $n$ locations in the bubble-chain with the same result, one can simply place it at the end of the bubble chain with a total number of $n$ bubbles. The bubble with one mass insertion reads
\begin{align}
& C_{m}(k^2)=-m_r^2\mu^{2\epsilon}\left(\frac{e^{\gamma_E}}{4\pi}\right)^{\epsilon}\int\frac{d^Dl}{(2\pi)^D}\frac{1}{(k-l)^2}\frac{1}{(l^2)^2}\nonumber \\
&=-m_r^2\mu^{2\epsilon}\left(\frac{e^{\gamma_E}}{4\pi}\right)^{\epsilon}\frac{1}{(4\pi)^{1-\epsilon}}\int \rho \rho^{-1+\epsilon} d\rho (\rho x)e^{-k^2\rho x(1-x)} \nonumber \\ 
&=\frac{m_r^2}{ k^4}\left(\frac{\mu^2}{k^2}\right)^{\epsilon}\frac{(1+2\epsilon)f(\epsilon)}{2\pi \epsilon} \ ,
\end{align}
with the same $f(\epsilon)$ as in Eq.~(\ref{eq:fe}). Here we need to perform the IR subtraction to $C_m$. Clearly, the $\frac{1}{(k-l)^2}$ propagator can still be re-normalized using the $\frac{1}{4\pi \epsilon}$ UV pole for $\frac{1}{N}\Phi^2$, while the $\frac{-m_r^2}{(l^2)^2}$ propagator has to be renormalized by the UV pole for the $\frac{1}{N}\Phi(i\partial)^2\Phi$ operator insertion replacing the $\frac{1}{(l^2)^2}$ line, using the Eq.~(\ref{eq:Oninsert1}) for the small-$l$ expansion rule of $\frac{1}{(k-l)^2}$ and Eq.~(\ref{eq:constants2}) for the UV pole $a_1m_r^2$, which is exactly proportional to $m_r^2$. Therefore, the IR subtraction for the $C_m$ can be performed as
\begin{align}\label{eq:Cm}
&RC_m=\frac{m_r^2}{k^4}\left(\frac{\mu^2}{k^2}\right)^{\epsilon}\frac{(1+2\epsilon)f(\epsilon)}{2\pi\epsilon}+\frac{-m_r^2}{k^4}\times\frac{1}{4\pi\epsilon}+\frac{1}{k^4}\frac{1+\epsilon}{1-\epsilon}\times a_1m_r^2\nonumber \\ 
&=\frac{2m_r^2}{k^4}\frac{1}{4\pi\epsilon}\left(f(\epsilon)\left(\frac{\mu^2}{k^2}\right)^{\epsilon}-1\right)+\frac{2m_r^2}{k^4}\frac{1}{4\pi}\left(\frac{\mu^2}{k^2}\right)^{\epsilon}2f(\epsilon)-\frac{2m_r^2}{k^4}\frac{1}{4\pi}g(\epsilon) \ ,
\end{align}
where $g(\epsilon)$ is defined as 
\begin{align}
g(\epsilon)=-\bigg(\frac{1}{\epsilon}-\frac{1}{2\epsilon}-\frac{1+\epsilon}{2\epsilon(1-\epsilon)}\bigg) \equiv \frac{1}{(1-\epsilon)} \ .
\end{align}
It is again regular at $\epsilon=0$. Notice that the $\frac{1+\epsilon}{1-\epsilon}$ coefficient in front of $a_1m_r^2$ is crucial. It leads to the finite function $g(\epsilon)$ and affects the $\epsilon \rightarrow 0$ limit
\begin{align}
\lim_{\epsilon\rightarrow0}RC_m\rightarrow m_r^2\frac{1-\ln \frac{k^2}{\mu^2}}{2\pi (k^2)^2} \ .
\end{align}
At $\mu=m$, it exactly reproduces the next-to-leading power large $k$ expansion of the $F(k^2)$ define in Eq.~(\ref{eq:defF})
\begin{align}
F(k^2) \rightarrow\frac{1}{4\pi k^2}\ln \frac{k^2}{m^2}+m^2\frac{1-\ln\frac{k^2}{m^2}}{2\pi (k^2)^2 }+... \ .
\end{align}
If we use the $a_1m_r^2$ instead of $\frac{1+\epsilon}{1-\epsilon}a_1m_r^2$, then the constant term will be $2$ instead of $1$.  Now, knowing the IR subtraction rule for the mass-inserted bubble,  one can combine with the rest $n-1$ bubbles without the mass-insertion to obtain the bubble-chain-partly-renormalized diagram
\begin{align}
&R_b \hat C_{n+1;0,1}\frac{m_r^2g^{n+1}}{(p^2)^{n+1}}=\nonumber \\ 
&\frac{2nm_r^2g^{n+1}}{(4\pi)^n}\mu^{2\epsilon}\left(\frac{e^{\gamma_E}}{4\pi}\right)^{\epsilon}\frac{1}{\epsilon^n}\int\frac{d^dk}{(2\pi)^d}\frac{1}{(k^2)^{n+1}}\bigg(f(\epsilon)\left(\frac{\mu^2}{k^2}\right)^{\epsilon}-1\bigg)^n\frac{1}{(p-k)^2} \nonumber \\
&+\frac{2nm_r^2g^{n+1}}{(4\pi)^n}\mu^{2\epsilon}\left(\frac{e^{\gamma_E}}{4\pi}\right)^{\epsilon}\frac{2f(\epsilon)}{\epsilon^{n-1}}\int\frac{d^dk}{(2\pi)^d}\frac{1}{(k^2)^{n+1}}\left(\frac{\mu^2}{k^2}\right)^{\epsilon}\bigg(f(\epsilon)\left(\frac{\mu^2}{k^2}\right)^{\epsilon}-1\bigg)^{n-1}\frac{1}{(p-k)^2} \  \nonumber \\ 
&-\frac{2nm_r^2g^{n+1}}{(4\pi)^n}\mu^{2\epsilon}\left(\frac{e^{\gamma_E}}{4\pi}\right)^{\epsilon}\frac{g(\epsilon)}{\epsilon^{n-1}}\int\frac{d^dk}{(2\pi)^d}\frac{1}{(k^2)^{n+1}}\bigg(f(\epsilon)\left(\frac{\mu^2}{k^2}\right)^{\epsilon}-1\bigg)^{n-1}\frac{1}{(p-k)^2} \ .
\end{align}
We can further split the second term to obtain
\begin{align}
&\frac{f(\epsilon)}{\epsilon^{n-1}}\int\frac{d^dk}{(2\pi)^d}\frac{1}{(k^2)^{n+1}}\left(\frac{\mu^2}{k^2}\right)^{\epsilon}\bigg(f(\epsilon)\left(\frac{\mu^2}{k^2}\right)^{\epsilon}-1\bigg)^{n-1}\frac{1}{(p-k)^2}\nonumber \\
&=\frac{1}{\epsilon^{n-1}}\int\frac{d^dk}{(2\pi)^d}\frac{1}{(k^2)^{n+1}}\bigg(f(\epsilon)\left(\frac{\mu^2}{k^2}\right)^{\epsilon}-1\bigg)^{n}\frac{1}{(p-k)^2} \nonumber \\ 
&+\frac{1}{\epsilon^{n-1}}\int\frac{d^dk}{(2\pi)^d}\frac{1}{(k^2)^{n+1}}\bigg(f(\epsilon)\left(\frac{\mu^2}{k^2}\right)^{\epsilon}-1\bigg)^{n-1}\frac{1}{(p-k)^2} \ .
\end{align}
The bubble-chain-partly-renormalized $C_{n+1;0,1}$ therefore reads
\begin{align} \label{eq:RbCn1a}
R_{b}\hat C_{n+1;0,1}=&\frac{2n}{(4\pi)^{n+1}}\sum_{k=0}^n(-1)^k\binom{n}{k}\frac{G_1(n,\epsilon,(n+1-k)\epsilon)}{\epsilon^{n+2}(n+1-k)} \nonumber \\
&+\frac{2n}{(4\pi)^{n+1}}\sum_{k=0}^n(-1)^k\binom{n}{k}\frac{2G_1(n,\epsilon,(n+1-k)\epsilon)}{\epsilon^{n+1}(n+1-k)}\nonumber \\
&+\frac{2n}{(4\pi)^{n+1}}\sum_{k=0}^{n-1}(-1)^k\binom{n-1}{k}\frac{(2-g(\epsilon))G_1(n,\epsilon,(n-k)\epsilon)}{\epsilon^{n+1}(n-k)} \ .
\end{align}
Thus, one has
\begin{align}
&R_b\hat C_{n+1;0,1}=\frac{2n}{(4\pi)^{n+1}}\nonumber \\ 
&\bigg(\frac{(-1)^nG_1(n,\epsilon,0)}{(n+1)\epsilon^{n+2}}+\frac{1}{\epsilon(n+1)}\frac{d^{n+1}}{ds^{n+1}}G_1(n,\epsilon,s)+\frac{1}{2}\frac{d^{n+2}}{ds^{n+2}}G_1(n,\epsilon,s)\nonumber \\ 
&+\frac{(-1)^n2G_1(n,\epsilon,0)}{(n+1)\epsilon^{n+1}}+\frac{2}{(n+1)}\frac{d^{n+1}}{ds^{n+1}}G_1(n,\epsilon,s)\nonumber \\ 
&+\frac{(-1)^{n-1}\kappa(\epsilon)G_1(n,\epsilon,0)}{n\epsilon^{n+1}}+\frac{\kappa(\epsilon)}{\epsilon n}\frac{d^{n}}{ds^{n}}G_1(n,\epsilon,s)+\frac{\kappa(\epsilon)}{2}\frac{d^{n+1}}{ds^{n+1}}G_1(n,\epsilon,s)\bigg)_{s=0} \ , 
\end{align}
with $\kappa(\epsilon)=2-g(\epsilon)=1-\epsilon-\epsilon^2+...$, and the $G_1$ function is defined as 
\begin{align}
G_1(n,\epsilon,s)=e^{\gamma_E\epsilon}f(\epsilon)^{\frac{s}{\epsilon}-1}\left(\frac{\mu^2}{p^2}\right)^s\frac{\epsilon\Gamma (-\epsilon)  s\Gamma (-n-s) \Gamma (n+s+1)}{\Gamma (-\epsilon-n-s) \Gamma (-\epsilon+n+s+1)}\equiv G(n+1,\epsilon,s) \ .
\end{align}
The $\ln p$-dependent $\frac{1}{\epsilon}$ terms are generated only from the first and the last two terms in Eq.~(\ref{eq:RbCn1a}), which are
\begin{align}
&\text{\rm $\ln p$-dependent poles of }(R_b\hat C_{n+1;0,1})=\frac{2n}{(4\pi)^{n+1}\epsilon}\bigg(-\frac{1}{n+1}\frac{d^{n+1}}{ds^{n+1}}s\left(\frac{\mu^2}{p^2}\right)^s\bigg|_{s=0}\bigg) \nonumber \\
&+\frac{2n}{(4\pi)^{n+1}\epsilon}\bigg(-\frac{1}{n}\frac{d^{n}}{ds^{n}}s\left(\frac{\mu^2}{p^2}\right)^s\bigg|_{s=0}\bigg) =\frac{2n}{(4\pi)^{n+1}\epsilon}\left(-\ln^{n}\frac{\mu^2}{p^2}-\ln^{n-1}\frac{\mu^2}{p^2}\right)\ .
\end{align}
They has to be canceled by the counter-term $T_{p-k}R_b\hat C_{n+1;0,1}$ for the $p-k$ line. Indeed, one has
\begin{align}
&T_{p-k}R_b\hat C_{n+1;0,1}\nonumber \\ 
&=\frac{2n}{(4\pi)^{n+1}\epsilon}\bigg(\frac{1+2\epsilon}{\epsilon^n}\left(f(\epsilon)\left(\frac{\mu^2}{p^2}\right)^{\epsilon}-1\right)^n+\frac{\kappa(\epsilon)}{\epsilon^{n-1}}\left(f(\epsilon)\left(\frac{\mu^2}{p^2}\right)^{\epsilon}-1\right)^{n-1}\bigg) \ .
\end{align}
Expanding the brackets, the $\frac{1}{\epsilon}$ poles are
\begin{align}
\text{\rm poles of }(T_{p-k}R_b\hat C_{n+1;0,1})=\frac{2n}{(4\pi)^{n+1}}\bigg(\frac{1}{\epsilon}\ln^n\frac{\mu^2}{p^2}+\frac{1}{\epsilon}\ln^{n-1}\frac{\mu^2}{p^2}\bigg) \ ,
\end{align}
which indeed cancels that from $R_b\hat C_{n+1;0,1}$. Moreover, by collecting all the high-order derivatives, one obtains after cancellations
\begin{align}
&R_p\hat C_{n+1;0,1}=C^B_{n+1;0,1}+\frac{2n}{(4\pi)^{n+1}}\frac{(-1)^nG_1(n,\epsilon,0)}{\epsilon^{n+2}}\bigg(\frac{1+2\epsilon}{n+1}-\frac{\epsilon(2-g(\epsilon))}{n}\bigg)\ ,  \\ 
&C_{n+1;0,1}^B=\frac{-2n!}{(4\pi)^{n+1}}\oint \frac{ds}{2\pi is^{n+1}}\left(n+s\right)e^{-l_{\mu}s}\bigg(2\gamma_E-l_{\mu}+\psi(-n-s)+\psi(1+n+s)\bigg) \ . \label{eq:Cmassb}
\end{align}
Clearly, the remaining $\ln p$-independent poles have to be canceled by the operator. This time, the operator poles determined are the $g^{n+1}m_r^2$ terms for $O_{n+1}\rightarrow{\rm I}$ (or $g^{n}m_r^2$ term for $O_n\rightarrow{\rm I}$). The diagram for this operator pole can still be read from the upper of Fig.~\ref{fig: operator-counter}, in which $-\frac{Z_ng^{n+1}}{(4\pi)^{n+1}}$ should be replaced by the $m_r^2g^{n+1}$-order poles for the $O_{n+1}$. In fact, by staying at the same $n$ but considering all the $C_{n;0,k}$ with $0\le k\le n-1$, one determines all the operator poles for the $O_n \rightarrow{\rm I}$ mixing. On the other hand, the $C_{n+1;0,1}^B$ contribution from high-order derivatives leads to the factorial growth. By shifting the contour outside, the leading non-alternating factorial growth reads:
\begin{align}\label{eq:massfac1}
C_{n;0,1}\bigg|_{n\rightarrow\infty, +1}=\frac{2n!}{(4\pi)^n}\frac{\mu^2}{p^2}\bigg(1+\frac{\mu^2}{p^2}\frac{(n+1)}{2^{n+1}n}+.....\bigg)\ .
\end{align}
Thus, for generic values of $\frac{g}{m^2}$ it will not cancel the factorial growth of $C_{n,0}$. 

\begin{figure}[htbp]
    \centering
    \includegraphics[height=7.0cm]{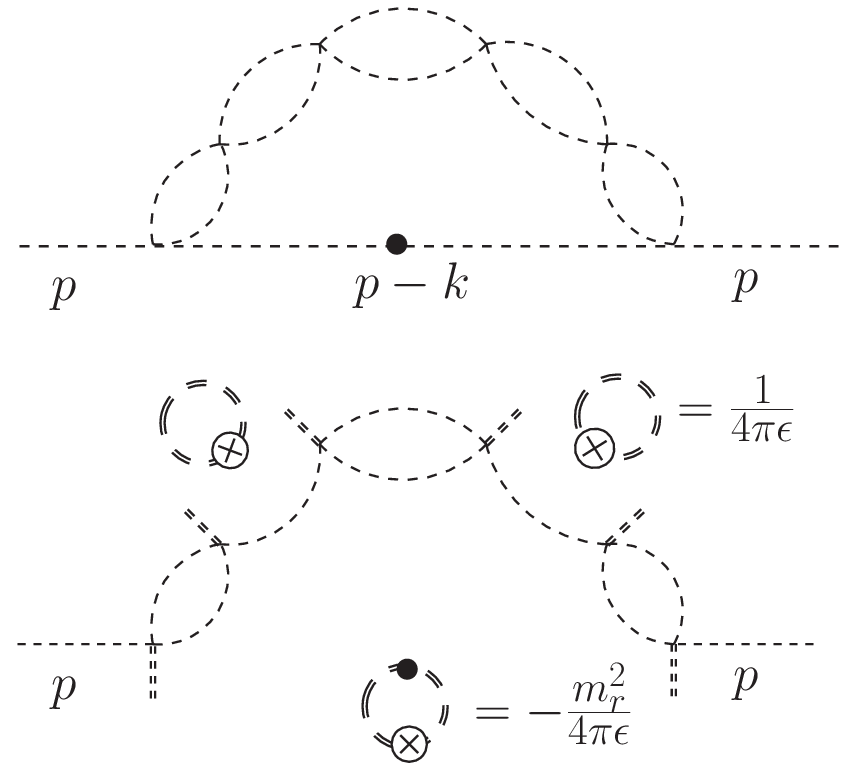}
    \caption{The massless bubble chain diagram contributing to $C_{n+1;1,0}$, with one mass insertion one the $p-k$ line $n=5$ (upper) and the related IR subtractions (lower). Notice that the IR subtraction for the mass-inserted $p-k$ propagator is again due to the $a_1$ in Eq.~(\ref{eq:constants2}). Notice that here although the IR counter-term $a_1$ do not correlate with IR counter-terms within the bubble chain, to perform the IR subtraction one must Taylor expand the IR subtracted bubble chain to the second order, see last term in Eq.~(\ref{eq:masssub}).}
    \label{fig:insert2}
\end{figure}

The mass-insertion on the $\frac{1}{(p-k)^2}$ line can also be calculated similarly, see Fig.~\ref{fig:insert2} for a depiction. This contribution is more illustrative than the mass-insertion within the bubble chain, since it requires the same operator poles as that of $O_n$ determined from the $C_{n,0}$ calculation, therefore it serves as a consistency check for the IR subtraction method. 
Here we can write directly the partly renormalized result (partly-renormalized-bubble-chain plus the IR counter-term on $p-k$)
\begin{align}
&R_p \hat C_{n+1;1,0}\frac{m_r^2g^{n+1}}{(p^2)^{n+1}}= \nonumber \\
&\frac{m_r^2g^{n+1}}{(4\pi)^n}\mu^{2\epsilon}\left(\frac{e^{\gamma_E}}{4\pi}\right)^{\epsilon}\frac{1}{\epsilon^n}\int\frac{d^dk}{(2\pi)^d}\frac{1}{(k^2)^{n}}\bigg(f(\epsilon)\left(\frac{\mu^2}{k^2}\right)^{\epsilon}-1\bigg)^n\frac{1}{((p-k)^2)^2} \nonumber \\ 
&+\frac{m_r^2g^{n+1}}{(4\pi)^{n+1}}\frac{1}{\epsilon^{n+1}}\left(\frac{\partial}{\partial p^2}+\frac{2p^2}{d}\frac{\partial^2}{\partial(p^2)^2}\right)\frac{1}{(p^2)^n}\bigg(f(\epsilon)\left(\frac{\mu^2}{p^2}\right)^{\epsilon}-1\bigg)^n \ . \label{eq:masssub}
\end{align}
Notice that the last term is due to the subtraction of IR divergence in the small $p-k$ region using the UV pole $a_1$ in Eq.~(\ref{eq:constants2}) for the $\frac{1}{N}\Phi(i\partial)^2\Phi$ operator. Clearly, in order to perform this subtraction, one must Taylor expand the IR subtracted bubble chain to the second order, explaining the second line of Eq.~(\ref{eq:masssub}). To further simplify, one needs the function 
\begin{align}
G_2(n,\epsilon,s)&=f(\epsilon)^{\frac{s}{\epsilon}-1}e^{\epsilon \gamma_E}\left(\frac{\mu^2}{p^2}\right)^s\frac{\epsilon\Gamma(-1-\epsilon)s\Gamma(1-n-s)\Gamma(1+n+s)}{\Gamma(-n-\epsilon-s)\Gamma(n+s-\epsilon)} \nonumber \\ 
&=\frac{(n+s)(n+s+\epsilon)}{1+\epsilon}G(n,\epsilon,s) \ .
\end{align}
The pre-factor $\frac{(n+s)(n+s+\epsilon)}{1+\epsilon}$ suggests to write 
\begin{align}
\frac{1}{(p^2)^{n+1}}G_2(n,\epsilon,s)=\frac{1-\epsilon}{1+\epsilon}\left(\frac{\partial}{\partial p^2}+\frac{2p^2}{d}\frac{\partial^2}{\partial(p^2)^2}\right)\frac{1}{(p^2)^n}G(n,\epsilon,s) \ .
\end{align}
In terms of the above, one has 
\begin{align}\label{eq:partlyCn1}
\frac{1}{(p^2)^{n+1}}R_p \hat C_{n+1;1,0}=&-\frac{1-\epsilon}{1+\epsilon}\left(\frac{\partial}{\partial p^2}+\frac{2p^2}{d}\frac{\partial^2}{\partial(p^2)^2}\right)\frac{1}{(p^2)^n}R_b\hat C_{n,0}\nonumber \\ 
&-\left(\frac{\partial}{\partial p^2}+\frac{2p^2}{d}\frac{\partial^2}{\partial(p^2)^2}\right)\frac{1}{(p^2)^n}T_{p-k}R_b \hat C_{n,0} \ .
\end{align}
Here $R_b\hat C_{n,0}$ is given in Eq.~(\ref{eq:partlyCn0}) and $T_{p-k}R_b \hat C_{n,0}$ is given in Eq.~(\ref{eq:TRC0}). 

To proceed, one splits the second line of Eq.~(\ref{eq:partlyCn1}) to obtain
\begin{align}
\frac{1}{(p^2)^{n+1}}R_{p}\hat C_{n+1;1,0}=&-\frac{1-\epsilon}{1+\epsilon}\left(\frac{\partial}{\partial p^2}+\frac{2p^2}{d}\frac{\partial^2}{\partial(p^2)^2}\right)\frac{1}{(p^2)^n}\left(R_b\hat C_{n,0}+T_{p-k}R_b\hat C_{n,0}\right)\nonumber \\ 
&-\frac{2\epsilon}{1+\epsilon} \left(\frac{\partial}{\partial p^2}+\frac{2p^2}{d}\frac{\partial^2}{\partial(p^2)^2}\right)\frac{1}{(p^2)^n}T_{p-k}R_b \hat C_{n,0}
\ . \label{eq:RpCnplusim}
\end{align}
From this equation, it is clear that the $\frac{1}{\epsilon}$ terms from the high-order derivatives again cancel. This is consistent with the fact that after subtracting all the IR sub-divergences, one only needs $p$-independent operator poles. In fact, the required operator poles here are still the $g^{n+1}$ terms in $Z_n$ for $O_n\rightarrow{\rm I}$, but attached to the $m_r^2$ power coefficient function:
\begin{align}
&\frac{-Z_n}{(4\pi)^{n+1}}\times \frac{1}{(p^2)^{n+1}}\frac{-n\Gamma(1-\epsilon)}{\Gamma(2+\epsilon)}\frac{\Gamma(n+\epsilon+1)}{\Gamma(n-\epsilon)}\nonumber \\ 
&=\frac{1-\epsilon}{1+\epsilon}\left(\frac{\partial}{\partial p^2}+\frac{2p^2}{d}\frac{\partial^2}{\partial(p^2)^2}\right)\frac{1}{(p^2)^n}\frac{Z_n}{(4\pi)^{n+1}}\frac{\Gamma(1-\epsilon)}{\Gamma(1+\epsilon)}\frac{\Gamma(n+\epsilon)}{\Gamma(n-\epsilon)} \ .
\end{align}
See lower part of Fig.~\ref{fig: operator-counter} for a depiction of this operator pole. 
Therefore, the contribution obtained by combining the operator poles and the first line of Eq.~(\ref{eq:RpCnplusim}) in the $\epsilon \rightarrow 0$ limit, exactly leads to $C_{n,0}$ inside the derivatives
\begin{align}
&-\frac{1-\epsilon}{1+\epsilon}\left(\frac{\partial}{\partial p^2}+\frac{2p^2}{d}\frac{\partial^2}{\partial(p^2)^2}\right)\frac{1}{(p^2)^n}\left(R_b\hat C_{n,0}+T_{p-k}R_b \hat C_{n,0}\right) \nonumber \\ 
&+\frac{1-\epsilon}{1+\epsilon}\left(\frac{\partial}{\partial p^2}+\frac{2p^2}{d}\frac{\partial^2}{\partial(p^2)^2}\right)\frac{1}{(p^2)^n}\frac{Z_n}{(4\pi)^{n+1}}\frac{\Gamma(1-\epsilon)}{\Gamma(1+\epsilon)}\frac{\Gamma(n+\epsilon)}{\Gamma(n-\epsilon)} \nonumber \\ 
&\rightarrow-\left(\frac{\partial}{\partial p^2}+p^2\frac{\partial^2}{\partial(p^2)^2}\right)\frac{C_{n,0}}{(p^2)^n}  \ .
\end{align}
Here we have used Eq.~(\ref{eq:Cn0def}) for fully renormalized $C_{n,0}$. Therefore, using the $Z_n$ already given in Eq.~(\ref{eq:Z}), we have seen that it automatically renormalizes the coefficient function with one more mass insertion on the $p-k$ line. This is a crucial consistency check for our IR subtraction. On the other hand, the $\frac{2\epsilon}{1+\epsilon}$ term in the second line of Eq.~(\ref{eq:RpCnplusim}) combines with the $\frac{1}{\epsilon}$ term in the $T_{p-k}R_b \hat C_{n,0}$ given in Eq.~(\ref{eq:TRC0}) to form a finite contribution. 

Combining all above, the fully renormalized $C_{n+1;1,0}$ reads
\begin{align}
&\frac{1}{(p^2)^{n+1}}C_{n+1;1,0}=-\left(\frac{\partial}{\partial p^2}+p^2\frac{\partial^2}{\partial(p^2)^2}\right)\frac{1}{(p^2)^n}\left(C_{n,0}-\frac{2}{(4\pi)^{n+1}}\ln^{n}\frac{\mu^2}{p^2}\right) . \label{eq:Cnplus110}
\end{align}
Clearly, the factorial growth is from the $C_{n,0}$ term. Using the previous results Eq.~(\ref{eq:CnA}) and Eq.~(\ref{eq:CBn}) for $C_{n,0}$, one finally obtains
\begin{align}
&C_{n+1;1,0}+n^2C_{n,0}^A=C_{n+1;1,0}^B\nonumber \\
&=-\frac{n!}{(4\pi)^{n+1}}\oint\frac{ds(n+s)^2}{2\pi is^{n+1}}e^{-l_{\mu}s}\bigg(2\gamma_E-2-l_{\mu}+\psi(-n-s)+\psi(1+n+s)\bigg) \ .
\end{align}
Therefore,the leading non-alternating asymtotics can be extracted as 
\begin{align}\label{eq:massfac2}
C_{n;1,0}\bigg|_{n\rightarrow\infty, +1}=\frac{1}{(4\pi)^n}\frac{\mu^2}{p^2} n\times n!\bigg(1+\frac{\mu^2}{p^2}\frac{(n+1)^2}{2^{n+1}n^2}+....\bigg)  \ .
\end{align}
Clearly, not only it will not cancel the contribution in $C_{n,0}$, but it has a factor-$n$ enhancement in comparison to $C_{n,0}$. The enhancement is clearly due to the fact that the mass-insertion is roughly equivalent to two derivatives. One can proceed further without difficulty and show that the leading factorial growth of $C_{n;l,0}$, namely, all the $l$ mass-insertions are on the $p-k$ line, reads
\begin{align}
C_{n;l,0}\bigg|_{n\rightarrow\infty, +1}=-\frac{1}{(4\pi)^{n+1-l}}\frac{\mu^2}{p^2}\frac{(-1)^{l} \Gamma (n+1)^2}{\Gamma (l+1)^2 \Gamma (n-l+1)} \ .
\end{align}
Later we will see that the above summed over $l$ is almost the leading large-$n$ asymptotics of $C_{\rm I,NL}^n$, up to an over-all multiplication constant $e^{-\frac{8\pi m_r^2}{g}}$ that is {\it exponentially small in $g$}. For now, simply notice that the maximal value of the asymptotics above is achieved around $l=\sqrt{n}$, with is roughly $n!\frac{\sqrt{\frac{1}{n}}e^{2\sqrt{n}-\frac{1}{2}}}{2\pi}$.

\subsection{Cancellation of factorial growth between coefficient functions and operators}
Given the discussion in the previous subsection, it is clear from now that the factorial growths in the coefficient functions can not be canceled within the coefficient functions. Instead, they are expected to cancel with the operators. How to see the factorial growth in the operators? We first consider the alternating asymptotics of $C_{n,0}$ in Eq.~(\ref{eq:Cnasym2}). It looks like an operator contribution from $O_{n-1}$ but with $n$-bubbles. Let' show it is indeed the case. For this we need the $n$-bubble contribution to $O_{n-1}$ 
\begin{align}
\langle O_{n-1} \rangle_{g^{n+1}}=(-1)^{n+1}\int\frac{d^2k}{(2\pi)^2}(k^2)^{n-2}g(gF(k^2))^n \ .
\end{align}
Notice that this contribution is the leading large $N$ contribution to $\langle O_{n-1}\rangle$ with $n\ge 3$. Here we focus on the large-momentum region, where one has the logarithmic behavior
\begin{align}
F(k^2)\rightarrow \frac{1}{4\pi k^2}\ln\frac{k^2}{m^2} \ ,
\end{align}
thus 
\begin{align}
\langle O_{n-1} \rangle_{g^{n+1}} &\rightarrow (-1)^{n+1}\frac{g^{n+1}}{(4\pi)^{n}}\int_{k^2\gg \alpha m^2}\frac{d^2k}{(2\pi)^2}\frac{1}{(k^2)^2}\ln^{n}\frac{k^2}{m^2} \nonumber 
 \\ &\rightarrow \frac{(-1)^{n+1}g^{n+1}}{(4\pi)^{n+1}}\frac{n!}{m^2} \ . \label{eq:operatorals}
\end{align} 
Interestingly, it exactly cancels the leading alternating factorial growth in Eq.~(\ref{eq:Cnasym2}) after setting $\mu=m$. The momentum region responsible for the factorial growth is $k^2\sim m^2e^n$, which moves to the UV as $n$ becomes large. The crucial point here is, for the operator $O_{n-1}$ with $n-2$ $\partial^2$s in the definition, the bubble-chain diagrams with the bubble number around $n$ just canceled the large algebraic power in $k^2$ from the denominators, while the large logarithmic power can not be canceled and survives, leading to factorial enhancements. In fact, for $O_n$ with very large $n$, the factorial contributions peak around the bubble number $n$ and decay quickly for very large or very small bubble numbers. This correlation between the number of derivatives and the perturbative order is a unique feature of those supper-renormalizable field theories in which large logarithms in single scales could be iterated and amplified along bubble-chain-like structures.

Here we comment on the scale $\mu$ dependency of the factorial growth. Notice that although $O_{n-1}$ in the $n$ bubble diagram is UV finite, for
smaller bubble numbers there are UV divergences and sub-divergences, which must be subtracted. The subtractions will contain not only the $O_{n-1}\rightarrow{\rm I}$ mixing, but will contain mixing to those ``tree-level'' operators that also contribute to the OPE at a generic $\mu^2$. Thus, one expects that the factorial growth of $O_{n-1}$ should cancel the total factorial growth accumulated by the tree-level operators plus the identity operator. Equivalently, the factorial growth of $O_{n-1}$ can be decomposed  and cancel with the tree-level operators separately. One possible way to perform this is to notice that in the large $k^2$ asymptotics of $F(k^2)$, one can write
\begin{align} \label{eq:Ddecomp}
\frac{1}{4\pi k^2}\ln \frac{k^2}{m^2}\rightarrow\frac{1}{4\pi k^2}\ln \frac{k^2}{\mu^2}+\frac{1}{4\pi k^2}\ln \frac{\mu^2}{m^2} \ .
\end{align}
The first term $\ln \frac{k^2}{\mu^2}$ can be interpreted as the ``purely hard'' contribution to the logarithm in which the $\mu^2$ plays the role of an IR cutoff and is responsible for the $l^2 \gg \mu^2$ region of the loop momentum in Eq.~(\ref{eq:defF}). The second term $\ln \frac{\mu^2}{m^2}$ can be interpreted as an operator insertion of $\frac{1}{N}\Phi^2$ replacing one of two lines of the bubble, responsible for the $m^2\ll l^2 \ll \mu^2$ region of the loop momentum. Thus, if one expand the $F(k^2)^{n}$, the $\ln^p \frac{\mu^2}{m^2}|_{p\ge 1}$ terms are also expected to cancel with coefficient functions of the tree-level operators in the OPE formed by replacing one of the two lines of $p$ massive bubbles with $\frac{1}{N}\Phi^2$ insertions. The $\ln^n \frac{k^2}{\mu^2}$ term is then expected to cancel with the coefficient function of the identity operator:
\begin{align}
\langle O_{n-1} \rangle_{g^{n+1},{\rm hard}} &\rightarrow (-1)^{n+1}\frac{g^{n+1}}{(4\pi)^{n}}\int_{k^2\gg \alpha m^2}\frac{d^2k}{(2\pi)^2}\frac{1}{(k^2)^2}\ln^{n}\frac{k^2}{\mu^2} \nonumber 
 \\ &\rightarrow \frac{(-1)^{n+1}g^{n+1}}{(4\pi)^{n+1}}\frac{n!}{\mu^2} \ . \label{eq:operatoralshard}
\end{align}
As expected, it exactly matches the $\frac{1}{\mu^2}$ in the Eq.~(\ref{eq:Cnasym2}). In conclusion, the alternating ``UV renormalons'' in $C_{n,0}$ have to be canceled by lower-dimension operators. 

We then move to the non-alternating factorial growths Eq.~(\ref{eq:Cnasym1}). They resemble operator contributions from $O_{n+1}$ but still with $n$-bubbles, thus there are UV divergences which should be renormalized according to our renormalization schemes. There is a contribution of the order $m^2g^{n+1}$, while the logarithms could lead to $g^{n+1}m^2\times\frac{\mu^2}{m^2}n!=g^{n+1}\mu^2n!$. The fact that $\mu^2$ is in the numerator indicates that this is a UV renormalon for the operator. To show this, one can use the following two tricks. The first way is to inspect the large momentum behavior for $\langle O_{n+1}\rangle_{g^{n+1}}$. Then, by decomposing the massive bubbles as in Eq.~(\ref{eq:Ddecomp}), the term that is expected to cancel with the factorial growth of $C_{n,0}$ is
\begin{align}
\langle O_{n+1} \rangle_{g^{n+1},{\rm hard}} &\rightarrow (-1)^{n+1}\frac{g^{n+1}}{(4\pi)^{n}}\mu_0^{2-d}\int_{ \alpha m^2\ll k^2}\frac{d^{d}k}{(2\pi)^d}\ln^{n}\frac{k^2}{\mu^2} \nonumber 
 \\ &\rightarrow \frac{g^{n+1}}{(4\pi)^{n+1}}n!\mu^2 \ . \label{eq:operatoralshard1}
\end{align}
The point is, in DR, the $d^{2-2\epsilon}k$ integral, although quadratically divergent in naive power counting, is in fact finite as $d\rightarrow 2$ and exactly cancels the factorial growth in the coefficient function Eq.~(\ref{eq:Cnasym1}). To further consolidate this cancellation pattern, let's consider the point-splitting regularization of the following two-point function
\begin{align}\label{eq:anotherOPE}
\frac{g^2}{4N}\langle\Phi^2(z)(i\partial)^{2n}\Phi^2(0)\rangle_{g^{n+1}}=\frac{1}{z^2}(C_{{\rm I},0}(L)+...+m^2z^2C_{{\rm I},1}(L)+...)+\langle O_{n+1} \rangle_{g^{n+1}}+.... \ .
\end{align}
A crucial fact is, for any finite $z^2$, the correlator on the left in the $n$ bubble diagram is free from factorial growth, since the high scale region is {\it exponentially suppressed} by the finite distance $z$. This point is not entirely transparent and is demonstrated in the following manner. First notice that $F^n(\omega)$ is an analytic function in $\omega$ with a branch-cut singularity along $(-\infty,-4m^2)$, while decays at infinity at a large algebraic speed. Therefore, one can establish the following spectral representation
\begin{align}
(-1)^{n+1}F^n(k^2)=\int_{8m^2}^\infty \frac{\rho(s)ds}{k^2+s}+\int_{C_1} \frac{d\omega}{2\pi i} \frac{(-1)^{n}F^n(-\omega)}{k^2+\omega} \ , 
\end{align}
where $C_1$ is a bounded closed contour in the {\it right half-plane}  enclosing the singularity point $4m^2$ and closes at $8m^2\pm i0$ on the right side, but avoided the $\omega=0$ point on the left side. The $\sqrt{\omega}$ is well defined on $C_1$ and is non-vanishing. The spectral function is given by
\begin{align}
\rho(s)=\frac{(-1)^{n}}{2\pi i}\left(F^n(-s+i0)-F^n(-s-i0)\right) \ ,
\end{align}
which is a real regular function away from the $s=4m^2$ point. At large $s$, the spectral function decays as $\frac{n}{(4\pi s)^n}\ln^{n-1} \frac{s}{m^2}$. Given the above, one can establish 
\begin{align}
&\frac{g}{4N}\langle\Phi^2(z)\Phi^2(0)\rangle\bigg|_{g^{n+1}}\nonumber \\ 
&=g^{n+1}\int \frac{d^2k}{(2\pi)^2}\int_{8m^2}\frac{ds \rho(s)}{k^2+s}e^{ik\cdot z}+g^{n+1}\int \frac{d^2k}{(2\pi)^2}\int_{C_1}\frac{d\omega}{2\pi i}\frac{(-1)^{n} F^n(-\omega)}{k^2+\omega}e^{ik\cdot z} \nonumber \\ 
&=\frac{g^{n+1}}{2\pi}\int_{8m^2}^\infty ds \rho(s) K_0(\sqrt{s}z)+\frac{g^{n+1}}{2\pi}\int_{C_1}\frac{d\omega}{2\pi i}(-1)^{n}F^n(-\omega)K_0(\sqrt{\omega}z) \ .
\end{align}
Now, one applies the derivative operator. Since $z\ne 0$ and $\omega \ne 0$, one can use the equation of motion for massive free field $((i\partial)^2+s)K_0(\sqrt{s}z)=0$ and $((i\partial)^2+\omega)K_0(\sqrt{\omega}z)=0$ to reach
\begin{align}
&\frac{g}{4N}\langle\Phi^2(z)(i\partial)^{2n}\Phi^2(0)\rangle\bigg|_{g^{n+1}} \nonumber \\ 
&=\frac{g^{n+1}}{2\pi}\int_{8m^2}^\infty ds (-1)^ns^n\rho(s) K_0(\sqrt{s}z)+\frac{g^{n+1}}{2\pi}\int_{C_1}\frac{d\omega}{2\pi i}\omega^nF^n(-\omega)K_0(\sqrt{\omega}z) \ .
\end{align}
At this step, it is clear that the second term contains no factorial growth in $n$, since $\omega$ encloses a bounded path away from $0$ and $4m^2$, along which $F(-\omega)$ and $K_0(\sqrt{\omega}z)$ are all bounded by finite constants, therefore the integral can grow at worst exponentially. So one focuses on the first term
\begin{align}
\bigg|\int_{8m^2}^\infty ds s^n\rho(s)K_0(\sqrt{s}z)\bigg| \le \frac{C^n}{\sqrt{z}} \int_{8m^2}^\infty \frac{ds}{s^\frac{1}{4}} \ln^{n-1}\frac{s}{m^2}e^{-\sqrt{s}z}\nonumber \\ 
=\frac{2^nC^nm^2}{\sqrt{zm}}\int_{\sqrt{8}}^{\infty} dy \sqrt{y}\ln^{n-1}y e^{-yzm} \ ,
\end{align}
where $0<C<\infty$ is a finite number and we used the inequality $|K_0(x)|\le \sqrt{\frac{\pi}{2x}}e^{-x}$ for $x>0$. The point is, for any $zm>0$, the large $y$ region of the integral 
\begin{align}
f(n)=\int_{\sqrt{8}}^\infty dy \sqrt{y} \ln^{n}ye^{-yzm} \ ,
\end{align}
is suppressed exponentially rather than algebraically, thus the growth speed of $f(n)$ as $n\rightarrow \infty$ will be much slower than $n!$. In fact, using the Laplace's method, one can show that in the large $n$ limit, $f(n)$ grows like $e^{n\ln \ln \frac{n}{zm}}$
\begin{align}
&f(n)\rightarrow\sqrt{\frac{2\pi T_n}{n}}\exp\left[n\ln T_n-\frac{n}{T_n}+\frac{3T_n}{2}\right]\left(1+{\cal O}\left(\sqrt{\frac{T_n}{n}}\right)\right) \ , \\ 
&T_n+\ln T_n=\ln \frac{n}{zm} \ .
\end{align}
As expected, as far as $zm>0$, the growth speed is slower than $(n!)^{\alpha}$ for any $\alpha>0$. 

Now, after showing that the point-splitting two point function at the order $g^{n+1}$ is free from factorial growth, we move back to its small-$z$ expansion in Eq.~(\ref{eq:anotherOPE}). From this equation, it is clear that by subtracting out all the singular contributions at $\frac{1}{z^2}$, as well as finite contributions from the mass-insertions, one reaches the $O_n$. On the other hand, although the two-point function for any finite $z$ is free from factorial growth, there is no guarantee for the small-$z$ expansion coefficients. If factorial growth is developed on $C_{{\rm I},0}$ that remains finite in $z^2\rightarrow 0$ limit, then by subtracting it out, the operator condensate $\langle O_{n+1}\rangle$ will also contain factorial growth. Let's check now this is indeed the case.  In fact,  the bared $n$ bubble-contribution to $C_{{\rm I},0}$ can be calculated as ($s=(n+1)\epsilon$)
\begin{align}
\frac{1}{z^2}\hat C_{{\rm I},0}(z^2)\bigg|_{g^{n+1}}&=-\frac{g^{n+1}e^{\epsilon\gamma_E}\mu^{2\epsilon}}{(4\pi)^{n+\epsilon}}\frac{f(\epsilon)^n}{\epsilon^n}\int\frac{d^dk}{(2\pi)^d} \left(\frac{\mu^2}{k^2}\right)^{n\epsilon}e^{ik\cdot z}
\nonumber \\&=-\frac{g^{n+1}e^{\gamma_E\epsilon}}{(4\pi)^{n+1}}\frac{4}{z^2}\left(\frac{z^2\mu^2
}{4}\right)^s\frac{f(\epsilon)^n\Gamma(1-s)}{\epsilon^n\Gamma(s-\epsilon)} \ ,
\end{align}
it contains exactly a $s=1$ renormalon singularity that is non-alternating. In the large $n$ limit, the factorial growth after the IR subtraction is therefore
\begin{align}
\frac{1}{z^2} C_{{\rm I},0}(L)\bigg|_{g^{n+1}}=-\frac{4n!g^{n+1}}{z^2(4\pi)^{n+1}}\oint\frac{ds}{2\pi i}\left(\frac{z^2\mu^2}{4}\right)^s\frac{\Gamma(1-s)}{s^{n+1}\Gamma(s)}\rightarrow -\frac{g^{n+1}}{(4\pi)^{n+1}}n!\mu^2 \ .
\end{align}
Therefore, after subtracting the $C_{{\rm I},0}(L)$ in Eq.~(\ref{eq:anotherOPE}), an opposite factorial growth is generated:
\begin{align}
\langle O_{n+1}\rangle\bigg|_{n\rightarrow\infty,+1}=\frac{g^{n+1}}{(4\pi)^{n+1}}n!\mu^2 \ .
\end{align}
Combining with its tree-level coefficient function, it cancels exactly the leading factorial growth of $C_{n,0}$ in Eq.~(\ref{eq:Cnasym1}). From the above, it is clear now that before the small-$z$ expansion, the two-point function is regulated by the mass in the IR while exponentially suppressed in the UV. The logarithms in $F(k^2)$ do not propagate to the IR region. After the small-$z$ expansion, the coefficient functions $C_{{\rm I},0}$ are given by IR subtracted massless integrals and acquire factorial enhancements due to the large logarithms in the IR region. This is consistent with the interpretation of the factorial enhancements in the coefficient functions as IR renormalons.

Not only the factorial growths in $C_{n,0}$ are canceled by the operators, one can show that the factorial growths for the $m_r^2$ terms in the coefficient functions are also canceled by operators. Let's consider the $C_{n;0,1}$ and $C_{n;1,0}$ with one mass insertions. First, the $n!$ growth of $C_{n;0,1}$ with one mass insertions in the bubble chain is canceled by the $O_{n}$ operator in $n-1$ bubble diagram. This time, the renormalon is proportional to $m_r^2$ and can be obtained in the point-splitting by subtracting the $m_r^2$ order coefficient function $C_{{\rm I},1}$ from the $\frac{1}{z^4}(C_{{\rm I},0}+m_r^2z^2C_{{\rm I},1})$. Second, the $n\times n!$ growth of $C_{n;1,0}$ with one $m_r^2$ insertion on the $p-k$ line can be explained by the operator $O_{n}$ in $n-1$ bubbles but attached to the $m_r^2$ order coefficient function. Indeed, the small-$k$ expansion of $-\frac{m_r^2}{(p-k)^4}$ introduces the factor of $-n^2m_r^2$. This combined with the operator's $(n-1)!\mu^2\frac{g^{n}}{(4\pi)^n}$ exactly cancels the factorial growth in Eq.~(\ref{eq:massfac2}). So the cancellation of leading factorial growth between coefficient functions and operators is checked up to $m_r^2$.

To conclude, the cancellation pattern is: alternating ``renomalons'' cancel with lower-dimensional operators, while non-alternating renormalons cancel with higher-dimensional operators. The renormalons always cancel at the same numbers of $g$, and the cancellation is always at the same values of $s$. For example, the $s=2$ subleading renormalon corresponding to the second term in Eq.~(\ref{eq:Cnasym1}) would cancel with the sub-leading UV renormalon of the operator $O_{n+2}$ with $n$-bubbles (and diagrams that can be obtained by deleting bubbles but adding tadpole-type insertions).

\subsection{OPE from Mellin-Barnes}
After discussing in detail the OPE based on short 
distance computations, in this subsection we show that the result obtained in the previous subsections can also be derived from Mellin-Barnes representations directly based on Eq.~(\ref{eq:fulloneverN}). 

The strategy is, one first introduces $0<{\rm Re }(s_1)=c<1$ and write 
\begin{align}
\frac{gF(k^2)}{1+gF(k^2)}=\int_{c-i\infty}^{c+i\infty} \frac{ds_1}{2\pi i} \frac{\pi}{\sin \pi s_1} (gF(k^2))^{s1} \ .
\end{align}
We then write 
\begin{align}
\ln^{s_1}\frac{A+1}{A-1}=\frac{1}{\Gamma(-s_1)}\int_{0}^{\infty} dt t^{-s_1-1} \left(\frac{A-1}{A+1}\right)^{t} \ ,
\end{align}
which should be understood as analytically continued from the $s_1<0$ region. Namely, for $0<s_1<1$ the above should be understood as 
\begin{align}
\frac{1}{\Gamma(-s_1)}\int_0^{\infty} dt t^{-s_1-1} F(t)=\frac{1}{s_1\Gamma(-s_1)}\int_0^{\infty} dt t^{-s_1} \frac{d}{dt} F(t) \ .
\end{align}
At this step, one can introduces another Mellin-Barnes parameters $s$ in a way similar to those in~\cite{Beneke:1998eq,Marino:2024uco,Liu:2024omb}
\begin{align}
&A^{-s_1}\left(\frac{A-1}{A+1}\right)^{t}=\int_{-t<\rm Re{s}<\frac{1}{2}s_1} \frac{ds}{2\pi i}\frac{  \, _2\tilde{F}_1(s_1-1,s+t;-s+s_1+t;-1) }{\sin 2\pi s \Gamma (2 s-s_1+1) \Gamma (-s-t+1)} \nonumber \\ 
& \times \pi ^2\csc (\pi  (s-s_1-t)) (\sin (\pi  s_1) \csc (\pi  (2 s-s_1))-\sin (\pi  (s-t)) \csc (\pi  (s+t))) \left(\frac{k^2}{m^2}\right)^s \ .
\end{align}
In particular, for $s_1=1$ the above reduces to the Mellin-Barnes representation of the effective coupling in the large-$N$ Gross-Neveu model~\cite{Marino:2024uco,Liu:2024omb}. Given the above, we then write
\begin{align}
&\frac{\pi}{\sin \pi s_1}\left(\frac{g}{4\pi}\right)^{s_1}g\int \frac{d^2k}{(2\pi)^2}\frac{1}{(k^2)^{s_1}}\left(\frac{k^2}{m^2}\right)^s\frac{1}{(p-k)^2+m^2} \nonumber \\ 
&=\left(\frac{g}{4\pi m^2}\right)^{s_1}\frac{g}{4\sin \pi s_1}\int_{0<{\rm Re}(u)<s_1-s} \frac{du}{2\pi i} \frac{\Gamma (u) \Gamma (s-s_1+1) \Gamma (-s+s_1-u)^2}{\Gamma (1-u) \Gamma (s_1-s)} \left(\frac{p^2}{m^2}\right)^{-u} \ .
\end{align}
Combining all the above, one obtains the following Mellin-Barnes representation of $S_{\rm NL}(p^2)(p^2+m^2)^2$ defined in Eq.~(\ref{eq:fulloneverN})
\begin{align}
& NS_{\rm NL}(p^2)(p^2+m^2)^2-\frac{g}{2}\langle :\Phi^2:\rangle=\int\frac{ds_1}{2\pi i}\int_0^{\infty} \frac{dt t^{-s_1}}{s_1}\frac{d}{dt}\int_{} \frac{ds}{2\pi i} \int \frac{du}{2\pi i}{\cal G}(t,s_1,s,u) \left(\frac{p^2}{m^2}\right)^{-u} \ , \\ 
& 0<{\rm Re}(s_1)<1 \ , \ 0<{\rm Re}(s)<\frac{1}{2}{\rm Re}(s_1) \ , \ 0<{\rm Re}(u)<{\rm Re}(s_1)-{\rm Re}(s) \  \ ,
\end{align}
where 
\begin{align}
&{\cal G}(t,s_1,s,u)=\frac{\Gamma (u) \Gamma (s-s_1+1) \Gamma (-s+s_1-u)^2}{\Gamma (1-u) \Gamma (s_1-s)}\times{\cal M}(t,s_1,s) \ ,  \\ 
& {\cal M}(t,s_1,s)=\left(\frac{g}{4\pi m^2}\right)^{s_1}\frac{g}{4\sin \pi s_1 \Gamma(-s_1)}\times  \frac{  \, _2\tilde{F}_1(s_1-1,s+t;-s+s_1+t;-1) }{\sin 2\pi s \Gamma (2 s-s_1+1) \Gamma (-s-t+1)}\nonumber \\ 
&\times \pi ^2\csc (\pi  (s-s_1-t)) (\sin (\pi  s_1) \csc (\pi  (2 s-s_1))-\sin (\pi  (s-t)) \csc (\pi  (s+t))) \ . 
\end{align}
It is interesting to notice that the $2s-s_1=-k$,  $k\in Z_{\ge 1}$ poles has been canceled by the denominator. It is also interesting to notice that the $s=s_1+t-k$ poles are also canceled among the sine functions. Thus, the only series of poles left from the ${\cal M}(t,s_1,s)$ when {\it shifting the $s$ to the left} are the $s=-t-k$, $k\ge 0$ poles.

Given the above, let's start to shift the contours to obtain the asymptotic expansions. The first step of performing the expansion is to pick up the $u=s_1-s+l$ poles with $l \ge $ 0. Clearly, $l$ has the interpretation as the number of mass insertions on the $p-k$ line. After this step, one obtains
     \begin{align}
     &NS_{\rm NL}(p^2)(p^2+m^2)^2-\frac{g}{2}\langle :\Phi^2:\rangle=\sum_{l=0}^{\infty}\left(\frac{m^2}{p^2}\right)^lS_l(p^2) \ , \\
     &S_l(p^2)=\int\frac{ds_1}{2\pi i}\int_0^{\infty} \frac{dt t^{-s_1}}{s_1}\frac{d}{dt}\int_{} \frac{ds}{2\pi i} {\cal M}(t,s_1,s) \left(\frac{p^2}{m^2}\right)^{s-s_1}  \Gamma (s-s_1+1)  \Gamma (l-s+s_1) \nonumber \\ 
     &\times \frac{\left(-\psi(-l+s-s_1+1)-\psi(l-s+s_1)+2 \psi(l+1)+\ln \left(\frac{p^2}{m^2}\right)\right)}{(l!)^2 \Gamma (s_1-s) \Gamma (-l+s-s_1+1)} \ . \label{eq:Sl}
     \end{align}
     In particular, one has
     \begin{align}
     &S_{l=0}(p^2)=-\int\frac{ds_1}{2\pi i}\int_0^{\infty} \frac{dt t^{-s_1}}{s_1}\frac{d}{dt}\int_{} \frac{ds}{2\pi i} {\cal M}(t,s_1,s) \left(\frac{p^2}{m^2}\right)^{s-s_1}   \nonumber \\ 
     &\times \bigg(-\ln \left(\frac{p^2}{m^2}\right)+\psi (s-s_1+1)+\psi(s_1-s)+2 \gamma_E\bigg) \ .
     \end{align}
 The second step of the expansion is to shift the $s$ contours to the left. There are two series of poles at $s=-t-k, k\ge 0$ from the function ${\cal M}(t,s_1,s)$ and $s=s_1-k, k\ge 1$. The first has to be interpreted as hard contributions with $k$ mass-insertions on the bubble chain, while the latter corresponds to the operators $O_k$. To show that the $s=s_1-k$ poles corresponds to the operator contribution $O_k$, one way is to show that for different $l$ we can reproduce their coefficient functions with $l$ mass insertions on $p-k$ line. Indeed, by picking up the $s=s_1-k$ pole in Eq.~(\ref{eq:Sl}), one obtains
\begin{align}
&S_l(p^2)\bigg|_{s=s_1-k}=\left(\frac{m^2}{p^2}\right)^k {\cal O}_{l,k} \ , \\
&{\cal O}_{l,k}=\frac{(-1)^{-l} \Gamma (k+l)^2}{\Gamma (k)^2 \Gamma (l+1)^2}\int \frac{ds_1}{2\pi i}\int_0^{\infty} \frac{dt t^{-s_1}}{s_1}\frac{d}{dt}\bigg[{\cal M}(t,s_1,s_1-k)\bigg]\equiv\frac{(-1)^{l} \Gamma (k+l)^2}{\Gamma (k)^2 \Gamma (l+1)^2}{\cal O}_{k} \ .
\end{align}
This is nothing but the operator $O_k$ attached to the 
coefficient function with $l$ insertions on the $p-k$ line, see Eq.~(\ref{eq:expansiongenral}) and Eq.~(\ref{eq:coefmassin}). Since different $l$ do not mix with each other in further expansion, below let's consider the $l=0$ contribution in detail, but there is no difficulty to generalize to all $l$s. 

For $l=0$, the expansion after shifting the $s$ contour  can be written explicitly as 
\begin{align}
&S_{l=0}(p^2)=\sum_{k=0}^{\infty}\left(\frac{m^2}{p^2}\right)^k{\cal H}_{0,k}(p^2)+\sum_{k=1}^{\infty}\left(\frac{m^2}{p^2}\right)^k{\cal O}_{k} \ .  
\end{align}
The hard contribution reads
\begin{align}
& {\cal H}_{0,k}(p^2)=g\int\frac{ds_1}{2\pi i}\left(\frac{g}{4\pi m^2}\right)^{s_1}\int_0^{\infty} dt \frac{t^{-s_1}}{s_1}\frac{d}{dt}\bigg[{\cal H}_{0,k}(t,s_1)\left(\frac{p^2}{m^2}\right)^{-s_1-t}\bigg] \ ,
\end{align}
with
\begin{align}
&{\cal H}_{0,k}(t,s_1)=\frac{ \pi  (-1)^{k+1}  \, _2\tilde{F}_1(-k,s_1-1;k+s_1+2 t;-1) }{4k!\sin(\pi  (s_1+2 t)) \sin (\pi  s_1) \Gamma (-s_1) \Gamma (-s_1-2t -2k+1)} \nonumber \\ 
& \times   \left(\psi(-k-s_1-t+1)+\psi (k+s_1+t)-\ln \left(\frac{p^2}{m^2}\right)+2 \gamma_E \right) \ . 
\end{align}
Similarly, the operator contribution reads
\begin{align}
 {\cal O}_{k}=g\int\frac{ds_1}{2\pi i}\left(\frac{g}{4\pi m^2}\right)^{s_1}\int_0^{\infty} dt \frac{t^{-s_1}}{s_1}\frac{d}{dt}\bigg[{\cal O}_{k}(t,s_1)\bigg] \ ,
\end{align}
where
\begin{align}
{\cal O}_{k}(t,s_1)=\frac{\pi ^2 (-1)^{k+1}  \, _2\tilde{F}_1(s_1-1,-k+s_1+t;k+t;-1)}{4\sin^2 (\pi s_1)\sin (\pi(s_1+t))\Gamma (-s_1) \Gamma (-2 k+s_1+1) \Gamma (k-s_1-t+1)} \ . \label{eq:defOkts}
\end{align}
The last and the most tricky step is to shift the $s_1$ contour to the right and pick up the contributions at various powers in $g$. There are poles at $s_1=n\ge 1$ due to the analytical continuation of $t$ integral. On the other hand, there are another series of poles at $s_1=-t+n$ with $n\ge 1$. The non-trivial fact that guarantees the correctness of the calculation is the following: the $s_1=-t+n$ poles all cancel between $\left(\frac{m^2}{p^2}\right)^{-s_1-t}{\cal H}_{0,k}(t,s_1)$ and $\left(\frac{m^2}{p^2}\right)^{n}{\cal O}_{k+n}(t,s_1)$:
\begin{align}
&{\rm Res} \ \left(\frac{m^2}{p^2}\right)^{-s_1-t} {\cal H}_{0,k}(t,s_1=-t+n)\nonumber \\ 
&=-\left(\frac{m^2}{p^2}\right)^n\frac{\pi  (-1)^{k+1}  \, _2\tilde{F}_1(-k,n-t-1;k+n+t;-1)}{4 k! \sin^2\pi t\Gamma (t-n) \Gamma (-2 k-n-t+1)} \ , \\ 
& \left(\frac{m^2}{p^2}\right)^n {\rm Res} \ {\cal O}_{k+n}(t,s_1=-t+n) \nonumber \\ 
&=\left(\frac{m^2}{p^2}\right)^n\frac{\pi  (-1)^{k+1}  \, _2\tilde{F}_1(-k,n-t-1;k+n+t;-1)}{4 k! \sin^2\pi t\Gamma (t-n) \Gamma (-2 k-n-t+1)} \ .
\end{align}
Notice that for ${\cal O}_{k}$ there are only $s_1=-t+n$ poles with $n\le k$. Given the above, all the $s_1=-t+n$ poles are canceled (thus no $g^{-t}$ type terms are generated), thus to finish the final step of the expansion, it remains to pick up the $s_1=n$ poles. 

Here we proceed the $s_1=n$ poles. For this purpose, it is convenient to use the representations
\begin{align}
&{\cal H}_{0,k}(p^2)\nonumber \\ 
&=g\int \frac{ds_1}{2\pi i} \left(\frac{g}{4\pi m^2}\right)^{s_1}\frac{1}{s_1(s_1-1)...(s_1-n)}\int_0^{\infty} dt t^{-s_1+n}\frac{d^{n+1}}{dt^{n+1}}\bigg[{\cal H}_{0,k}(t,s_1)\left(\frac{p^2}{m^2}\right)^{-s_1-t}\bigg] \ , 
\end{align}
and 
\begin{align}
&{\cal O}_{k}=g\int \frac{ds_1}{2\pi i} \left(\frac{g}{4\pi m^2}\right)^{s_1}\frac{1}{s_1(s_1-1)...(s_1-n)} \int_0^{\infty} dt t^{-s_1+n}\frac{d^{n+1}}{dt^{n+1}}\bigg[{\cal O}_{k}(t,s_1)\bigg] \ . \label{eq:operatorinter}
\end{align}
For ${\cal H}_{0,k}(t,s_1)$, since the function itself is finite at $s_1=n$, thus if the $dt$ integral is convergent, then one could pickup that pole by partial-integrating $n+1$ times and obtains
\begin{align}
{\cal H}_{0,k}(p^2)\bigg|_{s_1=n}=-g\left(\frac{g}{4\pi p^2}\right)^n\int_0^{\infty} dt \frac{d^{n+1}}{dt^{n+1}}\bigg[\frac{1}{n!}{\cal H}_{0,k}(t,n)\left(\frac{p^2}{m^2}\right)^{-t}\bigg] \ ,
\end{align}
and
\begin{align}
{\cal O}_{k,s_1=n}=-g\left(\frac{g}{4\pi m^2}\right)^n\int_0^{\infty} dt \frac{d^{n+1}}{dt^{n+1}}\bigg[\frac{1}{n!}{\cal O}_{0,k}(t,n)\bigg] \ .
\end{align}
The integrands read
\begin{align}
& \frac{1}{n!}{\cal H}_{0,k}(t,n)= \frac{  (-1)^{k+n}   \, _2\tilde{F}_1(-k,n-1;k+n+2 t;-1) }{4k! \sin(2\pi t) \Gamma (-n-2k-2t+1)} \nonumber \\ 
&\times \left(\psi(-k-n-t+1)+\psi(k+n+t)-\ln \left(\frac{p^2}{m^2}\right)+2 \gamma_E \right) \ ,
\end{align}
and
\begin{align}
\frac{1}{n!}{\cal O}_{k}(t,n)\bigg|_{n<2k}=-\frac{(-1)^{k+n} \Gamma (2 k-n) \, _2\tilde{F}_1(n-1,-k+n+t;k+t;-1)}{4 \sin (\pi  t) \Gamma (k-n-t+1)} \ .
\end{align}
Notice that for $n\ge 2k$ there are single-poles at $s_1=n$ for ${\cal O}_{k}(t,s_1)$ which has to be treated separately, we will come to these cases later. For now, the problem is, there are single poles on the positive $t$ axis for the integrands ${\cal H}_{0,k}(t,n)$ and ${\cal O}_k(t,n)|_{n\le k}$. Although the $t \ge 1$ poles could be get rid of by tilting the $t$ contour, the $t=0$ poles can not be avoided. As such, 
one expects that all the $t=0$ poles cancel between the $\{{\cal H}_{0,k}(t,n)\}$ and $\{{\cal O}_k(t,n)\}$ contributions. In fact, one can show more: not only the $t=0$ poles, but all the $t\in Z_{\ge 0}$ poles also completely cancel. The situation is in fact in a way similar to the renormalon-pole cancellation between Borel integrands in asymptotically-free theories~\cite{Beneke:1998eq,Marino:2024uco,Liu:2024omb}.

Indeed, one can show the following:
\begin{align}
&-{\rm Res} \ \frac{1}{n!}{\cal H}_{0,k}(t=q,n)\left(\frac{p^2}{m^2}\right)^{-t}  \nonumber \\ 
&= -\frac{(-1)^k  \, _2\tilde{F}_1(-k,n-1;k+n+2 q;-1) \Gamma (n+2 (k+q))}{4 \pi  k!}\left(\frac{m^2}{p^2}\right)^{q} \ , \label{eq:poleshard}\\ 
-&\left(\frac{m^2}{p^2}\right)^{q}{\rm Res} \frac{1}{n!} {\cal O}_{n+k+q}(t=q,n)\nonumber \\ 
&=\frac{(-1)^k  \, _2\tilde{F}_1(-k,n-1;k+n+2 q;-1) \Gamma (n+2 (k+q))}{4 \pi  k!}\left(\frac{m^2}{p^2}\right)^{q} \ . \label{eq:polesope}
\end{align}
Therefore, all the poles at $t=q\in Z$ under the constraint $2q\ge-2k-n$ completely cancel between the hard and operator contributions. Notice that poles that are cancel not only include $t\in Z_{\ge 0}$, but also contains all the negative poles that could lead to alternating factorial growths. Given the massive cancellation of poles, one finally obtains the hard contributions 
\begin{align}
&{\cal H}_{0,k}(p^2)\bigg|_{s_1=n}=g\left(\frac{g}{4\pi p^2}\right)^{n} \frac{d^{n}}{dt^{n}} \bigg[\frac{ (-1)^{k+n}   \, _2\tilde{F}_1(-k,n-1;k+n+2 t;-1) }{4k! \sin(2\pi t) \Gamma (-n-2k-2t+1)} \nonumber \\ 
&\times \left(\psi(-k-n-t+1)+\psi(k+n+t)-\ln \left(\frac{p^2}{m^2}\right)+2 \gamma_E \right)\left(\frac{p^2}{m^2}\right)^{-t} \bigg]_{t=0} \ , \label{eq:hardMellin}
\end{align}
and the operator contributions
\begin{align} \label{eq:operatorfull1}
&{\cal O}_{k,s_1=n<2k}\nonumber \\ 
&=-g\left(\frac{g}{4\pi m^2}\right)^n\frac{d^n}{dt^n}\bigg[\frac{(-1)^{k+n} \Gamma (2 k-n) \, _2\tilde{F}_1(n-1,-k+n+t;k+t;-1)}{4 \sin (\pi  t) \Gamma (k-n-t+1)}\bigg]_{t=0} \ .
\end{align}
In the above,  the $\frac{d^{n}}{dt^n} F(t)|_{t=0}$ is defined as the $n!$ times the $t^n$ coefficient of the Taylor expansion of $F(t)$ at $t=0$, even there are negative powers at $t=0$. Notice that the integrands decay sufficiently fast at large $t$ for both the hard and operator contributions and no $t=\infty$ boundary terms can be generated. In particular, for $k=0$ (no mass-insertion in the bubbles), the above gives 
\begin{align}\label{eq:hard00mellin}
&{\cal H}_{0,0}(p^2)\bigg|_{s_1=n}\nonumber \\ 
&=\frac{1}{(p^2)^n}\left(\frac{g}{4\pi}\right)^{n+1} \frac{d^n}{dt^n} \bigg[\left(\frac{m^2}{p^2}\right)^t\left(\psi(-n-t+1)+\psi(n+t)-\ln \left(\frac{p^2}{m^2}\right)+2 \gamma_E \right)\bigg]_{t=0} \nonumber \\ 
& \equiv \frac{g^{n+1}}{(p^2)^n}C_{n,0}^{B}(\mu^2=m^2) \ .
\end{align}
Interestingly, this is nothing but the $C^B_{n,0}$ term defined in Eq.~(\ref{eq:CBn}) that contains factorial growth calculated from the IR renormalized bubble chains, at the natural renormalization scale $\mu^2=m^2$ that kills the ``tree-level'' operators.
Similarly, with one mass insertion one also has
\begin{align}
&{\cal H}_{0,1}(p^2)\bigg|_{s_1=n}=-\frac{2}{(p^2)^n}\left(\frac{g}{4\pi}\right)^{n+1} \nonumber \\ 
&\times\frac{d^n}{dt^n} \bigg[\left(\frac{m^2}{p^2}\right)^t(n+t)\left(\psi(-n-t)+\psi(n+1+t)-\ln \left(\frac{p^2}{m^2}\right)+2 \gamma_E \right)\bigg]_{t=0} \nonumber \\ 
& \equiv \frac{g^{n+1}}{(p^2)^n}C_{n+1;0,1}^{B}(\mu^2=m^2) \ .
\end{align}
Again, it agrees with the $C_{n+1;0,1}^B$ contribution in Eq.~(\ref{eq:Cmassb}) obtained from short distance computation with one mass insertion within the bubble chain. The fact that the non-trivial coefficient functions obtained from the short distance OPE computations with IR subtractions can also be reproduced from the Mellin-Barnes serves as a strong consistency check for both approaches. Moreover, for the operator contributions in Eq.~(\ref{eq:operatorfull1}), we have also checked in special cases with $k<n<2k$ that the Eq.~(\ref{eq:operatorfull1}) agrees with direct evaluations of the corresponding $n$-bubble diagrams for $\frac{1}{m^{2k}}\langle O_{k}\rangle_{g^{n+1}}$ which are absolutely convergent when $n>k$. For example, with $k=2$ and $n=3$, both approaches produce the result $\frac{9\zeta_3}{2m^2}(\frac{g}{4\pi})^4$ for the operator condensate $\langle O_2\rangle_{g^4}=(m^2)^2{\cal O}_{2,s_1=3}$. This is another strong consistency check of our approaches. 

Using the Mellin Barnes, the renormalon cancellation pattern between operators and coefficient functions also becomes manifest. In fact, this simply corresponds to the cancellation of $t\in {Z}$ poles already demonstrated above in Eq.~(\ref{eq:poleshard}) and Eq.~(\ref{eq:polesope}). The pole at $t=q$ of ${\cal H}_{0,k}(t,s_1=n)$, after taking the $n$-th derivative at $t=0$, leads to the $\frac{n!}{q^{n+1}}$ subleading factorial growths of the coefficient function $C_{n+k;0,k}$. It is canceled by the ${\cal O}_{n+k+q,s_1=n}$, corresponding to the operator condensate $\langle O_{n+k+q}\rangle_{g^{n+1}}$ in the $n$-bubble diagram. In particular, for $q=0$ this is just the UV renormalization cancellation between $C_{n+k;0,k}$ and $\langle O_{n+k}\rangle_{g^{n+1}}$. For $q=1$, this is the cancellation of the leading non-alternating factorial growth between $C_{n+k;0,k}$ and $\langle O_{n+k+1}\rangle_{g^{n+1}}$, and for $q=-1$ this is the cancellation of the leading alternating factorial growth between $C_{n+k;0,k}$ and $\langle O_{n+k-1}\rangle_{g^{n+1}}$ . Moreover, the condition $2q\ge -2k-n$ translates to $n\le 2(n+k+q)$, implying that for the operator $O_{n+k+q}$, contributions from diagrams with at least $2(n+k+q)$ bubbles no-longer cancel with anything. As we will see now, these contributions no longer contain factorial growths.

We now come to the operator contributions ${\cal O}_{k,s_1\ge2n}$. As we have seen before, in such cases there are double poles in $s_1$ which must be treated separately. Since these contributions do not cancel with anything, we must show that they are totally finite and contain no factorial growth. For this purpose we can write the ${\cal O}_{k}(t,s_1)$ defined in Eq.~(\ref{eq:defOkts}) as 
\begin{align}
{\cal O}_{k}(t,s_1)=\frac{1}{s_1-2k-q}g_k(t,s_1) \ ,
\end{align}
where $g_k(t,s_1)$ is finite and non-vanishing at $s_1=2k+q$ with $q\in Z_{\ge 0}$. Now, using Eq.~(\ref{eq:operatorinter}),  the contribution from the $s_1=2k+q$ double pole can be extracted as
\begin{align}
&{\cal O}_{k,s_1=2k+q}\nonumber \\ =-&g\int_{0}^{\infty} dt \frac{d}{ds_1}\bigg[\left(\frac{g}{4\pi m^2}\right)^{s_1}\frac{t^{-s_1+2k+q}}{s_1(s_1-1)...(s_1-2k-q+1)} \bigg]_{s_1=2k+q}\frac{d^{2k+q+1}}{dt^{2k+q+1}}g_k(t,2k+q) \nonumber \\ 
&-g\left(\frac{g}{4\pi m^2}\right)^{2k+q}\frac{1}{(2k+q)!}\int_0^{\infty} dt\frac{d^{2k+q+1}}{dt^{2k+q+1}}\bigg[\frac{\partial}{\partial s_1}g_k(t,s_1)\bigg|_{s_1=2k+q}\bigg]_{t=0} \ . \label{eq:Oklarge}
\end{align}
The point is, the second line  vanishes since $g_k(t,2k+q)$ is a polynomial of order $q$ in $t$. Indeed, one has
\begin{align}
&g_k(t,2k+q)=(-1)^{k}\frac{(2k+q)!}{4q!}\frac{\,_2\tilde F_1(2k+q-1,k+q+t;k+t;-1)}{\sin \pi t \Gamma(-k-q-t+1)} \ , \\ 
&g_k(t,2k)=\frac{4^{-k} k \Gamma (2 k)}{\pi } \ , \\
&g_k(t,2k+1)=-\frac{4^{-k-1} t \Gamma (2 k+2)}{\pi } \ , \\
&g_k(t,2k+2)=-\frac{2^{-2 k-5} \left(k+1-2 t^2\right) \Gamma (2 k+3)}{\pi } \ .
\end{align}
In particular, no $\ln g$ terms can be generated. For the contribution in the third line of Eq.~(\ref{eq:Oklarge}), first notice that there are no singularities on the $t\ge0$ axis, and the integrand decays sufficiently fast. Thus one has
\begin{align}
{\cal O}_{k,s_1=2k+q}=g\left(\frac{g}{4\pi m^2}\right)^{s_1}\frac{1}{(2k+q)!}\frac{d^{2k+q}}{dt^{2k+q}}\bigg[\frac{\partial}{\partial s_1}g_k(t,s_1)\bigg|_{s_1=2k+q}\bigg]_{t=0} \ . \label{eq:operatorfull2}
\end{align}
In particular, for $k=1$ and $q=0$, one obtains
\begin{align}
&\frac{1}{2!}\frac{\partial}{\partial s_1}g_1(t,s_1)\bigg|_{s_1=2} \nonumber \\ 
=&\frac{1}{8\pi} \left(\frac{1}{t}+\psi(t)+2\,_2F_1^{(0,1,0,0)}(1,t+1;t+1;-1)+\frac{3}{2}-2\ln 2\right) \nonumber \\ 
=&\frac{1}{8\pi}\bigg(\frac{1}{t}+\psi(t)+\frac{1}{2} \psi \left(\frac{1}{2}+\frac{t}{2}\right)-\frac{1}{2} \psi \left(1+\frac{t}{2}\right)+\frac{3}{2}-2\ln 2\bigg)\ .
\end{align}
Expanding the above at $t=0$ to the quadratic order, one obtains 
\begin{align}
m^2{\cal O}_{1,s_1=2}=-g^2\times \frac{7\zeta_3}{4(4\pi)^3}\frac{g}{m^2}=-g^3\times\frac{I}{4}\equiv\langle O_1\rangle_{g^{3}} \ .
\end{align}
Here $I$ is the vacuum integral in Eq.~(\ref{eq:defI}). 
Clearly, this is nothing but the two-bubble contribution to the quartic operator $O_1=\frac{g^2}{4N}\left(\Phi^2\Phi^2\right)_c$.  For $k=2$ and $s_1=4$, we have also checked that one has
\begin{align}
(m^2)^2{\cal O}_{2,s_1=4}=\frac{72 \zeta_3-93 \zeta_5}{16(m^2)^2}\left(\frac{g}{4\pi }\right)^5\equiv \langle O_2\rangle_{g^5} \ , 
\end{align}
which also agrees with the four-bubble contribution to the operator $O_2=\frac{g^2}{4N}\Phi^2(-\partial^2)\Phi^2$. 
We should mention that since this is a rather indirect way of calculating these condensates, it serves as another strong consistency check of our calculation.

Given the finiteness of the ${\cal O}_{k,s_1=2k+q}$ contributions, let's show that they are in absent of factorial growths. This is expected, since these contributions are identified as the absolutely convergent $2k+q$-bubble contributions to the operators $O_k$, in which the denominators are sufficiently large to suppress the large logarithms. For simplicity we consider the $q=0$ case, but there is no difficulty to generalize to other $q\ge 1$. One can write
\begin{align}
&{\cal O}_{k,s_1=2k}=\frac{g}{4\pi}\left(\frac{g}{4\pi m^2}\right)^{2k}4^{-k}(2k)!{\cal I}_{2k}\nonumber \\ 
&{\cal I}_{2k}= \oint \frac{d\omega}{2\pi  i\omega^{2k+1}}\bigg(2 \psi(-k-\omega+1)-2\pi  \cot (\pi  \omega)+4^k\,_2F_1^{(0,1,0,0)}(2k-1,\omega+k;\omega+k;-1)\bigg) \ .
\end{align}
The contour is along a circle with radius smaller than $k$.
The point is, although there is an overall factor $(2k)!$, since the first pole of the integrand is at $\omega=-k$, by shifting the contour outside, when $|\omega| \sim k$, the $(2k)!$ will be suppressed by the factor $\frac{1}{\omega^{2k+1}}$. For example, let's try to shift the contour to $|\omega|=k+\frac{1}{2}=r_k$. The negative residue at $\omega=-k$ contributes to 
\begin{align}
{\cal I}_{2k,\omega=-k}=\frac{2}{(-k)^{2k+1}}+\frac{4^k (2 k-1)}{(-k)^{2k+1}}\oint_{|z|=1}\frac{dz}{2\pi i z} \, _2F_1(2 k,z+1;2;-1)=\frac{4^k}{(-k)^{2k+1}} \ ,
\end{align}
which cancels the factorial growth of the $(2k)!$ from the denominator. Notice that the $\oint dz$ integral is introduced to represent the derivative of the hypergoemetric function with respect to the second parameter.  The contribution from the deformed contour reads with $r_k=k+\frac{1}{2}$
\begin{align}
{\cal I}_{2k,|\omega|=r_k}=\frac{1}{(r_k)^{2k}}\int_0^{2\pi} \frac{d\theta}{2\pi}e^{-2ki\theta} \bigg(2\psi(1-k-r_ke^{i\theta})-2\pi \cot\pi(r_ke^{i\theta})\bigg) \nonumber \\ 
+\frac{4^k}{r_k^{2k}}\int_{0}^{2\pi} \frac{d\theta}{2\pi}e^{-2ki\theta}\int_0^{2\pi} \frac{d\varphi}{\pi}e^{-i\varphi}\,_2F_1\left(2k-1,k+r_ke^{i\theta}+\frac{1}{2}e^{i\varphi};r_ke^{i\theta}+k;-1\right) \ .
\end{align}
The point is, the integrands along the contour do not grow factorially. For example, the hypergeometric function peaks around $\theta=\varphi=\pi$ since the second and third arguments become small
\begin{align}
{\rm max} \bigg|\,_2F_1(2k-1,k+r_ke^{i\theta}+\frac{1}{2}e^{i\varphi};r_ke^{i\theta}+k;-1)\bigg|=\bigg|\,_2F_1(2k-1,-1;-\frac{1}{2};-1)\bigg|=|4k-3|\ .
\end{align}
Given the above, one finally obtains
\begin{align}
\bigg|{\cal O}_{k,s_1=2k}\bigg|=\frac{g}{4\pi}\left(\frac{g}{4\pi m^2}\right)^{2k}\frac{(2k)!}{4^k}\bigg|{\cal I}_{2k,\omega=k}+{\cal I}_{2k,|\omega|=r_k}\bigg|\le Cg\left(\frac{g}{4\pi m^2}\right)^{2k}k\times\frac{(2k)!}{k^{2k}} \ .
\end{align}
Therefore, as expected, the operator contribution $\langle O_{k}\rangle$ in diagrams with more than $2k-1$ bubbles are free from factorial growths. Moreover, the full contribution to the operator $\langle O_{k}\rangle$ beyond the $g^{2k}$ order can be bounded by the $2k$ bubble contribution. This is manifest from the momentum space representation in which the $(-1)^k(gF(k^2))^{k}$ factors always sum to geometric series $\frac{1}{1+gF(k^2)}$ thus bounded from above by one.

To summarize, in this subsection we have derived in detail the power expansion at $\mu^2=m^2$ from the Mellin-Barnes. In particular, up to finite scheme conversions (such as the $C_{n,0}^A$ in Eq.~(\ref{eq:CnA})) which contain no factorial growths and always cancel between operators and coefficient functions at the same power, one has the following identifications
\begin{align}
&\frac{g^{n+k+1}}{(p^2)^{n+k}}\left(\frac{m^2}{g}\right)^{k}C_{n+k;0,k}(l_m)=\left(\frac{m^2}{p^2}\right)^k{\cal H}_{0,k}(p^2)\bigg|_{s_1=n} \ , \label{eq:iden1} \\
&\langle O_k \rangle_{g^{n+1}}=(m^2)^k{\cal O}_{k,s_1=n} \ . \label{eq:iden2}
\end{align}
As a reminder, $C_{n+k;0,k}$ denotes the contribution to $C_{n+k}$ with $k$ mass insertions within the bubble chain. The hard contribution ${\cal H}_{0,k}$ is given in Eq.~(\ref{eq:hardMellin}), while the operator contributions ${\cal O}_{k,s_1=n}$ are given by Eq.~(\ref{eq:operatorfull1}) and Eq.~(\ref{eq:operatorfull2}). There is no difficulty to extend the analysis to higher values of $l$ in Eq.~(\ref{eq:Sl}), where $l$ corresponds to the number of mass insertions on the $p-k$ line. In particular, we have checked that for $l\ge 1$, the Mellin Barnes and the OPE computation lead to the same result for $C_{n;1,0}$.  One can show that for each $l$, the cancellation pattern remains essentially the same.

One should also notice that the hard and soft contributions obtained here from the Mellin-Barnes can be regarded as defined in the ``minimally-short-distance scheme'', which differs from the standard $\overline{\rm MS}$ scheme by finite renormalization that can not be calculated purely through the Mellin-Barnes, but are expected to cancel completely. In the Appendix.~\ref{sec:operatorms}, we have shown that the finite scheme conversion factors exactly cancel between the coefficient function $g^{n+1}C_{n;0,0}(\mu^2=m^2)$ and the operator contribution $\langle O_n(\mu=m) \rangle_{g^{n+1}}$, and their sum remains the same as obtained from the Mellin-Barnes. In fact, we have shown that for an arbitrary $\mu^2=m^2$, by adding to $C_{n;0,0}$ its ``associated'' tree-level condensates attached to their coefficient functions and then adding to the $\mu$-dependent operator condensate $\langle O_n(\mu) \rangle_{g^{n+1}}$, one obtains the same $\mu$-independent result as the Mellin-Barnes. In fact, this pattern will persists: tree-level condensates can be grouped in an inclusive and disjoint manner together with those truly non-trivial contributions (identity and the $O_n$ operators) in the OPE and leads to the Mellin-Barnes result.  Finally, the finite scheme conversions between the 
Mellin-Barnes and $\overline{\rm MS}$ schemes contain no factorial growths and will not change the factorial asymptotics presented in the next subsection.

\subsection{Large $n$ factorial asymptotics}
After introducing in details the Mellin-Barnes analysis, in this subsection we provide closed formulas for the large-$n$ factorial asymptotics of the coefficient functions and operator condensates. Clearly, such asymptotics are most efficiently obtained from the Mellin-Barnes approach, especially for the operator condensates.  For this purpose, first notice that the method discussed in the previous subsection allows to obtain the following leading IR-renormalon asymptotics for the contributions with $l$ mass insertions on the $p-k$ line, while with $k$ mass insertions within the bubble-chain
\begin{align}\label{eq:generalasym}
&\left(\frac{m^2}{g}\right)^{k+l}C_{n;l,k}(\mu^2=m^2)\bigg|_{n\rightarrow\infty,+1}=\nonumber \\ 
&-\frac{m^2}{p^2}\frac{(-1)^{k+l}}{(4\pi)^{n+1}}\left(\frac{4\pi m^2}{g}\right)^{k+l} \frac{(n!)^2  (n-k-l)! \Gamma (k-l+n+2)}{  k! (l!)^2 ((n-l)!)^2} \nonumber \\ &\times  \, _2\tilde{F}_1(-k,-k-l+n-1;-l+n+2;-1) \ .
\end{align}
Corrections to the above are exponentially small in $n$. For example, for $l=0$ the above can be derived using Eq.~(\ref{eq:hardMellin}) and the identification Eq.~(\ref{eq:iden1}). For other values of $l$, the derivation is a repetition of the procedures for $l=0$, thus is omitted to save space.  Using the fact that the largest contribution is clearly supported in the region $n\gg k$, one can perform the large $n$ expansion to obtain
\begin{align}
\left(\frac{m^2}{g}\right)^{k}\frac{C_{n;l,k}}{C_{n;l,0}}\bigg|_{n\rightarrow \infty, +1}=\frac{(-2)^k}{\Gamma (k+1)}\left(\frac{4\pi m^2}{g^2}\right)^k+\frac{(-1)^k 2^{k-1}}{ \Gamma (k-1)}\frac{1}{n}\left(\frac{4\pi m^2}{g^2}\right)^k+{\cal O}\left(\frac{l}{n^2}\right) \ .
\end{align}
Since the peak of $l$ is around $l=\sqrt{n}$, the terms neglected are subleading in $\frac{1}{n}$. Thus using the fact
\begin{align}
\left(\frac{m^2}{g}\right)^{l}C_{n;l,0}(\mu^2=m^2)\bigg|_{n\rightarrow \infty,+1}=-\frac{1}{(4\pi)^{n+1}}\left(\frac{4\pi m^2}{g^2}\right)^l\frac{m^2}{p^2}\frac{(-1)^{l} \Gamma (n+1)^2}{\Gamma (l+1)^2 \Gamma (n-l+1)} \ ,
\end{align}
one obtains after summing over $l$ and $k$, the following leading large $n$ asymptotics originated from the $s=1$ IR renormalons of the $C_n$ 
\begin{align}\label{eq:asymfinal1}
& C_{{\rm I},{\rm NL}}^{(n)}(\mu^2=m^2,p^2)\bigg|_{n\rightarrow \infty,+1}\ \nonumber \\ 
& \rightarrow -\frac{g^{n+1}}{(4\pi)^{n+1}(p^2)^{n}}\frac{m^2}{p^2}e^{-\frac{8\pi m^2}{g}}n!L_n\left(\frac{4\pi m^2}{g}\right)\bigg(1+\left(\frac{4\pi m^2}{g}\right)^2\frac{2}{n}+{\cal O}\left(\frac{1}{n^{\frac{3}{2}}}\right)\bigg) \ .
\end{align}
Here $L_n$ is the Laguerre polynomial with $L_n(0)=1$. The above asymptotics  has been checked numerically given Eq.~(\ref{eq:generalasym}). For generic $\mu^2$, the natural generalization is 
\begin{align}\label{eq:asymfinal2}
& C_{{\rm I},{\rm NL}}^{(n)}(\mu^2,p^2)\bigg|_{n\rightarrow \infty,+1}\ \nonumber \\ 
& \rightarrow -\frac{g^{n+1}}{(4\pi)^{n+1}(p^2)^{n}}\frac{\mu^2}{p^2}e^{-\frac{8\pi m_r^2}{g}}n!L_n\left(\frac{4\pi m_r^2}{g}\right)\bigg(1+\left(\frac{4\pi m_r^2}{g}\right)^2\frac{2}{n}+{\cal O}\left(\frac{1}{n^{\frac{3}{2}}}\right)\bigg) \ .
\end{align}
Clearly, the factorial growth will never be canceled and the power expansion for the coefficient function as well as for the momentum space two-point function will not be convergent. Notice that the over-all factor $e^{-\frac{8\pi m_r^2}{g}}$ is non-perturbative in the coupling constant $g$. 

Similarly, at $\mu^2=m^2$, from the Mellin Barnes one can derive the following leading non-alternating factorial asymptotics up to subleading renormalons for the operator condensates
\begin{align} \label{eq:generalasyo}
\bigg\langle O_{n;k}(\mu=m)\bigg\rangle_{n\rightarrow \infty,+1} & \rightarrow\frac{(-1)^{k+1}g^{n+1}}{(4\pi)^{n+1}} \left(\frac{4\pi m^2}{g}\right)^k\frac{ (n-k)! \Gamma (k+n)}{\Gamma (k)} \nonumber \\ 
& \times \, _2\tilde{F}_1(1-k,-k+n-1;n+1;-1) \ , 
\end{align}
where $O_{n;k}$ denotes the contribution to the operator $O_n$ in the $n-k$ bubble diagram with $k\ge 1$.  To derive the above, simply use the Eq.~(\ref{eq:operatorfull1}) and the identification Eq.~(\ref{eq:iden2}), which states $\langle O_{n;k}\rangle=\langle O_n\rangle_{g^{n-k+1}}=m^{2n}{\cal O}_{n,s_1=n-k}$. For $k\le 0$, there is no non-alternating factorial growth, consistent with the UV renormalon interpretation for operators. Combining Eq.~(\ref{eq:generalasym}) and Eq.~(\ref{eq:generalasyo}), one can see that the leading factorial growth of the operators $O_{n+1}$ in the $n-k$ bubble diagram exactly cancel that of $C_{n;0,k}$ with $k$ mass insertions within the $n-k$ bubbles. Now, if there are $l$ mass insertions on the $p-k$ line, one should expect that the $O_{n+1-l; k+1}$ combined with the $(m^2)^l$ term in the coefficient function cancel with $C_{n;l,k}$. Indeed, using results in Appendix.~\ref{sec:angular}, one obtains 
\begin{align}
& \frac{(-1)^lm^2}{(p^2)^{n+1}}\frac{g^{n+1}}{(4\pi)^{n+1}}\left(\frac{4\pi m^2}{g}\right)^{k+l}\frac{(n-l+k)!^2}{l!^2(n-l)!^2}\times \frac{(n-l-k)!\Gamma(n-l+k+2)}{k!} \nonumber \\ 
& \times \,_2\tilde F_1(-k,n-k-l-1;n-l+2;-1) \ ,
\end{align}
which exactly cancels Eq.~(\ref{eq:generalasym}) multiplying the overall factor $\frac{g^{n+1}}{(p^2)^n}$ for $C_n$.  This is completely consistent with the general cancellation pattern. Moreover, by summing over $k\ge 0$, the leading non-alternating factorial growth for the operator condensate reads 
\begin{align}\label{eq:asymfinal3}
\bigg\langle O_{n}(\mu=m)\bigg\rangle_{n\rightarrow \infty,+1}  \rightarrow\frac{m^2g^n}{(4\pi)^n}(n-1)!e^{-\frac{8\pi m^2}{g}}\bigg(1+\left(\frac{4\pi m^2}{g}\right)^2\frac{2}{n-1}+{\cal O}\left(\frac{1}{n^{\frac{3}{2}}}\right)\bigg) \ .
\end{align}
Again, the non-alternating factorial asymptotics is exponentially small in $g$. The Eq.~(\ref{eq:asymfinal1}), Eq.~(\ref{eq:asymfinal2}) and Eq.~(\ref{eq:asymfinal3}) represent one of the major results of this work.

Leading asymptotics from the alternating renormalons can also be calculated in similar manners. For example, at $\mu=m$ one has 
\begin{align}
&\bigg\langle O_{n;-1+k}(\mu=m)\bigg\rangle_{n\rightarrow\infty;(-1)^n} \rightarrow \frac{(-1)^ng^{n+2}}{(4\pi)^{n+2}}\left(\frac{4\pi m^2}{g}\right)^k \frac{(-k+n+1)! \Gamma (k+n-1)}{k!} \nonumber \\ 
& \times \, _2\tilde{F}_1(-k,n-k;n-1;-1) \ .
\end{align}
In particular, for $k=0$ the above exactly reduces to the Eq.~(\ref{eq:operatorals}) generated in the absolutely convergent $n+1$ bubble diagram. Notice that for $k\le -1$ there are no leading alternating renormalon, while the $s_1=-q$ subleading renormalon persists up to $k=-q+1$. Summing over $k$ one obtains 
\begin{align}
&\bigg\langle O_{n}(\mu=m)\bigg\rangle_{n\rightarrow\infty;(-1)^n} \nonumber \\ 
& \rightarrow \frac{(-1)^ng^{n+2}}{(4\pi)^{n+2}m^2}(n+1)!e^{\frac{8\pi m^2}{g}}\bigg(1-\left(\frac{4\pi m^2}{g}\right)^2\frac{(\frac{g}{\pi m^2}-2)}{n}+{\cal O}\left(\frac{1}{g^4n^2}\right)\bigg) \ .
\end{align}
Strikingly, the leading alternating factorial growth is exponentially large in $g$. This is not in contradiction with the fact that in the $g\rightarrow 0$ limit there are no factorial growths, since to reach the asymptotic region one needs $n\gg \frac{1}{g^2}$, which moves to infinity as $g\rightarrow 0$.  

Given the asymptotics of the operators and the coefficient functions, let's consider their effects on the power expansion. Then, when projected to any given power $n$ in $(p^2+m^2)^2S_{\rm NL}$, in the large $n$ limit one has 
\begin{align}
&(p^2+m^2)^2S_{\rm NL}^{(n)}\bigg|_{n\rightarrow\infty,+1} \nonumber \\ 
&\rightarrow m^2e^{-\frac{2m^2}{\hat g}} \bigg(\left(\frac{\hat g}{p^2}\right)^n(n-1)!L_{n-1}\left(\frac{m^2}{\hat g}\right)-\left(\frac{\hat g}{p^2}\right)^{n+1}n!L_{n}\left(\frac{m^2}{\hat g}\right)\bigg) \ ,
\end{align}
where $\hat g=\frac{g}{4\pi }$. Clearly, the first contribution is due to the operators $O_{n-l},l \ge 0$, while the second contribution is due to the coefficient function $C_{{\rm I},{\rm NL}}^{(n)}$. Given the factorial behavior, the power expansion clearly failed to converge. If one introduces the Borel transforms
\begin{align}
&B_s[t]=m^2e^{-\frac{2m^2}{\hat g}}\sum_{n=0}^{\infty}t^n
L_n\left(\frac{m^2}{\hat g}\right)=m^2e^{-\frac{2m^2}{\hat g}}\frac{\exp \left(-\frac{ m^2}{\hat g}\frac{t}{1-t}\right)}{1-t} \ , \\ 
&B_h[t]=-\frac{\hat g}{p^2}m^2e^{-\frac{2m^2}{\hat g}}\sum_{n=0}^{\infty}t^n(n+1)L_{n+1}\left(\frac{m^2}{\hat g}\right)=-\frac{\hat g}{p^2}\frac{d}{dt}B_s[t] \ , 
\end{align}
then there is a singularity at $t=1$ which is of infinite order. The above implies that the power-expansion for the non-trivial coefficient functions, as well as the sum over non-trivial operators are not only divergent, but are  Borel non-summable. Despite this divergence, since $B_s[t]$ do not grow at infinity, one clearly has for $\frac{\hat g}{p^2}>0$
\begin{align}
\int_{0}^{(1+i\epsilon)\infty}dt e^{-\frac{p^2}{\hat g}t} \bigg(B_s[t]-\frac{\hat g}{p^2}\frac{d}{dt}B_s[t]\bigg)=\int_{0}^{(1-i\epsilon)\infty}dt e^{-\frac{p^2}{\hat g}t} \bigg(B_s[t]-\frac{\hat g}{p^2}\frac{d}{dt}B_s[t]\bigg) \ .
\end{align}
Namely, the full power expansion, although divergent and Borel non-summable in the strict sense, contain no ``Borel ambiguity'' if only $(1\pm i0)R$ paths are allowed. In fact, this is a general property for sums of the following form $a_{k+q}x^{k+q}-a_kx^k$, as far as the Borel transform of $\{a_k\}$ do not grow too fast at infinity. 

\section{Conclusion and Comments}\label{sec:conclu}
In this work, we have investigated the OPE/power expansion of a two-point function in a 2D large-$N$ quartic $O(N)$ model. Factorial enhancements to the power-expansion/OPE are observed that is severe enough to prevent the convergence of the large $p^2$ power-expansion of two-point function, an expansion that is well defined at any given power. The reason of the divergence can be summarized as follows. 

The large $p^2$ expansion is naturally performed in the form of OPE in terms of {\it short-distance-scheme} (such as the $\overline{\rm MS}$-scheme) operators and their coefficient functions.  Since short-distance-scheme operators are too point like, there are increased UV sensitivity as their dimensions increase. Due to super-renormalizability, this is not sufficient to introduce renormalon ambiguities that affect their existences. They are all well-defined. But also since the theory still contains  a lot of logarithms, the UV fluctuations are still strong enough to lead to factorial enhancements to condensates: when the increased denominators (due to super-renormalizability) are just canceled by the numerators (introduced by the operators), large-logarithms could survive and lead to factorial enhancements.

On the other hand, since the two-point function is a well-defined function and since at fixed order in the large-$N$ expansion multi-particle singularities are bounded from the above, the factorial enhancements must be canceled by coefficient functions. Since the factorial enhancements are due to UV region for operators, we expect them to cancel with coefficient functions with many hard propagators, thus for lower-dimensional operators. For example, for the operator $O_n$, the region that leads to the largest factorial growths are very similar to $O_n\rightarrow{\rm I}$ operator mixing, thus we expect them to cancel with high-power terms in the coefficient function of the identity operator. But there are two-many $\ln p^2$s in such terms, and a non-trivial property is that the factorial growths at different $\ln p^2$ orders also cancel with each-other, and the net factorial growths are always at least one power higher or lower. The factorial asymptotics in the coefficient functions are power-changing. Then, the net factorial growths at any fixed power can not be canceled, thus the power expansion failed to converge. 

Here we must emphasize that before this work, factorial enhancements to the large $p^2$ expansion due to unbounded multi-particle thresholds are anticipated and has been explicitly mentioned in references such as~\cite{Beneke:1998eq, Marino:2023epd}. In fact, in the 2D planar QCD, due to the presence of infinitely-many stable mesons, the large $p^2$ expansion for the polarization function is very likely to be divergent as well~\cite{Boito:2017cnp,deRafael:2010ac}. In many 2D integrable QFTs, one usually has unbounded particle numbers in the form-factor expansions of local operators~\cite{Smirnov:1992vz} (such that the $\langle\sigma(z)\sigma(0)\rangle$~\cite{Wu:1975mw,Berg:1978sw} propagator in the $m\bar\psi\psi$-perturbed 2D Ising). In all such cases, we expect that the large $p^2$ expansions are divergent. But the divergence of the large $p^2$ expansion at fixed orders in the large $\frac{1}{N}$ expansion where multi-particle thresholds are bounded from the above, to the knowledge of the author, was not noticed before in the literature. Compared to the divergence of the large $p^2$ expansion due to multi-particle thresholds, the factorial enhancements found in this work are less sensitive to large distance effects, but reflect better the subtle role played by the bubble-chain-amplified logarithms, in modifying quantitative properties measuring how the short distance and large distance scales are coupled/decoupled when one deviates/approaches the fixed point.   

We should notice that the same $O(N)$ quartic model has also been studied in the context of weak-coupling expansion in the negative mass formulation~\cite{Marino:2019fvu,Marino:2025ido}, corresponding to the strong coupling expansion in the positive mass formulation adopted in this work. The focus of these works is to show how the perturbative expansion around the $O(N)$-broken false vacuum could still lead to the correct asymptotic expansions of physical quantities with modified high-order behavior due to the presence of massless particles in the computation. In particular, in the recent work~\cite{Marino:2025ido}, the author has shown that the weak-coupling expansion of the two-point function in the negative mass formulation can also be performed around the $O(N)$-broken vacuum with IR safety. Since the OPE remains the same form for all stable coupling constants, it is interesting to compare the results in~\cite{Marino:2025ido} and the OPE in this work to gain better understanding of both. It is possible to show that at the power $\frac{1}{p^2}$ for the self-energy, by combining the condensates ${\cal O}_{1,s_1\ge 2}$ with the coefficient function, one can recover the results in~\cite{Marino:2025ido} at the same momentum power. The OPE of this work could also be used to study exponentially-small corrections to~\cite{Marino:2025ido}, at least in the large-$p^2$ expanded form. More details will be presented in a future work.

Before ending the work, let's further make the following comments.
\begin{enumerate}
    \item First, we emphasize that due to the off-diagonal cancellation pattern across different powers, the naive power-expansion in $\frac{1}{(p^2)^{n+2}}$ for the two point function, although free from any ambiguity at any given power, failed to converge. The OPE is also not convergent, unless one redefines all the operators by subtracting out all the factorial growths and adding back to coefficient functions. Even after this procedure, the coefficient function still can not be summed as a convergent power expansion. 
    \item Second, we should also notice that the renormalization factors in the $\overline{\rm MS}$ scheme for operators do not introduce factorial enhancements,  see Eq.~(\ref{eq:Z}) for example. This is consistent with the general wisdom that the $\overline{\rm MS}$ scheme renormalization data such as anomalous dimensions do not introduce renormalons~\cite{Palanques-Mestre:1983ogz}. The factorial growths of $O_n$ in our opinion therefore reflect their genuine properties. For further analysis of the operator condensates in the $\overline{\rm MS}$ scheme and the scheme conversion factors, see Appendix.~\ref{sec:operatorms}.

    \item We should emphasize that behind the factorial enhancements of operators is a correlation between operator's derivative numbers and perturbative orders. Indeed, for the operator $O_n$ with $n-1$ $\partial^2$ inside, it is the bubble-chain diagrams with bubble numbers around $n$ that cause the factorial growth of $O_n$ in $n$: for these diagrams, there is a very large power of the logarithms, while a relatively small algebraic power. For bubble numbers $n+k \gg n$, the UV region is suppressed by $\frac{1}{k^n}$ and will no-longer cause factorial growth when $k\sim{\cal O}(n)$. This is consistent with the fact that $O_n$ themselves have no Borel ambiguities when regarded as perturbative expansions in $\frac{g}{m^2}$.  
    
    The situation is different from marginal theory: as least based on lessons from large-$N$ expansion~\cite{David:1982qv,David:1983gz,David:1985xj,Beneke:1998eq,Marino:2024uco,Liu:2024omb}, in asymptotically free theories there are ambiguities to define high-dimensional operators, but these ambiguities can be resolved easily in an ``one-for-all'' manner using a Borel prescription. Once the Borel prescription has been fixed, there are no further factorial enhancements of ``high-twist'' operators. 
   \item We should also mention that unlike the renormalons in marginal theories, the factorial enhancements observed here are rather subtle and are generated from $\frac{1}{\epsilon} \times \epsilon$ effects. As such, it is less transparent to see its presence using naive $d=2$ running-coupling-type arguments. To see if it works, let's consider the IR renormalized mass-less bubble chain with 
   \begin{align}
   F(k^2)=\frac{1}{4\pi k^2}\ln\frac{k^2}{\mu^2} \ .
   \end{align}
   Plug it in the diagram with $n$ bubbles, one obtains
   \begin{align}
   \frac{-g^{n+1}}{(4\pi)^n}\int \frac{d^dk}{(2\pi)^d}\frac{\mu_0^{2-d}}{(k^2)^n}\left(\frac{\mu^2}{k^2}\right)^s\frac{n!}{(p-k)^2}\bigg|_{s^n}-\frac{g^{n+1}}{(4\pi)^{n+1} \epsilon}\frac{1}{(p^2)^{n}}\ln^n\frac{\mu^2}{p^2} \ .
   \end{align}
   Here $|_{s^n}$ denotes the $s^n$ power Taylor expansion coefficient at $s=0$. Notice the last term corresponds to the IR subtraction on the $p-k$ line. In this manner, one obtains
   \begin{align}
&-\frac{g^{n+1}n!}{(4\pi)^{n+1}(p^2)^n}\left(\frac{
\mu^2
}{p^2}\right)^s\frac{\exp (\gamma_E \epsilon) \Gamma (-\epsilon) \Gamma (-\epsilon-n-s+1) \Gamma (\epsilon+n+s)}{\Gamma (n+s) \Gamma (-2 \epsilon-n-s+1)}\bigg|_{s^n} \nonumber \\ 
&-\frac{g^{n+1}}{(4\pi)^{n+1} \epsilon}\frac{1}{(p^2)^{n}}\ln^n\frac{\mu^2}{p^2} \nonumber \\ 
& \rightarrow \frac{g^{n+1}}{(4\pi)^{n+1}(p^2)^n}\left(\frac{\mu^2}{p^2}\right)^s n!\left(-l_{\mu}+\psi (-n-s+1)+\psi(n+s)+2 \gamma_E \right) \bigg|_{s^n}\ .
   \end{align}
   Again, the $\ln p$ in the $\frac{1}{\epsilon}$ is removed by the IR subtraction on $p-k$, but one is left with a pole at $s=0$. This pole is related to the operator renormalization for $O_n$ that contains no dependency on $p$ and can not be calculated accurately in this method. Neglecting this pole, one obtains exactly the Eq.~(\ref{eq:CBn}) for $C_{n,0}^B$ that contains the factorial growth in $n$. Therefore, similar to the 4D bubble chain in QCD~\cite{Beneke:1998ui}, the naive bubble chain approximation here also produces the correct large $n$ factorial behavior, up to a scheme dependent constant part  that is related to operator renormalization and contains no factorial growth in $\overline{\rm MS}$-like schemes. 

   The factorial growth in fact allows a simpler explanation: if one consider the integrand
   \begin{align}
   \frac{(-1)^{n+1}g^{n+1}}{(4\pi)^n}\int_{k^2 \ll \mu^2}\frac{d^2k}{(2\pi)^2}\frac{1}{(k^2)^n}\ln^{n}\frac{k^2}{\mu^2}\frac{1}{(p-k)^2} \ ,
   \end{align}
   then for small-$k$, although it has a high degree of IR divergence, the large-power of $k^2$ seems to have balanced out the large power of the logarithm and no-factorial behavior seems possible. However, if one Taylor expand the integrand to the first power where the integral converges in the IR, then one has
\begin{align}
\frac{(-1)^{n+1}g^{n+1}}{(4\pi)^n}\int_{k^2 \ll \mu^2}\frac{d^2k}{(2\pi)^2}\frac{1}{(k^2)^n}\ln^{n}\frac{k^2}{\mu^2}\frac{(k^2)^n}{(p^2)^{n+1}} \rightarrow -\frac{g^{n+1}}{(4\pi)^n}\frac{\mu^2}{p^2}\frac{n!}{(p^2)^n} \ ,
\end{align}
which is just the leading non-alternating factorial growth in Eq.~(\ref{eq:Cnasym1}) after multiplying the overall factor $\frac{g^{n+1}}{(p^2)^n}$. From the above, it is clear that the factorial growth in the coefficient function is produced in the region where $k^2\ll \mu^2$ in the renormalized bubbles. Naively, in the IR subtraction, $\mu$ serves as the separation point between UV and IR and coefficient functions should not go the momentum region below than $\mu$. The point is, here the subtraction is {\it local} in nature (they are just the UV poles of operators) and can not see the momenta within the massless bubble. Even when the $\mu$ is actually larger than the $k^2$ of the bubble, one still subtracts up to the point $\mu$ which is actually much harder than the massless bubble. This is completely consistent with the interpretation of the observed factorial growth as an {\it IR} renormalon, and the fact that renormalons are generated due to over-subtractions~\cite{Feldman:1985th,Rivasseau:2014zpm}. 

On the other hand, we have seen that the $\overline{\rm MS}$-scheme operators $O_n$, despite being finite after renormalization, do receive contributions from the momentum region $k^2\sim m^2e^{n}$ which moves deeply into the UV as $n$ becomes large. Combining the above, it is clear that the factorial growths arise because in $\overline{\rm MS}$-like schemes, one subtracts too much in the coefficient functions, while subtracts too little in the operators. However, since it is exactly through the short-distance-scheme operators such that the expansion of correlation functions can be performed up to an arbitrary power accuracy on a detailed level, factorial growths in the power expansion are unavoidable.

\item Finally, we must emphasize that the factorial enhancement is only $n!$. This is sufficient to make the momentum space OPE divergent. However, in the coordinate space expansion, there is always an additional $\frac{1}{(n!)^2}$ suppression factor for the $(z^2)^n$ power, thus a single $n!$ can not affect the convergence of coordinate space OPE or power expansion at large $N$. Moreover, beyond the $\frac{1}{N}$ expansion, there can be factorial growth in the numbers of Feynman-diagrams. But as far as the number of Feynman diagrams at order $g^n$ is bounded by $n!\times C^n$, the factorial enhancement for the $n$-the power in the coefficient functions would at worst lead to $\frac{n!\times n!}{(n!)^2}\tilde C^n (gz^2)^{n\pm 1}$, which still has a finite convergence radius. This is consistent with the statements in~\cite{Lukyanov:1996jj}. 
\end{enumerate} 

\acknowledgments
The author thanks Marcos Marino for the constructive discussions and the notes sent. The author also thanks him for sharing his recent work~\cite{Marino:2025ido} and the interesting topics therein. The author thanks Zoltán Bajnok for the constructive discussions and notification of the reference~\cite{Zamolodchikov:1990bk}. The author thanks Yushan Su, Leszek Motyka, and Maciej Nowak for the valuable discussions.

\appendix
\section{Expansion of the sunrise diagram}\label{sec:sunrise}
We consider the $g^2$ order sunset diagram for the scalar two point function (this is the only diagram at order $g^2$)
\begin{align}
S(z^2m^2)=\frac{g^2}{6}\int \frac{d^2p}{(2\pi)^2}\frac{e^{ip\cdot z}}{(p^2+m^2)^2}\Sigma(p^2) \ , \\
\Sigma(p^2)=\int \frac{d^2k d^2l}{(2\pi)^4}\frac{1}{(k^2+m^2)(l^2+m^2)((p-k-l)^2+m^2)} \ .
\end{align}
In the appendix without otherwise mentioning we use the letter $S$ only to denote the order $g^2$ contribution to the full two-point function.
The simplest way of expanding $S(z^2m^2)$ is to use twice the equation of motion to obtain
\begin{align}
(-\partial^2+m^2)^2S(z^2m^2)=\frac{g^2}{6}G_0^3(z^2m^2) \ , 
\end{align}
Given the first two terms in the small $z^2$ expansion, coefficients for higher order terms can all be determined from this equation. For example, we can write
\begin{align}
S(z^2m^2)=\frac{g^2}{6m^4}\frac{1}{(4\pi)^3}\bigg(\alpha+\beta \frac{z^2m^2}{4}+\left(\frac{z^2m^2}{4}\right)^2\left(aL^3+bL^2+cL+d\right) \bigg)\ .
\end{align}
Using $\partial^2=\frac{d^2}{dz^2}+\frac{1}{z}\frac{d}{dz}$, and $G_0=-\frac{1}{4\pi}L+{\cal O}(z^2m^2L)$, by matching all the logarithms and constants at the order $z^0$, one can solve
\begin{align}
4a=-1 \ , 4b+36a=4c+78a+24b=4d+36a+26b+12c+\alpha-2\beta=0 \ .
\end{align}
Thus
\begin{align}
a=-\frac{1}{4}, \ b=\frac{9}{4}, \  c=-\frac{69}{8} , \ d=\frac{27}{2}-\frac{\alpha}{4}+\frac{\beta}{2} \ . 
\end{align}
By calculating the diagrams for the $\phi^2$, $\frac{z^2}{4}\phi\partial^2\phi$ contributions as in Sec.~\ref{sec:OPEintro}, one can also determine $\alpha=\frac{7\zeta_3}{4}$, $\beta=-\frac{21\zeta_3}{4}$, thus $d=-\frac{7\zeta_3}{16}-\frac{42\zeta_3}{16}=-\frac{49\zeta_3}{16}$.
However, for illustration purpose, here we present the direct approach based on Mellin-Barnes.

The first step of the Mellin-Barnes is to introduce the Schwinger parameters for the self energy
\begin{align}
\Sigma(p^2)=\frac{1}{(4\pi)^2}\int \frac{d\alpha_1d\alpha_2d\alpha_3}{\alpha_{1}\alpha_2+\alpha_{1}\alpha_3+\alpha_{2}\alpha_3}\exp \bigg[-p^2\frac{\alpha_{1}\alpha_2\alpha_3}{\alpha_{1}\alpha_2+\alpha_{1}\alpha_3+\alpha_{2}\alpha_3}-m^2\alpha_{123}\bigg] \ .
\end{align}
Now, we perform Mellin transform with respect to $\frac{p^2}{m^2}$
\begin{align}
&\Sigma(p^2)=\frac{1}{(4\pi)^2 m^2}\int \frac{ds}{2\pi i} \left(\frac{p^2}{m^2}\right)^{-s}{\cal M}(s) \ , \\
&{\cal M}(s)=\Gamma(s)\int_{0}^{\infty} \frac{(\alpha_1\alpha_2\alpha_3)^{-s}d\alpha_1d\alpha_2d\alpha_3}{(\alpha_{1}\alpha_{2}+\alpha_1\alpha_3+\alpha_{2}\alpha_3)^{1-s}}\exp \bigg[-(\alpha_1+\alpha_{2}+\alpha_3)\bigg] \ .
\end{align}
Here $0<{\rm Re}(s)<1$. It is a meromorphic function with poles at integers. Right to the convergence strip $0<{\rm Re}(s)<1$, it contains an overall $s=1$ pole corresponding to the $\alpha_1\sim\alpha_2\sim\alpha_3 \rightarrow 0$ region, and also contains poles from the sub-regions $\alpha_1 \gg \alpha_2\sim\alpha_3$ or $\alpha_1\sim \alpha_2 \gg \alpha_3$. Using $\alpha_1=\beta_0$, $\alpha_2=\beta_0\beta_1$, $\alpha_3=\beta_0\beta_1\beta_2$, one can de-singularize all these poles and see the analyticity structure directly
\begin{align}
{\cal M}(s)=6\Gamma(s)\Gamma(1-s)\int_0^1 d\beta_1\int_0^1 d\beta_2(\beta_1\beta_2)^{-s}(1+\beta_1+\beta_1\beta_2)^{s-1}(1+\beta_2+\beta_1\beta_2)^{s-1} \ .
\end{align}
From the above, it is clear that all the singularities have been factorized to the overall factor $(\beta_1\beta_2)^{-s}$, while the rest of the integrand are never singular. Moreover, by partial integrating, ${\cal M}(s)$ becomes a meromorphic function in $s$ with poles at $s=n \in Z$. Furthermore, the ${\cal M}(s)$ decays exponentially as ${\cal O}(e^{-\pi|{\rm Im}(s)|})$ in the vertical strip, while grows in ${\rm Re}(s)$ generically no faster than ${\cal O}\left(9^{{\rm Re}(s)}\right)$. This is consistent with the general pattern of Mellin transform for momentum space correlation function in fixed order perturbation theory.

Although the parametric representation can be useful to establish general properties of ${\cal M}(s)$, for calculation purposes this representation is rather inconvenient. Instead of using the parametric representation, the expansion of ${\cal M}(s)$ around its poles can be systematically calculated using another layer of Mellin Barnes. One can first introduce the Mellin-Barnes representation for $\tilde \Sigma(p^2)=\int \frac{d^2k}{(k^2+m^2)((p-k)^2+m^2)}$, then obtain that for $\Sigma(p^2)=\int \frac{d^2k}{(2\pi)^2}\frac{\tilde \Sigma(k^2)}{(p-k)^2+m^2}$. This leads to the Barnes representation
\begin{align}
{\cal M}(s)=\int \frac{dz}{2\pi i}\frac{\Gamma (1-z)^4 \Gamma (s) \Gamma (z-s)^2}{\Gamma (2-2 z) \Gamma (1-s)} \ ,
\end{align}
where $0<{\rm Re}(z)<1$ and $0<{\rm Re}(s)<{\rm Re}(z)<1$. Thus, by shifting the contour of $z$ to the left and pick up the poles at $z=s-k, k\ge 0$, one obtains the following absolutely-convergent expansion
\begin{align}
&{\cal M}(s)=\sum_{k=0}^{\infty}{\cal M}_k(s) \ , \\
&{\cal M}_k(s)=\frac{2 \Gamma (s) \left(-2 H_{k-s}+H_{2 k+1-2 s}+H_k\right) \Gamma (k+1-s)^4}{(k!)^2 \Gamma (1-s) \Gamma (2 k+2-2s)} \ .
\end{align}
Notice that ${\cal M}_{k \ge n}(s)$ vanishes at $s=n\ge 1$, thus only ${\cal M}_{0 \le k \le n-1}$ are required to calculate the expansion of ${\cal M}(s)$ up to constant terms at $s=n, n\ge 1$. In particular, the expansion of ${\cal M}(s)$ for $s=1,2,3$ up to constant term read
\begin{align}
&{\cal M}(s)\bigg|_{s\rightarrow 1} = -\frac{6}{(s-1)^3}+4\zeta_3 \ , \\
&{\cal M}(s)\bigg|_{s\rightarrow 2} \rightarrow -\frac{6}{s-2}+\frac{24}{(s-2)^2}+\frac{18}{(s-2)^3}-12 (\zeta_3+1) \ , \\ 
&{\cal M}(s)\bigg|_{s\rightarrow 3} \rightarrow\frac{3}{2 (s-3)}-\frac{156}{(s-3)^2}-\frac{90}{(s-3)^3}+\left(60 \zeta_3+\frac{183}{2}\right) \ .
\end{align}
And poles at $s=n\ge4$ can be calculated systematically.

Given the Mellin representation of $\Sigma(p^2)$, it is straightforward to obtain the Mellins representation of the full two-point function using
\begin{align}
&\int \frac{d^2p}{4\pi^2(p^2+m^2)^2(p^2/m^2)^s}e^{ip\cdot z}=\frac{1}{\Gamma(s)}\frac{1}{4\pi m^2}\int \rho d\rho  \rho x (\rho(1-x))^{s-1}\int dxe^{-\rho  x-\frac{z^2m^2}{4\rho}} \nonumber \\
&=\frac{1}{4\pi m^2} \int \frac{du}{2\pi i} \left(\frac{z^2m^2}{4}\right)^{-u}\frac{\Gamma (u)\Gamma (1-s-u) \Gamma (1+s+u)}{\Gamma (1-u)} \ .
\end{align}
Here the contour is $0<{\rm Re}(u)<1-{\rm Re}(s)$. Thus
\begin{align}
S(z^2m^2)=\frac{g^2}{6}\frac{1}{(4\pi)^3m^4}\int \frac{ds}{2\pi i}\int\frac{du}{2\pi }\frac{\Gamma (u)\Gamma (1-s-u) \Gamma (1+s+u){\cal M}(s)}{\Gamma (1-u)} \left(\frac{z^2m^2}{4}\right)^{-u} \ .
\end{align}
Now, one can obtain the expansion of $S(z^2)$ by shift the $u$ contours to the left, and then shift the $s$ contours to the right. The poles at $u=-l\le 0$ will be the ``soft'' contributions at powers $(z^2)^l$, while the pole at $u=-l-s, l\ge 1$ then $s=n\ge 1$ will contain the ``hard contributions'' at powers $(z^2)^{n+l}$, namely, contribution from the regions where at least one line is hard.  

We first consider the hard contributions. For our purpose one needs 
\begin{align}
{S}_{u=-1-s}(z^2m^2)=\frac{g^2}{6}\frac{1}{(4\pi)^3m^4}\int \frac{ds}{2\pi i}\frac{ {\cal M}(s) \Gamma (-1-s) }{\Gamma (s+2)}\left(\frac{z^2m^2}{4}\right)^{s+1} \ .
\end{align}
 Using the expansion of ${\cal M}(s)$ at $s=1$ one obtains 
\begin{align}
&\Sigma(p^2)\bigg|_{p^2\rightarrow \infty}\rightarrow \frac{3}{(4\pi)^2 p^2} \ln^2\frac{p^2}{m^2}+{\cal O}\left(\frac{m^2}{p^4}\right) \ , \\ 
&S_{u=-1-s}\bigg|_{z^2\rightarrow0}\rightarrow \frac{1}{6(4\pi)^3}\bigg(\frac{z^2g}{4}\bigg)^2\bigg(-\frac{L^3}{4}+\frac{9 L^2}{4}-\frac{69 L}{8}+\frac{27}{2}\bigg) \ ,
\end{align}
with $L=\ln \frac{z^2m^2e^{2\gamma_E}}{4}$. 
Moreover, using the expansion of ${\cal M}(s)$ at $s=2,3$ allows the expansion for hard parts to be performed up to $z^8$. 

We then consider purely soft contributions from $u=0,-1,-2,...$ poles. Up to $z^4$, there are three numbers which are due to $\langle \phi^2\rangle_{g^2}$, $\langle\phi \partial^2\phi\rangle_{g^2}$ and $\langle \phi \partial^4\phi\rangle_{g^2}$. These numbers only affect expansion in coordinate space. The expansion for the $u=-n$ poles are
\begin{align}
S_{u \in Z_{\le 0}}(z^2m^2)=\frac{g^2}{6(m^2)^2(4\pi)^3}\sum_{n=0}^{\infty}\left(\frac{z^2m^2}{4}\right)^ns_n \ , \\
s_n=\frac{1}{(n!)^2}\int \frac{ds}{2\pi}\frac{\pi  (s-n)}{ \sin \pi s}{\cal M}(s) \ .
\end{align}
To evaluate the above, it is convenient to use again the representation 
\begin{align}
{\cal M}(s)=\int \frac{dz}{2\pi i}\frac{\Gamma (1-z)^4 \Gamma (s) \Gamma (z-s)^2}{\Gamma (2-2 z) \Gamma (1-s)} \ ,
\end{align}
where $0<{\rm Re}(z)<1$ and $0<{\rm Re}(s)<{\rm Re}(z)<1$. Thus, one can perform first the $ds$ integral within $dz$ using the first Barnes lemma, which leads to
\begin{align}
& a=\int \frac{ds}{2\pi i}\frac{\pi s}{\sin \pi s}{\cal M}(s)=\int \frac{dz}{2\pi i}\frac{\pi ^3 z\cot (\pi  z) \csc ^2(\pi  z)}{1-2 z} \ , \\
& b=\int \frac{ds}{2\pi i}\frac{\pi }{\sin \pi s}{\cal M}(s)=\int \frac{dz}{2\pi i} \frac{2 \pi ^3 \cot (\pi  z) \csc ^2(\pi  z)}{1-2 z} \ .
\end{align}
The above can be evaluated by shifting the contour to the right
\begin{align}
& a=\sum_{k=1}^{\infty}\frac{2}{(2k-1)^3}=\frac{7}{4}\zeta_3
 \ ,  \\ 
 & b=\sum_{k=1}^{\infty}\frac{8}{(2k-1)^3}=7\zeta_3 \ .
 \end{align}
This determines the soft part of the expansion as
\begin{align}
S_{u \in Z_{\le 0}}(z^2m^2)=\frac{g^2}{6m^4}\frac{7\zeta_3}{(4\pi)^3}\sum_{n=0}^{\infty} \frac{1}{(n!)^2}\left(\frac{1}{4}-n\right)\left(\frac{z^2m^2}{4}\right)^n \ .
\end{align}
In particular, the contribution at $(z^2m^2)^4$ reads
\begin{align}
S_{u =-2}=-\frac{g^2z^4}{6}\frac{49\zeta_3}{256(4\pi)^3} \ .
\end{align}
Moreover, the soft part converges absolutely for any $z^2$ and can be summed into  Bessel $I$ function. Notice that the soft contribution grows exponentially at large $z$. Both soft and hard contributions are required to guarantee exponential decay. This is consistent with the
statements in~\cite{Parisi:1978az} on the breakdown of clustering property of massless perturbation theory. The coefficient functions obtained from massless computations must be combined with very specific condensates in order to satisfy the clustering property.

To summarize, the full expansion of the two point function up to power $z^4$ and ${\cal O}(g^2)$ reads
\begin{align}
&S_{g^2}(z^2m^2,\frac{g}{m^2})=\nonumber \\&-\frac{1}{4\pi}L+\frac{7\zeta_3}{24(4\pi)^3}\frac{g^2}{m^4}\nonumber \\
& +\bigg(\frac{2-L}{4\pi}-\frac{7\zeta_3}{8(4\pi)^3}\frac{g^2}{m^4}\bigg)\frac{z^2m^2}{4} \nonumber \\ 
&+ \bigg(\frac{3-L}{16\pi}+\frac{g^2}{6(4\pi)^3m^4}\left(-\frac{L^3}{4}+\frac{9L^2}{4}-\frac{69L}{8}+\frac{27}{2}-\frac{49\zeta_3}{16}\right)\bigg)\left(\frac{z^2m^2}{4}\right)^2 \ .
\end{align}
Here we have added the expansion of the tree-level diagram. 

\section{Angular averages}\label{sec:angular}
In this appendix we derive the angular averages for the small-$k$ expansion of the tree-level propagator $\frac{1}{(p-k)^2+m^2}$ in an arbitrary $d$-dimension. 

The simplest way to perform this expansion is to use the following trick. Consider
\begin{align}
I(z,k)=\int \frac{d^Dp}{(2\pi)^D}\frac{e^{ip\cdot z}}{(p-k)^2+m^2}= \int \frac{d^Dp}{(2\pi)^D}\frac{e^{ip\cdot z}}{p^2+m^2}e^{ik\cdot z} \ .
\end{align}
Clearly, in order to obtain the small $k$ expansion from the region where $p$ is hard, it is sufficient to keep only the hard contribution for the coordinate-space two-point function and then Fourier-transform back. For this purpose, one can write
\begin{align}
&\int \frac{d^Dp}{(2\pi)^D}\frac{e^{ip\cdot z}}{p^2+m^2}\bigg|_{\rm hard}=\sum_{l=0}^{\infty}\int\frac{d^Dp}{(2\pi)^D}\frac{(-1)^lm^{2l}e^{ip\cdot z}}{(p^2)^{l+1}} \nonumber \\ 
&=\sum_{l=0}^{\infty}(-1)^lm^{2l}\frac{\Gamma\left(\frac{D}{2}-l-1\right)}{(4\pi)^{\frac{D}{2}}\Gamma(l+1)}\left(\frac{z^2}{4}\right)^{1+l-\frac{d}{2}} \ .
\end{align}
Now, for the angular average of $e^{ik\cdot z}$, one has
\begin{align}
\frac{(-1)^nk^{2n}z^{2n}}{(2n)!}\frac{\int_0^{\pi} d\theta \sin^{D-2}\theta\cos^{2n} \theta}{\int_0^{\pi} d\theta \sin^{D-2}\theta}=\frac{\left(-\frac{1}{4}\right)^n \Gamma \left(\frac{D}{2}\right) k^{2 n}z^{2n}}{\Gamma (n+1) \Gamma \left(\frac{D}{2}+n\right)} \ .
\end{align}
Given the above, by Fourier transforming back from the coordinate space to the momentum space again, one obtains after angular average over $k$
\begin{align}
&\frac{1}{(p-k)^2+m^2}=\sum_{n=0}^{\infty}\frac{k^{2n}}{(p^2)^{n+1}}{\cal C}_n \left(\frac{m^2}{p^2}\right) \ , \\ 
&{\cal C}_n\left(\frac{m^2}{p^2}\right)=\sum_{l=0}^{\infty}\left(\frac{m^2}{p^2}\right)^l\frac{\Gamma \left(1-\epsilon\right) (-1)^{l+n} \Gamma \left(-l-\epsilon\right)  (l+n)!}{l! n! \Gamma \left(1-\epsilon+n\right) \Gamma \left(- l-n-\epsilon\right)}\nonumber \\ 
&=\sum_{l=0}^{\infty}\left(\frac{m^2}{p^2}\right)^l(-1)^l\frac{(l+n)!}{l!n!}\frac{\Gamma(1-\epsilon)}{\Gamma(l+1+\epsilon)}\frac{\Gamma(n+l+1+\epsilon)}{\Gamma(n+1-\epsilon)} \nonumber \\ 
&=\frac{\Gamma (1-\epsilon) \Gamma (n+1+\epsilon) }{\Gamma (1+\epsilon) \Gamma (n+1-\epsilon)}\, _2F_1\left(n+1,n+1+\epsilon;1+\epsilon;-\frac{m^2}{p^2}\right) \ . \label{eq:coefmassin}
\end{align}
In the large $n$ limit, $C_n$ is absent of factorial growth. For example, at $p^2=m^2$ and $\epsilon=0$, one has
\begin{align}
{\cal C}_n\bigg|_{p^2=m^2,D=2}=\frac{\Gamma\left(\frac{n}{2}+\frac{3}{2}\right)}{\Gamma(n+2)\Gamma\left(-\frac{1}{2}-\frac{n}{2}\right)} \ .
\end{align}
It vanishes at $n=2k+1$, while for $n=2k$ it decays exponentially as ${\cal O}\left(2^{-n}\right)$. 

\section{Operator condensates in the $\overline{\rm MS}$ scheme}\label{sec:operatorms}
From the Mellin-Barnes, one obtains the operator condensates as high-order derivatives at $t=0$ of analytic functions. For ${\cal O}_{n,s_1\le n}$ given by Eq.~(\ref{eq:operatorfull1}), corresponding to $\langle O_n \rangle_{g^{k\le n+1}}$, the corresponding analytic functions are singular at $t=0$. This is due to the UV divergences of these contributions. From the Mellin-Barnes, such singularities are canceled by the corresponding contributions of the coefficient functions. In fact, UV singularity of $\langle O_n\rangle_{s_1=k}$ is canceled by the ${\cal H}_{0,n-k}\bigg|_{s_1=k}$ contribution (coefficient functions with $n-k$ mass-insertions within the bubble chain at the $g^{k+1}$ order) and their summation must remain scheme independent and agrees with the result in the $\overline{\rm MS}$ scheme at $\mu=m$. In this appendix we show this for the  $\langle O_n\rangle_{s_1=n}$ contributions. The $\langle O\rangle_{n,s_1<n}$ contributions can be analyzed similarly with more subtraction terms.

Before moving to the non-trivial condensates, one must add to the coefficient function of the identity operator, the tree-level condensates attached to their coefficient functions. Since it is the $C_{n,0}$ that correlates with $O_{n,n}$ at logarithmic level, one must add to $C_{n,0}$ those tree-level condensates without derivatives. Moreover, since the $\frac{1}{N}\Phi^2$ inserted along hard propagators should be combined with $m_r^2$ insertions to change the $m_r^2$ to $m^2$, the tree-level condensates associated to $C_{n,0}$ must be of {\it replacing-type}, namely, the $\frac{1}{N}\Phi^2$ insertions replacing one of two lines within a bubble, and replacing the $p-k$ line.  In Sec.~\ref{sec:OPEdiverge}, we have shown that (see Eq.~(\ref{eq:Rbc}))
\begin{align}
\frac{g^{n+1}}{(p^2)^n}R_{b}\hat C_{n,0}=-\frac{g^{n+1}}{(4\pi)^n \epsilon^n}\left(\frac{\mu^2 e^{\gamma_E}}{4\pi}\right)^{\epsilon}\int \frac{d^dk}{(2\pi)^d}\bigg(f(\epsilon)\left(\frac{\mu^2}{k^2}\right)^{\epsilon}-1\bigg)^n\frac{1}{(k^2)^n(p-k)^2} \ .
\end{align}
We would like to add to this the {\it associated } tree-level condensates that are powers in $I_r=I_0-\frac{1}{4\pi \epsilon}$
\begin{align}
&\frac{1}{(p^2)^n}R_b\mathfrak{H}_{n,0}=-\frac{1}{(4\pi)^n \epsilon^n}\left(\frac{\mu^2 e^{\gamma_E}}{4\pi}\right)^{\epsilon}\int \frac{d^dk}{(2\pi)^d}\bigg(f(\epsilon)\left(\frac{\mu^2}{k^2}\right)^{\epsilon}-1-4\pi \epsilon I_r\bigg)^n\frac{1}{(k^2)^n(p-k)^2}\nonumber \\ 
&=-\frac{1}{(4\pi)^n\epsilon^n}\left(\frac{\mu^2 e^{\gamma_E}}{4\pi}\right)^{\epsilon}\sum_{i=0}^n\binom{n}{i}(-1)^i(4\pi \epsilon I_r)^i \int \frac{d^dk}{(2\pi)^d}\bigg(f(\epsilon)\left(\frac{\mu^2}{k^2}\right)^{\epsilon}-1\bigg)^{n-i}\frac{1}{(k^2)^n(p-k)^2}  \ . 
\end{align}
Here, the term ``associated'' means the following:
when we obtain the bubble-chain-renormalized coefficient function $R_b\hat C_{n,0}$, we have added only the pole part of the $\langle \frac{1}{N}\Phi^2\rangle_0=I_0=I_r+\frac{1}{4\pi \epsilon}$ insertions (of replacing type) within the bubble chain to renormalize the IR divergences of momentum integrals. Then, if we also include the finite part $I_r$ of these operator insertions, one obtains all the tree-level condensates (including the identity) inserted within the bubble chain {\it associated} to $C_{n,0}$.  Similarly, to obtain the subtraction $T_{p-k}R_b \hat C_{n,0}$ Eq.~(\ref{eq:contertermpminusk}) on the $p-k$ line, we only included the pole part of the $I_0$ insertion replacing the $p-k$ line, as well as the pole parts of the $I_0$ insertions within the bubble chain. The finite parts can all be added back and one obtains
\begin{align}
T_{p-k}R_b\mathfrak{H}_{n,0}=-\left(\frac{1}{4\pi \epsilon}+I_r\right)\sum_{i=0}^{n}(-1)^i\binom{n}{i}I_r^i \frac{1}{(4\pi\epsilon)^{n-i}}\bigg(\left(\frac{\mu^2}{p^2}\right)^{\epsilon}-1\bigg)^{n-i} \ .
\end{align}
The crucial combination is clearly
\begin{align}
R_p\mathfrak{\tilde H}_{n}(i)=&-\left(\frac{\mu^2 e^{\gamma_E}}{4\pi}\right)^{\epsilon}\frac{1}{(4\pi \epsilon)^{n-i}}\int \frac{d^dk}{(2\pi)^d}\bigg(f(\epsilon)\left(\frac{\mu^2}{k^2}\right)^{\epsilon}-1\bigg)^{n-i}\frac{(p^2)^n}{(k^2)^n(p-k)^2} \nonumber \\ 
&-\left(\frac{1}{4\pi \epsilon}+I_r\right)\frac{1}{(4\pi\epsilon)^{n-i}}\bigg(\left(\frac{\mu^2}{p^2}\right)^{\epsilon}-1\bigg)^{n-i} \ .
\end{align}
In terms of the above, the partly renormalized $R_p\hat C_{n,0}$ plus all its associated tree-level contributions attached to their partly renormalized coefficient functions can be written as
\begin{align}
&R_p\mathfrak{H}_{n,0}=R_b\mathfrak{H}_{n,0}+T_{p-k}R_b\mathfrak{H}_{n,0}=\nonumber \sum_{i=0}^{n}(-1)^i\binom{n}{i}I_r^iR_p\mathfrak{\tilde H}(i) \\ 
&=\sum_{i=0}^{n}I_r^iR_p\mathfrak{h}_n(i)-(-1)^nI_r^{n+1} +{\cal O}(\epsilon)\ .
\end{align}
Notice that the $R_p\mathfrak{h}_n(0)=R_p\hat C_{n,0}$ is just the partly renormalized coefficient function for the identity operator, while the $R_p\mathfrak{h}_n(i)$ can all be regarded as the partly renormalized coefficient functions for the renormalized tree-level condensates $I_r^i=\langle\left(\frac{1}{N}\Phi^2\right)^n\rangle_r$ (at LO in $\frac{1}{N}$). The computation of $R_p\mathfrak{h}_n(i)$ is similar to that of $R_bC_{n,0}$ and the results read
\begin{align}
R_p\mathfrak{h}_n(i)=&R_p\mathfrak{h}^A_n(i)+R_p\mathfrak{h}^B_n(i) \ , \\ 
R_p\mathfrak{h}^A_n(i)=&-\binom{n}{i}\frac{\Gamma(1-\epsilon)\Gamma(n+\epsilon)}{\Gamma(1+\epsilon)\Gamma(n-\epsilon)}\bigg(\frac{1}{(4\pi \epsilon)^{n+1-i}}\frac{2 (-1)^n \sin (\pi  \epsilon) \Gamma (-2 \epsilon)}{(n+1-i)\pi \Gamma^2 (1-\epsilon)}\bigg) \ , \\ 
R_p\mathfrak{h}^B_n(i)=&(-1)^i\binom{n}{i-1}\frac{1}{(4\pi)^{n-i+1}}\ln ^{n-i+1}\frac{\mu^2}{p^2}\nonumber \\ 
+(-1)^i\binom{n}{i}&\frac{(n-i)!}{(4\pi)^{n+1}}\oint\frac{ds (4\pi s)^i}{2\pi i s^{n+1}}e^{-sl_{\mu}}\bigg(2\gamma_E-l_{\mu}+\psi(1-n-s)+\psi(n+s)\bigg) \ .
\end{align}
Notice that $l_{\mu}=\ln \frac{p^2}{\mu^2}$. 
Given the above, we can write 
\begin{align}
&\sum_{i=0}^nI_r^iR_p\mathfrak{h}^A_n(i)=-\frac{(-1)^n}{(4\pi)^{n+1}(n+1)}(4\pi  I_r)^{n+1} \nonumber \\ 
&-(1+4\pi \epsilon I_r)^{n+1}\frac{1}{(4\pi)^{n+1}}\frac{\Gamma(1-\epsilon)\Gamma(n+\epsilon)}{\Gamma(1+\epsilon)\Gamma(n-\epsilon)}\bigg(\frac{1}{\epsilon^{n+1}}\frac{2 (-1)^{n}  \sin (\pi  \epsilon) \Gamma (-2 \epsilon)}{(n+1)\pi \Gamma^2 (1-\epsilon)}\bigg) +{\cal O}(\epsilon) \ ,
\end{align}
where we have used the binomial theorem. 
On the other hand, the $B$ part is
\begin{align}
&\sum_{i=0}^nI_r^iR_p\mathfrak{h}^B_n(i)-(-1)^nI_r^{n+1}=-\frac{1}{(4\pi)^{n}}I_r \ln^n\frac{m^2}{p^2}\nonumber \\ 
&+\sum_{i=1}^n\frac{n!}{(4\pi)^{n+1} i!}\oint\frac{ds (-4\pi sI_r)^ie^{-sl_{\mu}}}{2\pi i s^{n+1}}\bigg(2\gamma_E-l_{\mu}+\psi(1-n-s)+\psi(n+s)\bigg) \ .
\end{align}
Notice that the first term is due to the $I_r$ insertion on the $p-k$ line. We write
\begin{align}
\sum_{i=1}^n\frac{1}{i!}(-4\pi I_r s)^i=\bigg(e^{-4\pi I_rs}-\frac{(-1)^{n+1}}{(n+1)!}(4\pi I_r)^{n+1} s^{n+1}+{\cal O}(s^{n+2})\bigg) \ .
\end{align}
Now, the ${\cal O}(s^{n+2)}$ term drops due to the residue theorem, while the $s^{n+1}$ term will only pick up the $s=0$ pole of $\psi(1-n-s)$. Thus we have
\begin{align}
&\sum_{i=0}^nI_r^iR_p\mathfrak{h}^B_n(i)-(-1)^nI_r^{n+1}=-\frac{1}{(4\pi)^{n}}I_r \ln^n\frac{m^2}{p^2} +\frac{(-1)^n}{(n+1)(4\pi)^{n+1}}(4\pi I_r)^{n+1} \nonumber \\ 
&+\frac{n!}{(4\pi)^{n+1}}\oint\frac{ds }{2\pi i s^{n+1}}e^{-sl_{m}}\bigg(2\gamma_E-l_{\mu}+\psi(1-n-s)+\psi(n+s)\bigg) +{\cal O}(\epsilon) 
&\nonumber \\ 
&\equiv\frac{(-1)^n}{(n+1)(4\pi)^{n+1}}(4\pi I_r)^{n+1}\nonumber \\ 
&+\frac{n!}{(4\pi)^{n+1}}\oint\frac{ds }{2\pi i s^{n+1}}e^{-sl_{m}}\bigg(2\gamma_E-l_{m}+\psi(1-n-s)+\psi(n+s)\bigg) +{\cal O}(\epsilon)  \ .
\end{align}
Given the above, one finally has
\begin{align}
&R_p\mathfrak{H}_{n,0}={\color{red}-(1+4\pi \epsilon I_r)^{n+1}\frac{1}{(4\pi)^{n+1}}\frac{\Gamma(1-\epsilon)\Gamma(n+\epsilon)}{\Gamma(1+\epsilon)\Gamma(n-\epsilon)}\bigg(\frac{1}{\epsilon^{n+1}}\frac{2 (-1)^{n}  \sin (\pi  \epsilon) \Gamma (-2 \epsilon)}{(n+1)\pi \Gamma^2 (1-\epsilon)}\bigg)} \nonumber \\ 
&{\color{blue}+\frac{n!}{(4\pi)^{n+1}}\oint\frac{ds }{2\pi i s^{n+1}}e^{-sl_{m}}\bigg(2\gamma_E-l_{m}+\psi(1-n-s)+\psi(n+s)\bigg) }+{\cal O}(\epsilon) \ . \label{eq:cnall}
\end{align}
Notice that the second line (shown in blue) is just the hard coefficient ${\cal H}_{0,0}\bigg|_{s_1=n}$ given by Eq.~(\ref{eq:hard00mellin}) obtained from the Mellin-Barnes. As expected, it is exactly the $C_{n,0}^B$ given by Eq.~(\ref{eq:CBn}) at $\mu^2=m^2$. On the other hand, the first line (shown in red) are the scheme dependent contributions. Below we show that this scheme dependent part cancels exactly with the bare operator contributions 
\begin{align}
O_{n,n}(\epsilon)=(-1)^{n+1}\left(\frac{\mu^2e^{\gamma_E}}{4\pi}\right)^{\epsilon}\int \frac{d^dk}{(2\pi)^d}(k^2)^{n-1}F^n(k^2,\epsilon) \ ,
\end{align}
namely, the sum 
\begin{align}\label{eq:invariantsum}
\frac{g^{n+1}}{(p^2)^n}\bigg(R_p\mathfrak{H}_{n,0}+\frac{\Gamma(1-\epsilon)\Gamma(n+\epsilon)}{\Gamma(1+\epsilon)\Gamma(n-\epsilon)}O_{n,n}(\epsilon)\bigg)\rightarrow {\cal H}_{0,0}(p^2)\bigg|_{s_1=n}+\frac{(m^2)^n}{(p^2)^n}{\cal O}_{n,s_1=n} \ , 
\end{align}
remains the same as the result directly obtained from the Mellin-Barnes, where ${\cal O}_{n,s_1=n}$ is given by Eq.~(\ref{eq:operatorfull1}). 
Notice that the factor $\frac{\Gamma(1-\epsilon)\Gamma(n+\epsilon)}{\Gamma(1+\epsilon)\Gamma(n-\epsilon)}$ is due to the coefficient function of $O_n$ in the $2-2\epsilon$ dimension. The analysis of the condensate is not entirely trivial, since $F(k^2,\epsilon)$ can only be expressed in terms of a hyperglycemic function. But the technique in Ref.~\cite{Marino:2021six} made this possible. 

For simplicity, from now on we set $m^2=1$, as there are no issue to restore the $m^2$ dependency. The first step of the analysis is to obtain the explicit representation of the function $F(k^2,\epsilon)$~\cite{Marino:2021six}
\begin{align}
F(k^2,\epsilon)=&\left(\frac{\mu^2e^{\gamma_E}}{4\pi}\right)^{\epsilon} \frac{1}{8\pi (4\pi)^{-\epsilon}}\int_0^1 dx\int_0^{\infty}\rho^{\epsilon}d\rho e^{-\rho(x(1-x)k^2+1)} \nonumber \\ 
&=(\mu^2e^{\gamma_E})^{\epsilon}\frac{\Gamma(1+\epsilon)}{8\pi}\left(\frac{k^2+4}{4}\right)^{-1-\epsilon}\, _2F_1\left(1+\epsilon,\frac{1}{2};\frac{3}{2};\frac{k^2}{k^2+4}\right) \ .
\end{align}
Given the above, one has
\begin{align}
O_{n,n}(\epsilon)=&\frac{(-1)^{n+1}}{(4\pi)^{n+1}}(\mu^2e^{\gamma_E})^{(n+1)\epsilon}\frac{\Gamma^n(1+\epsilon)}{2^n\Gamma(1-\epsilon)}\nonumber \\ 
&\times\int_0^{\infty} dx x^{n-1-\epsilon}\left(\frac{x+4}{4}\right)^{-(1+\epsilon)n}\, _2F_1^n\left(1+\epsilon,\frac{1}{2};\frac{3}{2};\frac{x}{x+4}\right) 
\end{align}
We further use the formula for the bare tree-level condensate $\frac{1}{N}\langle \Phi^2_0\rangle$
\begin{align}
I_0=\frac{(\mu^2e^{\gamma_E})^{\epsilon}}{(4\pi)}\frac{\Gamma(1+\epsilon)}{\epsilon}=I_r+\frac{1}{4\pi \epsilon} \ ,
\end{align}
to write
\begin{align}
O_{n,n}(\epsilon)=\frac{(-1)^{n+1}}{(4\pi)^{n+1}}(1+4\pi \epsilon I_r)^{n+1}\frac{1}{\Gamma(1+\epsilon)\Gamma(1-\epsilon)}\frac{1}{2^n}{\cal I}_{n,1} \ .
\end{align}
Here the crucial integral reads
\begin{align}
{\cal I}_{n,1}=4^{n-\epsilon}\int_0^1 dz z^{n-1-\epsilon}(1-z)^{-1+(n+1)\epsilon}\bigg[\, _2F_1\left(1+\epsilon,\frac{1}{2};\frac{3}{2};z\right)\bigg]^n \ .
\end{align}
This integral appeared earlier  in~\cite{Marino:2021six} and the notation ${\cal I}_{n,1}$ is also adopted from it. This integral can be analyzed in the following way, which is essentially how it was calculated in~\cite{Marino:2021six}. First, notice that there are $z\rightarrow 1$ UV singularities at $\epsilon=0$. Thus one needs the $z\rightarrow 1^-$ expansion of the hypergeometric function~\cite{Marino:2021six} 
\begin{align}
&f(\epsilon,z)=_2F_1\left(1+\epsilon,\frac{1}{2};\frac{3}{2};z\right) \nonumber \\ 
&=-\frac{1}{2\epsilon}\frac{\Gamma^2(1-\epsilon)}{4^{\epsilon}\Gamma(1-2\epsilon)} \bigg(1-\frac{4^{\epsilon}\Gamma(1-2\epsilon)}{\Gamma^2(1-\epsilon)}(1-z)^{-\epsilon}\bigg)+{\cal O}((1-z)^{1-\epsilon}) \ .
\end{align}
Thus, if we introduce the tail function
\begin{align}
f_{t}(\epsilon,z)=-\frac{1}{2\epsilon}\frac{\Gamma^2(1-\epsilon)}{4^{\epsilon}\Gamma(1-2\epsilon)} \bigg(1-\frac{4^{\epsilon}\Gamma(1-2\epsilon)}{\Gamma^2(1-\epsilon)}(1-z)^{-\epsilon}\bigg) \ ,
\end{align}
then the following difference has a finite $\epsilon\rightarrow0$ limit
\begin{align}
&\hat {\cal I}_{n,1}=4^{n-\epsilon}\int_0^1 dz z^{n-1-\epsilon}(1-z)^{-1+(n+1)\epsilon} \bigg(f(\epsilon,z)^n-f_t(\epsilon,z)^n\bigg) \nonumber \\ 
& \rightarrow 4^n\int_0^1 dz z^{n-1}(1-z)^{-1}\bigg(f(0,z)^n-f_t(0,z)^n\bigg) \ .
\end{align}
This is because the difference $\bigg|f(\epsilon,z)-f_t(\epsilon,z)\bigg|$ can be bound by $A(\epsilon_0)(1-z)^{1-\epsilon_0}$ for $\epsilon \in (0,\epsilon_0)$ and $z\in(0.5,1)$, thus the $\epsilon \rightarrow0$ limit can be taken within the integral. Given the above, one has
\begin{align}
\frac{1}{2^n}{\cal I}_{n,1}=\frac{1}{2^n}\hat{\cal I}_{n,1}+\frac{(-1)^n4^{-\epsilon}}{\epsilon^n}\int_0^1 dz z^{n-1-\epsilon}\bar z^{-1+(n+1)\epsilon}\bigg(\frac{\Gamma^2(1-\epsilon)}{4^{\epsilon}\Gamma(1-2\epsilon)} -\bar z^{-\epsilon}\bigg)^n \ .
\end{align}
Here $\bar z=1-z$. Thus, after introducing the function
\begin{align}
\mathfrak{O}(n,\epsilon,s)=4^{-\epsilon}\bigg(\frac{\Gamma^2(1-\epsilon)}{4^{\epsilon}\Gamma(1-2\epsilon)}\bigg)^{\frac{s}{\epsilon}-1}\frac{\Gamma(n-\epsilon)\Gamma(s+1)}{\Gamma(n-\epsilon+s)} \ ,
\end{align}
by expanding the bracket using the binomial theorem one has
\begin{align}
&\frac{(-1)^n4^{-\epsilon}}{\epsilon^n}\int_0^1 dz z^{n-1-\epsilon}\bar z^{-1+(n+1)\epsilon}\bigg(\frac{\Gamma^2(1-\epsilon)}{4^{\epsilon}\Gamma(1-2\epsilon)} -\bar z^{-\epsilon}\bigg)^n \nonumber \\ 
&=(-1)^n\sum_{k=0}^n\binom{n}{k}(-1)^k\frac{\mathfrak{O}(n,\epsilon,(n+1-k)\epsilon)}{(n+1-k)\epsilon^{n+1}} \nonumber \\ 
&=(-1)^n \bigg(\frac{(-1)^n}{(n+1)\epsilon^{n+1}}\mathfrak{O}(n,\epsilon,0)+\frac{1}{n+1}\frac{d^{n+1}}{ds^{n+1}}\mathfrak{O}(n,0,s)\bigg|_{s=0}\bigg)+{\cal O}(\epsilon) \ ,
\end{align}
where we have used again the standard summation formulas.
Thus, one can write
\begin{align}
&O_{n,n}(\epsilon)=\frac{(-1)^{n+1}}{(4\pi)^{n+1}}(1+4\pi \epsilon I_r)^{n+1}\frac{1}{\Gamma(1+\epsilon)\Gamma(1-\epsilon)}\frac{1}{2^n}{\cal I}_{n,1}  \nonumber \ \\
&\rightarrow{\color{red}\frac{(-1)^{n+1}}{(4\pi)^{n+1}}(1+4\pi\epsilon I_r)^{n+1}\frac{1}{\Gamma(1+\epsilon)\Gamma(1-\epsilon)}\frac{1}{(n+1)\epsilon^{n+1}}\mathfrak{O}(n,\epsilon,0) }\nonumber \\ 
&{\color{blue}-\frac{1}{(4\pi)^{n+1}}\frac{1}{n+1}\frac{d^{n+1}}{ds^{n+1}}\mathfrak{O}(n,0,s)\bigg|_{s=0} +\frac{(-1)^{n+1}}{(4\pi)^{n+1}}\frac{1}{2^n}\hat {\cal I}_{n,1} }+{\cal O}(\epsilon)\nonumber \\ 
&\equiv {\color{red} O^A_{n,n}(\epsilon)}+{\color{blue}O^B_{n,n}}+{\cal O}(\epsilon) \ .
\end{align}
In the above, the divergent part reads
\begin{align}
&O_{n,n}^A(\epsilon)=\frac{(-1)^{n+1}}{(4\pi)^{n+1}}(1+4\pi\epsilon I_r)^{n+1}\frac{1}{\Gamma(1+\epsilon)\Gamma(1-\epsilon)}\frac{1}{(n+1)\epsilon^{n+1}}\mathfrak{O}(n,\epsilon,0)\nonumber \\ 
&=-\frac{(1+4\pi\epsilon I_r)^{n+1}}{(4\pi)^{n+1}}\frac{(-1)^n}{(n+1)\epsilon^{n+1}}\frac{\Gamma(1-2\epsilon)}{\Gamma^3(1-\epsilon)\Gamma(1+\epsilon)} \ .
\end{align}
After multiplying the coefficient function $\frac{\Gamma(1-\epsilon)\Gamma(n+\epsilon)}{\Gamma(1+\epsilon)\Gamma(n-\epsilon)}$, it exactly cancels the divergent contribution shown in red in Eq.~(\ref{eq:cnall}) 
\begin{align}
&-\frac{(1+4\pi\epsilon I_r)^{n+1}}{(4\pi)^{n+1}}\frac{\Gamma(1-\epsilon)\Gamma(n+\epsilon)}{\Gamma(1+\epsilon)\Gamma(n-\epsilon)}\bigg(\frac{1}{\epsilon^{n+1}}\frac{2 (-1)^n  \sin (\pi  \epsilon) \Gamma (-2 \epsilon)}{(n+1)\pi \Gamma^2 (1-\epsilon)}\bigg) \nonumber \\
&+\frac{\Gamma(1-\epsilon)\Gamma(n+\epsilon)}{\Gamma(1+\epsilon)\Gamma(n-\epsilon)}O^A_{n,n}(\epsilon) \equiv 0 \ .
\end{align}
On the other hand, the convergent part can be written as
\begin{align}
O_{n,n}^B=\frac{(-1)^{n+1}}{(4\pi)^{n+1}}\frac{1}{2^n}\hat {\cal I}_{n,1}-\frac{1}{(4\pi)^{n+1}}\frac{1}{n+1}\frac{d^{n+1}}{ds^{n+1}}\bigg(\frac{4^{-s}\Gamma(n)\Gamma(1+s)}{\Gamma(n+s)}\bigg)_{s=0} \ .
\end{align}
Here we show that in the $\epsilon \rightarrow 0$ limit, this is exactly the $\frac{(m^2)^n}{g^{n+1}}{\cal O}_{n,s_1=n}$ contribution obtained from the Mellin Barnes.  This is manageable since we can use the $\epsilon\rightarrow0$ limits
\begin{align}
f(0,z)=\frac{1}{2\sqrt{z}}\ln \frac{1+\sqrt{z}}{1-\sqrt{z}} \ ,  \ 
f_t(0,z)=\frac{1}{2}\ln \frac{4}{1-z} \ .
\end{align}
Given the above, one has
\begin{align}
&\frac{(-1)^{n+1}}{2^n}\hat {\cal I}_{n,1}\bigg|_{\epsilon\rightarrow0}=\frac{d^ng(t)}{dt^n}\big|_{t=0} \ , \\
&g(t)=\int_0^1 dz z^{n-1}\bar z^{-1}\bigg(\left(\frac{1-z}{4}\right)^t-z^{-\frac{n}{2}}\left(\frac{1-\sqrt{z}}{1+\sqrt{z}}\right)^t\bigg) \nonumber \\ 
&=\frac{4^{-t}\Gamma(n)\Gamma(t)}{\Gamma(n+t)}-\frac{\Gamma (n) \Gamma (t) \, _2{F}_1(t,n-1;n+t;-1)}{\Gamma(n+t)} \ .
\end{align}
The result should be compare with Eq.~(B.2) of~\cite{Marino:2021six}, in which a different subtraction function $\frac{1}{z^{\frac{1}{2}}}f_t(\epsilon,z)$ has been used. Given the above, one finally has
\begin{align}
O_{n,n}^B=-\frac{1}{(4\pi)^{n+1}}\frac{d^n}{dt^n}\bigg[\frac{\Gamma (n) \Gamma (t) \, _2{F}_1(t,n-1;n+t;-1)}{\Gamma(n+t)}\bigg]_{t=0} \ ,
\end{align}
where the $n$-th derivative is defined as $n!$ times the coefficient of $t^n$. 
The above completely agrees with the $\frac{(m^2)^n}{g^{n+1}}{\cal O}_{n,s_1=n}$ given by Eq.~(\ref{eq:operatorfull1}). The crucial identity Eq.~(\ref{eq:invariantsum}) is therefore proven.

Notice that in the above, the operator condensate $O_{n,n}(\epsilon)$ remains unrenormalized. Similarly, the coefficient function $R_pC_{n,0}=R_p\mathfrak{h}_n(0)$ and $R_p\mathfrak{h}_n(i)$ for the tree-level operators are {\it partly renormalized}.  To obtain the fully renormalized OPE in the $\overline{\rm MS}$ scheme, one can simply subtract out all the $\frac{1}{\epsilon}$ poles at any power in $I_r$ for $O_{n,n}(\epsilon)$
\begin{align}
&O_{n,n}(\mu)={\color{red}-\oint\frac{d\epsilon}{2\pi i}\frac{(1+4\pi\epsilon I_r)^{n+1}}{(4\pi)^{n+1}}\frac{(-1)^n}{(n+1)\epsilon^{n+2}}\frac{\Gamma(1-2\epsilon)}{\Gamma^3(1-\epsilon)\Gamma(1+\epsilon)}} \nonumber \\ 
&-{\color{blue}\frac{n!}{(4\pi)^{n+1}}\oint\frac{ds }{2\pi i s^{n+1}}\bigg(\frac{\Gamma (n) \Gamma (s) \, _2{F}_1(s,n-1;n+s;-1)}{\Gamma(n+s)}\bigg) }\ .
\end{align} 
Notice that in the formula above, $I_r$ should be regarded as $\epsilon$-independent. 
Similarly, the tree-level condensates including the identity associated to $C_{n,0}$ attached to their renormalized coefficient functions sum to
\begin{align}
\mathfrak{H}_{n,0}(\mu,p^2)=& {\color{blue}\frac{n!}{(4\pi)^{n+1}}\oint\frac{ds }{2\pi i s^{n+1}}e^{-sl_{m}}\bigg(2\gamma_E-l_{m}+\psi(1-n-s)+\psi(n+s)\bigg)}\\ 
&+{\color{red}\oint\frac{d\epsilon}{2\pi i}\frac{(1+4\pi\epsilon I_r)^{n+1}}{(4\pi)^{n+1}}\frac{(-1)^n}{(n+1)\epsilon^{n+2}}\frac{\Gamma(1-2\epsilon)}{\Gamma^3(1-\epsilon)\Gamma(1+\epsilon)}} \ .
\end{align}
Notice that the largest logarithm in $O_{n,n}(\mu)$ is $\frac{(-1)^{n+1}}{(4\pi)^{n+1}(n+1)}\ln^{n+1}\frac{\mu^2}{m^2}$ and this term is generated due to $\frac{1}{\epsilon^{n+1}}\times \epsilon^{n+1}$ effects. Also notice that the red parts in the large $n$ limit are free from any factorial growths, and the factorial enhancements are entirely due to the ``genuine'' contributions shown in blue.

We should also mention that the divergent contributions $O_{n,k<n}$ can also be analyzed using similar methods with more subtraction terms. The results in~\cite{Marino:2021six} for 
${\cal I}_{n,0}$ can be used for $O_{n,s_1=n-1}$. The computation would be similar to the ones given for the $C_{n;0,1}$, as among the $n$ factors of $f(\epsilon,z)^n$, only one has to be subtracted to the second order and resembles the distinguished bubble with the mass insertion. On the coefficient function side, with mass insertions the $C_{n;l,k}$ contributions can also be combined with their associated tree-level condensates. For example, the $C_{n-1;0,1}$ with one mass insertion within the bubble-chain has to be combined with tree-level insertions such that one $\frac{1}{N}\Phi(-\partial^2) \Phi$ of replacing type is inserted within the bubble chain, one $\frac{1}{N}\Phi^2$ of non-replacing-type is inserted together with the mass insertion, while the rest are replacing-type $\frac{1}{N}\Phi^2$ insertions like the $C_{n,0}$. In fact, we even used the UV pole $a_1m_r^2$ for $\frac{1}{N}\Phi(-\partial^2) \Phi$ in Eq.~(\ref{eq:Cm}) for IR subtraction. To obtain the full result, one should first add the $a_2gI_r$ pole to the $a_1m_r^2$ pole, and add one non-replacing $\frac{1}{N}\Phi^2$ insertion at the same place with the $m_r^2$ insertion. After this two steps, the overall $m_r^2$ has been changed to $m^2$. Then, one add the finite terms for the $\frac{1}{N}\Phi(-\partial^2) \Phi$ insertion and the remaining $I_0$ insertions as before. The fact that the tree-level condensates should be combined together in a natural manner with the identity operator is the major reason why from the Mellin-Barnes one only sees the non-trivial condensates. 

\section{The four-quark condensate in the large-$N$ Gross-Neveu}
The analysis of operator condensates in the previous appendix can be performed in the 2D large-$N$ gross Neveu as well. Here we consider the connected four-quark condensate at NLO in the large-$N$ expansion, or equivalently, the $\langle \sigma^2\rangle_c=\langle \sigma^2\rangle-\langle\sigma\rangle^2$ condensate at LO. This condensate appears in the large-$p^2$ expansion of the self-energy at the next-to-leading power~\cite{Marino:2024uco}. Moreover, it contains an UV renormalon at $t=1$ which cancels the $t=1$ IR renormalon in the coefficient function of the identity operator. Here we demonstrate how the $\overline{\rm MS}$ scheme analysis of this operator-condensate can be performed, at least on a perturbative basis. We use the $m^2=1$ convention, where $m$ is the leading-order saddle-point mass. 

The integral we would like to analyze is similar to that of the $O_{n,n}$, but with the hypergeometric function in the {\it denominator}. The bare sigma condensate at the NLO in $\frac{1}{N}$ reads
\begin{align}
&2N\langle \sigma^2\rangle_c=\int\frac{d^dk}{(2\pi)^d}\frac{1}{(k^2+4)F(k^2,\epsilon)} \\ 
&=\frac{1}{(4\pi)^{1-\epsilon}\Gamma(1-\epsilon)}\int_0^1 dz\left(\frac{4z}{1-z}\right)^{-\epsilon}\frac{4}{(1-z)^2}\frac{2\pi(4\pi)^{-\epsilon}(1-z)^{-\epsilon}}{\Gamma(1+\epsilon)\,_2F_1\left(1+\epsilon,\frac{1}{2},\frac{3}{2},z\right)} \nonumber \\ 
&=\frac{1}{2\Gamma(1+\epsilon)\Gamma(1-\epsilon)}4^{1-\epsilon}\int_0^1 dz\frac{z^{-\epsilon}}{(1-z)^2\,_2F_1\left(1+\epsilon,\frac{1}{2},\frac{3}{2},z\right)} \ .
\end{align}
Here we introduced the variable $z=\frac{k^2}{k^2+4}$~\cite{Marino:2021six}. At $\epsilon=0$, this integral is badly divergent at $z=1$. Only for $\epsilon>1$, the integral is convergent at $z=1$, but then it diverges at $z=0$. For generic $\epsilon$ it could be defined by splitting the integral in the middle and then analytically continue from different regions. The analytical continuation could also lead to branch cut singularities which is hard to see. 
We would like to study the integral's behavior as $\epsilon \rightarrow 0$. Since in perturbation theory, we expanded the coefficient functions in the $\mu$-dependent running coupling, here we would also like to expand the condensate perturbatively. 

For that purpose, we use the following trick. Notice that in the Gross-Neveu, the bare and the $\overline{\rm MS}$-scheme renormalized couplings at large $N$ read (the $g_0$ here is half of the $g_0N$ in~\cite{Marino:2024uco}, we use this convention merely for convenience)
\begin{align}
\frac{1}{g_0}=\frac{1}{(4\pi)^{1-\epsilon}}\Gamma(\epsilon) \ , \  \frac{1}{g_r}=\frac{1}{4\pi}(\mu^2e^{\gamma_E})^{\epsilon}\Gamma(\epsilon)-\frac{1}{4\pi \epsilon} \ , \ 
g_0=\left(\frac{\mu^2e^{\gamma_E}}{4\pi }\right)^{\epsilon}\frac{g_r}{1+\frac{g_r}{4\pi \epsilon}} \ .
\end{align}
Given the above, $1/2$ times the $\epsilon$-dependent inverse $\sigma$ propagator, as appeared in the effective action obtained by expanding the saddle point solution reads 
\begin{align}
&\frac{1}{g_0}-\frac{1}{(4\pi)^{1-\epsilon}}\Gamma(\epsilon)+\frac{\Gamma(1+\epsilon)}{2\pi(4\pi)^{-\epsilon}}(1-z)^{\epsilon}\,_2F_1\left(1+\epsilon,\frac{1}{2},\frac{3}{2},z\right)\nonumber \\ 
&\frac{(\mu^2e^{\gamma_E})^{-\epsilon}}{(4\pi)^{-\epsilon}}\bigg(\frac{1}{g_r}+\frac{1}{4\pi \epsilon}\bigg)-\frac{1}{(4\pi)^{1-\epsilon}}\Gamma(\epsilon)+\frac{\Gamma(1+\epsilon)}{2\pi(4\pi)^{-\epsilon}}(1-z)^{\epsilon}\,_2F_1\left(1+\epsilon,\frac{1}{2},\frac{3}{2},z\right) \nonumber \\ 
&=\left(\frac{\mu^2 e^{\gamma_E}}{4\pi}\right)^{-\epsilon}(g_r)^{-1}\bigg(1+\frac{g_r}{4\pi} Q(\epsilon,z)\bigg) \ ,
\end{align}
with 
\begin{align}
Q(\epsilon,z)= \frac{1}{\epsilon}-(\mu^2e^{\gamma_E})^{\epsilon}\Gamma(\epsilon)+2(\mu^2e^{\gamma_E})^{\epsilon}\Gamma(1+\epsilon)\bar z^{\epsilon}\,_2F_1\left(1+\epsilon,\frac{1}{2},\frac{3}{2},z\right)  \ .
\end{align}
Thus, the ``$n$-bubble'' contribution in the perturbation theory expanded in $g_r$ is
\begin{align}
2N\langle \sigma^2\rangle_c(n)=\frac{(-1)^ng_r^{n+1}}{(4\pi)^{n+1}\Gamma(1-\epsilon)}(\mu^2e^{\gamma_E})^{\epsilon}4^{1-\epsilon}\int_0^1 dz z^{-\epsilon}(1-z)^{-2+\epsilon} Q(z,\epsilon)^n \ .
\end{align}
The integral can be analyzed in the following manner, which is similar to the one for ${\cal I}_{n,0}$ in~\cite{Marino:2021six} . First, we write the first two tails of $Q(\epsilon,z)$ in the $z\rightarrow 1$ limit
\begin{align}
&Q_0(\epsilon,z)=\frac{1}{\epsilon}-\frac{\pi ^{1/2} (1-z)^{\epsilon}\Gamma(\epsilon)\Gamma(1-\epsilon)}{\Gamma \left(\frac{1}{2}-\epsilon\right)}(\mu^2e^{\gamma_E})^{\epsilon}=\frac{1}{\epsilon}\bigg(1-4^{-\epsilon}\mu^{2\epsilon}(1-z)^{\epsilon}f(\epsilon)\bigg) \  , \\
&Q_1(\epsilon,z)=\frac{1-2\epsilon}{2(1-\epsilon)}\Gamma (\epsilon)(\mu^2e^{\gamma_E})^{\epsilon} -(1-z)^{\epsilon}\frac{\pi ^{1/2} \Gamma(\epsilon)\Gamma(1-\epsilon)}{2 \Gamma \left(\frac{1}{2}-\epsilon\right)}(\mu^2e^{\gamma_E})^{\epsilon} \ .
\end{align}
Notice $f(\epsilon)$ is defined in Eq.~(\ref{eq:fe}). Also notice that the $-(\mu^2e^{\gamma_E})^{\epsilon}\Gamma(\epsilon)$ contribution is exactly canceled by one of the two terms after expanding the hypergeometric function. Then, the following subtracted $n$-bubble integrals are finite in the $\epsilon \rightarrow 0$ limit
\begin{align}
&\lim_{\epsilon\rightarrow 0}4^{-\epsilon}\int_0^1 dz z^{-\epsilon}(1-z)^{-2+\epsilon} \bigg(Q(z,\epsilon)^n-Q_0(z,\epsilon)^n-n(1-z)Q_0^{n-1}(z,\epsilon)Q_1(z,\epsilon)\bigg) \nonumber \\ 
&=\int_0^1 dz (1-z)^{-2} \bigg(Q(z,0)^n-Q_0(z,0)^n-n(1-z)Q_0^{n-1}(z,0)Q_1(z,0)\bigg) \ \nonumber \\ 
& \equiv \hat {\cal I}(n).
\end{align}
On the other hand, the contributions that are divergent order by order in the $g_r$ expansion, are contained in the integrals of the tails $Q_0^n$ and $n(1-z)Q_1Q_0^{n-1}$, which are simpler to evaluate at finite $\epsilon$. The total result for the sigma condensate at $n$-bubble level then reads
\begin{align}
& 2N\langle \sigma^2\rangle_c(n)\nonumber \\ 
&=4\left(\frac{g_r}{4\pi}\right)^{n+1}\frac{1}{\Gamma(1-\epsilon)} \bigg({\cal I}_A(n)+{\cal I}_B(n)+{\cal I}_C(n)\bigg)+4\left(\frac{g_r}{4\pi}\right)^{n+1}(-1)^n\hat {\cal I}(n) \ .\\ 
&{\cal I}_A(n)=4^{-\epsilon}  \frac{(\mu^2e^{\gamma_E})^{\epsilon}}{\epsilon^n}\int_0^1 dz z^{-\epsilon}\bar z^{-2+\epsilon}\bigg(\frac{\mu^{2\epsilon}}{4^{\epsilon}}f(\epsilon)\bar z^{\epsilon}-1\bigg)^n \ , \\
&{\cal I}_B(n)=-n4^{-\epsilon} \frac{d(\epsilon)(\mu^2e^{\gamma_E})^{\epsilon}}{2\epsilon^{n-1}}\int_0^1 dz z^{-\epsilon}\bar z^{-1+\epsilon}\bigg(\frac{\mu^{2\epsilon}}{4^{\epsilon}}f(\epsilon)\bar z^{\epsilon}-1\bigg)^{n-1} \ , \\
&{\cal I}_C(n)=n4^{-\epsilon} \frac{(\mu^2e^{\gamma_E})^{\epsilon}}{2\epsilon^n}\int_0^1 dz  z^{-\epsilon}\bar z^{-1+\epsilon}\bigg(\frac{\mu^{2\epsilon}}{4^{\epsilon}}f(\epsilon)\bar z^{\epsilon}-1\bigg)^{n} \ , \\
& d(\epsilon)=\frac{1}{\epsilon}\bigg(\frac{1-2\epsilon}{(1-\epsilon)}\Gamma(1+\epsilon)(\mu^2e^{\gamma_E})^{\epsilon}-1\bigg)\ .
\end{align}
The tail contributions ${\cal I}_{A, \ B, \ C}$ can then be evaluated using the standard summation formulas. For example, the ${\cal I}_A$ can be evaluated as
\begin{align}
&{\cal I}_A(n)=4^{-\epsilon} \frac{(\mu^2e^{\gamma_E})^{\epsilon}}{\epsilon^n}\sum_{k=0}^n(-1)^k\binom{n}{k}\left(\frac{\mu^{2\epsilon}f(\epsilon)}{4^{\epsilon}}\right)^{\frac{(n+1-k)\epsilon}{\epsilon}-1}\int_0^1 dz z^{-\epsilon}\bar z^{-2+(n+1-k)\epsilon} \nonumber \\ 
&=\frac{(-1)^n\epsilon e^{\gamma_E \epsilon}}{(n+1)\epsilon^{n+1}f(\epsilon)}+\frac{1}{n+1}\frac{d^{n+1}}{ds^{n+1}}\bigg(\frac{1}{4^s}\frac{s}{s-1}(\mu^2)^s\bigg)_{s=0}+{\cal O}(\epsilon) \ .
\end{align}
As expected, since this part is due to the power-divergence subtraction, it contains an UV renormalon at $s=1$, but no other singularities.  Similarly, one has
\begin{align}
&{\cal I}_C(n)=\frac{e^{\gamma_E \epsilon}n(-1)^n}{2(n+1)f(\epsilon)\epsilon^{n+1}}+\frac{n}{2(n+1)}\frac{d^{n+1}}{ds^{n+1}}\bigg(\frac{1}{4^s}(\mu^2)^s\bigg)_{s=0}+{\cal O}(\epsilon) \ , \\
&{\cal I}_B(n)=-\frac{n(-1)^{n-1}e^{\gamma_E \epsilon}}{2nf(\epsilon)\epsilon^{n}}d(\epsilon)-\frac{n}{2n}\frac{d^{n}}{ds^{n}}\bigg(\frac{1}{4^s}(\mu^2)^s\bigg)_{s=0}d(0)+{\cal O}(\epsilon) \ .
\end{align}
Notice that the above are correct at $n=0$ as well. The above combined with the finite $\hat {\cal I}(n)$ are therefore the $n$-bubble contribution to the sigma condensate. The results above are required, if one would like to check that after adding the non-trivial condensates to contributions with non-trivial perturbative contents, the full result of the two-point function at the next-to-leading power remains scheme-independent.

Now, instead of comparing with the perturbative computations, we would like to investigate the bare condensate itself in the $\epsilon \rightarrow 0$ limit. Defined purely in a $\mu$-independent way, we must show that by summing over $n$, the full result is independent of the $\mu$. This can be performed, if we use
\begin{align}
d(\epsilon)=\frac{1-2\epsilon}{1-\epsilon}\frac{4\pi}{g_r}-\frac{1}{1-\epsilon} \rightarrow \ln \mu^2-1\ ,
\end{align}
then after straightforward calculations, one has ($\hat g=\frac{g_r}{4\pi}\rightarrow\frac{1}{\ln \mu^2}$)
\begin{align}
& \sum_{n=0}^{\infty}\hat g^{n+1}\frac{n}{2(n+1)}\frac{d^{n+1}}{ds^{n+1}}\bigg(\frac{1}{4^s}(\mu^2)^s\bigg)_{s=0}=\frac{ \ln (\hat g \ln (4))}{2}+\frac{1}{2\hat g \ln (4)}-\frac{1}{2} \ , \\
&-\frac{d(0)}{2}\sum_{n=0}^{\infty}\hat g^{n+1}\frac{d^{n}}{ds^{n}}\bigg(\frac{1}{4^s}(\mu^2)^s\bigg)_{s=0}=\frac{1}{2\ln 4}-\frac{1}{2\hat g \ln (4)} \ , 
\end{align}
while all the divergent parts sums to
\begin{align}
&\frac{e^{\gamma_E\epsilon}}{f(\epsilon)}\sum_{n=0}^{\infty}(-1)^n\hat g^{n+1}\bigg(\frac{1}{(n+1)\epsilon^n}+\frac{n}{2(n+1)\epsilon^{n+1}}+\frac{d(\epsilon)}{2\epsilon^n}\bigg)\nonumber \\ 
&\rightarrow \frac{1}{2} (\ln (\epsilon)-\ln (\hat g)+1)+{\cal O}(\epsilon\ln \epsilon) \ .
\end{align}
Notice that after summing over $n$, the power $\frac{1}{\epsilon}$ divergences receive no enhancements, but are smeared to $\ln \epsilon$ divergences. We therefore expect that the ${\cal O}(\epsilon)$ terms we thrown way in the finite-order perturbative computations also do not receive $\frac{1}{\epsilon}\times\epsilon$ enhancements after summing to all orders. Given the above, the sum of the tail contributions reads
\begin{align}
&\frac{1}{\Gamma(1-\epsilon)}\sum_{n=0}^{\infty}\hat g^{n+1}\bigg({\cal I}_A(n)+{\cal I}_B(n)+{\cal I}_C(n)\bigg)\nonumber \\ 
&\rightarrow \frac{1}{2}\bigg(\ln \epsilon+\ln \ln 4\bigg)+\frac{1}{2\ln 4}+\sum_{n=0}^{\infty}\frac{\hat g^{n+1}}{(n+1)}\frac{d^{n+1}}{ds^{n+1}}\bigg(\frac{se^{\left(\frac{1}{\hat g}-\ln4\right)s}}{s-1}\bigg)_{s=0} \nonumber \\ 
& \sim  \frac{1}{2}\bigg(\ln \epsilon+\ln \ln 4\bigg)+\frac{1}{2\ln 4}+\int_0^{\infty} dt \frac{1}{4^t(t-1)}\ .
\end{align}
As expected, all the $\mu$ dependency in $\hat g$ cancel. On the other hand, the $\ln \epsilon$ is a reflection of the branch cut in the $\epsilon$ plan. We found that the $\pm i\pi$ ambiguity between the $\epsilon>0$ and $\epsilon<0$ choices is exactly canceled by the Borel ambiguity of the last integral
\begin{align}
\int_0^{(1+i0)\infty} dt \frac{1}{4^t(t-1)}-\int_0^{(1-i0)\infty} dt \frac{1}{4^t(t-1)}=-\frac{i\pi}{2} \ .
\end{align}
At the moment, we are not sure if this is merely a coincidence or a generic feature of dimensional regularization in the presence of renormalon ambiguities. 

On top of the tail contributions, the finite parts can be summed as
\begin{align}
\sum_{n=0}^{\infty}(-1)^n\hat g^{n+1}\hat{\cal I}(n)=-\frac{\gamma_E }{2}-\frac{1}{2} \ln \ln 4-\frac{1}{2 \ln 4}-{\cal PV}\int_0^{\infty} dt \frac{1}{4^t(t-1)} \ .
\end{align}
The sum is well defined and the principle value is not due to any ambiguity. To see this, notice that one has
\begin{align}
\sum_{n=0}^{\infty}(-1)^n\hat g^{n+1}\hat{\cal I}(n)\rightarrow \int_0^1 dz (1-z)^{-2} \hat Q(z) \ , \\
\hat Q(z)=\sqrt{z}\frac{1}{\ln \frac{1+\sqrt{z}}{1-\sqrt{z}}}-\frac{1}{\ln \frac{4}{1-z}}-(z-1)\frac{\ln\frac{4}{1-z}-1}{2\ln^2\frac{4}{1-z}} \ .
\end{align}
Introducing the Borel parameter $t$ for the inverse-logarithms, one has
\begin{align}
\int_0^1(1-z)^{-2}\hat Q(z)=\int_0^{\infty} dt \bigg(\frac{1}{2 t(t-1)(t+1)}-\frac{1}{4^t(t-1)}-\frac{1}{4^t}\frac{(t-1)}{2 t}\bigg) \ , 
\end{align}
which leads to the closed expression as above.
Thus, the full result for the bare condensate 
 obtained by summing the perturbative expansion, reads
\begin{align}
2N\langle \sigma^2\rangle_c \rightarrow 2\ln \epsilon-2\gamma_E \pm i\pi \ . 
\end{align}
Notice that all the $\ln 4$s completely cancel, but the $\gamma_E$ survives.

\bibliographystyle{apsrev4-1} 
\bibliography{bibliography}

\end{document}